%%%%%%%%%%%%%%%%%%%%
%%THIS IS A PLAIN TEX FILE. IT IS SELF CONTAINED 
%%%%%%%%%%%%%%%%%%%%%%%%%%%%%%%%%%%%%%%%%%%%%%%%%%%
%%CORRECTIONS OF THE VERSION OF AUG. 06 1998 
%%MADE IN ACCORD WITH THE GALLEY PROOFS
%%%%%%%%%%%%%%%%%%%%%%%%%%%%%%%%%%%%%%%%%%%%%%%%%%

\magnification=\magstephalf
%\magnification=\magstep1
\baselineskip=3ex
\raggedbottom
\overfullrule=0pt
\font\fivepoint=cmr5
%\headline={\hfill{\fivepoint  EHLJY 05/Jan/99}}
\input epsf.sty
\def\d{{\rm d}}

\def\I{{\cal I}}
\def\sr{{\cal R}}
\def\R{{\bf R}}
\def\S{{\cal S}}
\def\simt{\mathrel{\rlap{\hbox{$\sim$}}\raise.9ex\hbox{{\fivepoint
$\,$T}}}}
\def\sima{\mathrel{\rlap{\hbox{$\sim$}}\raise.95ex\hbox{{\fivepoint
$\,$A}}}}
\def\lanbox{\hbox{$\, \vrule height 0.25cm width 0.25cm depth 0.01cm 
\,$}}
\def\uprho{\raise1pt\hbox{$\rho$}}
\def\mfr#1/#2{\hbox{${{#1} \over {#2}}$}}

\font\subsubt=cmtt10 scaled \magstep1
\font\subt=cmbx10 scaled \magstep1
\font\tit=cmbx10 scaled \magstep2 
\font\fivepoint=cmr5
\font\sixpoint=cmr6
\font\ninepoint=cmr9
\font\sevenpoint=cmr7

\catcode`@=11
\def\eqalignii#1{\,\vcenter{\openup1\jot \m@th
\ialign{\strut\hfil$\displaystyle{##}$&
        $\displaystyle{{}##}$\hfil&
        $\displaystyle{{}##}$\hfil\crcr#1\crcr}}\,}
\catcode`@=12

%this is for automatic equation numbering
\def\eqlbl#1{\global\advance\equno by 1
  \global\edef#1{{\number\chno.\number\equno  }}
  (\number\chno.\number\equno  )}
%\eqno\eqlbl\fermat This is an example of usage
\newcount\chno    \chno=0
\newcount\equno   \equno=0

%%%%%%%%%%%%%%%%%%%%%%%%%%%%%%%%%%

\centerline{\tit THE PHYSICS AND MATHEMATICS OF}
\medskip
\centerline{\tit THE SECOND LAW OF THERMODYNAMICS}
\bigskip
\bigskip

\centerline{Elliott H. Lieb\footnote{$^*$}{\sixpoint Work partially
supported by U.S. National Science Foundation grant PHY95-13072A01.} }
\centerline{\it Departments of Physics and Mathematics, Princeton 
University}
\centerline{\it Jadwin Hall,  P.O. Box 708, Princeton, NJ  08544, USA}
\bigskip

\centerline{Jakob Yngvason
\footnote{$^{**}$}{\sixpoint Work partially
supported by the Adalsteinn Kristjansson Foundation, 
University of Iceland.} }
\centerline{\it Institut f\"ur Theoretische Physik, Universit\"at Wien,}
\centerline{\it Boltzmanngasse 5, A 1090 Vienna, Austria}
\footnote{}{\baselineskip=0.6\baselineskip\hskip -\parindent\sixpoint
\copyright  1997 by the authors.
Reproduction of this article, by any means, is permitted for 
non-commercial
purposes.\par}
\bigskip
\bigskip
{\narrower\smallskip\noindent

\bigskip\bigskip\noindent 

{\subt Abstract:} The essential postulates of classical thermodynamics 
are formulated, from which the second law is deduced as the principle 
of increase of entropy in irreversible adiabatic processes that take 
one equilibrium state to another.  The entropy constructed here is 
defined only for equilibrium states and no attempt is made to define 
it otherwise.  Statistical mechanics does not enter these 
considerations.  One of the main concepts that makes everything work 
is the comparison principle (which, in essence, states that given any 
two states of the same chemical composition at least one is 
adiabatically accessible from the other) and we show that it can be 
derived from some assumptions about the pressure and thermal 
equilibrium.  Temperature is derived from entropy, but at the start 
not even the concept of `hotness' is assumed.  Our formulation offers 
a certain clarity and rigor that goes beyond most textbook discussions 
of the second law.  }
\bigskip
\bigskip
\bigskip
1998 PACS: \  05.70.-a
\smallskip
Mathematical Sciences Classification (MSC) 1991 and 2000: 80A05,\ 80A10
%\centerline{\tit (DRAFT)}

\bigskip\bigskip\bigskip\bigskip\bigskip\bigskip\bigskip\bigskip\bigskip
This paper is scheduled to appear in Physics Reports {\bf 310}, 1-96 (1999)
\vfill\eject

{\vbox
{\ninepoint 
\baselineskip=0.9\baselineskip

\noindent
I. INTRODUCTION
\item{A.} The basic Questions\dotfill 3
\item{B.} Other approaches\dotfill 6
\item{C.} Outline of the paper\dotfill 10
\item{D.} Acknowledgements\dotfill 11\break

\noindent 
II. ADIABATIC ACCESSIBILITY AND CONSTRUCTION OF ENTROPY   
\item{A.} Basic concepts\dotfill 12
\item\item{1.} Systems and their state spaces\dotfill 13
\item\item{2.} The order relation\dotfill 16
\item{B.}  The entropy principle\dotfill 18
\item{C.} Assumptions about the order relation\dotfill 20
\item{D.} The construction of entropy for a single system\dotfill 23
\item{E.} Construction of a universal entropy in the absence
           of mixing\dotfill 27
\item{F.} Concavity of entropy\dotfill 30
\item{G.} Irreversibility and Carath\'eodory's principle\dotfill 32
\item{H.} Some further results on uniqueness\dotfill 33\break

\noindent 
III. SIMPLE SYSTEMS
\item{{\phantom{A.}}} Preface\dotfill 36
\item{A.} Coordinates for simple systems\dotfill 37
\item{B.} Assumptions about simple systems\dotfill 39
\item{C.} The geometry of forward sectors\dotfill 42\break 

\noindent 
IV. THERMAL EQUILIBRIUM
\item{A.} Assumptions about thermal contact\dotfill 51
\item{B.} The comparison principle in compound systems\dotfill 55
\item\item{1.} Scaled products of a single simple system\dotfill 55
\item\item{2.} Products of different simple systems\dotfill 56
\item{C.} The role of transversality\dotfill 59\break

%\vfill\eject
\noindent 
V. TEMPERATURE AND ITS PROPERTIES
\item{A.} Differentiability of entropy and the definition of 
          temperature\dotfill 62
\item{B.} The geometry of isotherms and adiabats\dotfill 68
\item{C.} Thermal equilibrium and the uniqueness of entropy\dotfill
             69\break

\noindent 
VI. MIXING AND CHEMICAL REACTIONS
\item{A.} The difficulty in fixing entropy constants\dotfill 72
\item{B.} Determination of additive entropy constants\dotfill 73\break

\noindent
VII.  SUMMARY AND CONCLUSIONS\dotfill 83\break

\noindent
LIST OF SYMBOLS \dotfill 88\break

\noindent
INDEX OF TECHNICAL TERMS
\dotfill 89\break

\noindent
REFERENCES\dotfill 91\break

}}

\vfill\eject
%%%%%%%%%%%%%%%%%%%%%%%%%%%%%%%%%%%%
\noindent
{\tit I. INTRODUCTION}
\bigskip

The second law of thermodynamics is, without a doubt, one of the most 
perfect laws in physics.  Any {\it reproducible} violation of it, 
however small, would bring the discoverer great riches as well as a 
trip to Stockholm.  The world's energy problems would be solved at one 
stroke.  It is not possible to find any other law (except, perhaps, 
for super selection rules such as charge conservation) for which a 
proposed violation would bring more skepticism than this one.  Not 
even Maxwell's laws of electricity or Newton's law of gravitation are 
so sacrosanct, for each has measurable corrections coming from quantum 
effects or general relativity.  The law has caught the attention of 
poets and philosophers and has been called the greatest scientific 
achievement of the nineteenth century.  Engels disliked it, for it 
supported opposition to dialectical materialism, while Pope Pius XII 
regarded it as proving the existence of a higher being (Bazarow, 1964, 
Sect.  20).

\bigskip\noindent
{\subt A. The basic questions}
\bigskip

In this paper we shall attempt to formulate the essential elements of 
{\it classical } thermodynamics of equilibrium states and deduce from 
them the second law as the principle of the increase of entropy.  
`Classical' means that there is {\it no mention of statistical 
mechanics here} and `equilibrium' means that we deal only with states 
of systems in equilibrium and do not attempt to define quantities such 
as entropy and temperature for systems not in equilibrium.  This is 
not to say that we are concerned only with `thermostatics' because, as 
will be explained more fully later, arbitrarily violent processes are 
allowed to occur in the passage from one equilibrium state to another.

Most students of physics regard the subject as essentially perfectly 
understood and finished, and concentrate instead on the statistical 
mechanics from which it ostensibly can be derived.  But many will 
admit, if pressed, that thermodynamics is something that they are sure 
that someone else understands and they will confess to some misgiving 
about the logic of the steps in traditional presentations that lead to 
the formulation of an entropy function.  If classical thermodynamics 
is the most perfect physical theory it surely deserves a solid, 
unambiguous foundation free of little pictures involving unreal Carnot 
cycles and the like.  [For examples of `un-ordinary' Carnot cycles see 
(Truesdell and Bharatha 1977, p.~48).]

There are two aims to our presentation.  One is frankly pedagogical, 
i.e., to formulate the foundations of the theory in a clear and 
unambiguous way.  The second is to formulate equilibrium 
thermodynamics as an `ideal physical theory', which is to say a theory 
in which there are well defined mathematical constructs and well 
defined rules for translating physical reality into these constructs; 
having done so the mathematics then grinds out whatever answers it can 
and these are then translated back into physical statements.  The 
point here is that while `physical intuition' is a useful guide for 
formulating the mathematical structure and may even be a source of 
inspiration for constructing mathematical proofs, it should not be 
necessary to rely on it once the initial `translation' into 
mathematical language has been given.  These goals are not new, of 
course; see e.g., (Duistermaat, 1968), (Giles, 1964, Sect.  1.1) and 
(Serrin, 1986, Sect.  1.1).

Indeed, it seems to us that many formulations of thermodynamics, 
including most textbook presentations, suffer from mixing the physics 
with the mathematics.  Physics refers to the real world of experiments 
and results of measurement, the latter quantified in the form of 
numbers.  Mathematics refers to a logical structure and to rules of 
calculation; usually these are built around numbers, but not always.  
Thus, mathematics has two functions: one is to provide a transparent 
logical structure with which to view physics and inspire experiment.  
The other is to be like a mill into which the miller pours the grain 
of experiment and out of which comes the flour of verifiable 
predictions.  It is astonishing that this paradigm works to perfection 
in thermodynamics.  (Another good example is Newtonian mechanics, in 
which the relevant mathematical structure is the calculus.)  Our 
theory of the second law concerns the mathematical structure, 
primarily.  As such it starts with some axioms and proceeds with rules 
of logic to uncover some non-trivial theorems about the existence of 
entropy and some of its properties.  We do, however, explain how 
physics leads us to these particular axioms and we explain the 
physical applicability of the theorems.

As noted in I.C below, we have a total of 15 axioms, which might seem
like a lot. We can assure the reader that any other mathematical
structure that derives entropy with minimal assumptions will have at
least that many, and usually more. (We could roll several axioms into
one, as others often do, by using sub-headings, e.g., our A1-A6 might
perfectly well be denoted by A1(i)-(vi).) The point is that we leave
nothing to the imagination or to silent agreement; it is all laid out.

It must also be emphasized that our desire to clarify the structure of
classical equilibrium thermodynamics is not merely pedagogical and not
merely nit-picking. If the law of entropy increase is ever going to be
derived from statistical mechanics---a goal that has so far eluded the
deepest thinkers---then it is important to be absolutely clear about what
it is that one wants to derive.

Many attempts have been made in the last century and a half to
formulate the second law precisely and to quantify it by means of an
entropy function. Three of these formulations are classic (Kestin,
1976), (see also Clausius (1850), Thomson (1849)) and they can be paraphrased as 
follows:
\smallskip

{\sl Clausius:\/} No process is possible, the sole result of which is
that heat is transferred from a body to a hotter one.

{\sl Kelvin (and Planck):\/} No process is possible, the sole result 
of  which is that a body is cooled and work is done.

{\sl Carath\'eodory:\/} In any neighborhood of any state there are
states that cannot be reached from it by an adiabatic process.
\smallskip

The crowning glory of thermodynamics is the quantification  of these
statements by means of a precise, measurable quantity called entropy. 
There are two kinds of  problems, however. One is to give a precise
meaning to the words above. What is `heat'? What is `hot' and `cold'?
What is `adiabatic'?  What is a `neighborhood'? Just about the only word
that is relatively unambiguous is `work' because it comes from
mechanics. 

The second sort of problem involves the rules of logic that lead from
these statements to an entropy. Is it really necessary to draw pictures,
some of which are false, or at least not self evident?  What are all the
hidden assumptions that enter the derivation of entropy? For instance,
we all know that discontinuities can and do occur at phase transitions,
but almost every presentation of classical thermodynamics is based on
the differential calculus (which presupposes continuous derivatives),
especially (Carath\'eodory, 1925) and (Truesdell-Bharata, 1977, p.xvii).

We note, in passing, that the Clausius, Kelvin-Planck and 
Carath\'eodory formulations are all assertions about {\it impossible} 
processes.  Our formulation will rely, instead, mainly on assertions 
about {\it possible} processes and thus is noticeably different.  At 
the end of Section VII, where everything is succintly summarized, the 
relationship of these approaches is discussed.  This discussion is 
left to the end because it it cannot be done without first presenting 
our results in some detail.  Some readers might wish to start by 
glancing at Section VII.

Of course we are neither the first nor, presumably, the last to present
a derivation of the second law (in the sense of an entropy principle)
that pretends to remove all confusion and, at the same time, to achieve
an unparalleled  precision of logic and structure. Indeed, such attempts
have multiplied in the past three or four decades.  These other
theories, reviewed in Sect. I.B, appeal to their creators as much as
ours does to us and we must therefore conclude that ultimately a
question of `taste' is involved.

It is not easy to classify other approaches to the problem that
concerns us.  We shall attempt to do so briefly, but first let us state
the problem clearly.  Physical systems have certain states (which
always mean equilibrium states in this paper) and, by means of certain
actions, called {\it adiabatic processes}, it is possible to change the
state of a system to some other state.  (Warning: The word `adiabatic'
is used in several ways in physics. Sometimes it means `slow and
gentle', which might conjure up the idea of a quasi-static process, but
this is certainly not our intention.  The usage we have in the back of
our minds is `without exchange of heat', but we shall avoid defining
the word `heat'.  The operational meaning of `adiabatic' will be
defined later on, but for now the reader should simply accept it as
singling out a particular class of processes about which certain
physically interesting statements are going to be made.) Adiabatic
processes do not have to be very gentle, and they certainly do not have
to be describable by a curve in the space of equilibrium states. One is
allowed, like the gorilla in a well-known advertisement for luggage, to
jump up and down on the system and even dismantle it temporarily,
provided the system returns to some equilibrium state at the end of the
day.  In thermodynamics, unlike mechanics, not all conceivable
transitions are adiabatic and it is a nontrivial problem to
characterize the allowed transitions.  We shall characterize them as
transitions that have no {\it net} effect on other systems except that
energy has been exchanged with a mechanical source.  The truly
remarkable fact, which has many consequences, is that for every system
there is a function, $S$, on the space of its (equilibrium) states,
with the property that one can go adiabatically from a state $X$ to a
state $Y$ if and only if  $S(X) \leq S(Y)$. This, in essence,  is the
`entropy principle' (EP) (see subsection II.B).

The $S$ function can clearly be multiplied by an arbitrary constant 
and still continue to do its job, and thus it is not at all obvious 
that the function $S_1$ for system $1$ has anything to do with the 
function $S_2$ for system $2$.  The second remarkable fact is that the 
$S$ functions for all the thermodynamic systems in the universe can be 
simultaneously calibrated (i.e., the multiplicative constants can be 
determined) in such a way that the entropies are {\it additive}, i.e., 
the $S$ function for a compound system is obtained merely by adding 
the $S$ functions of the individual systems, $S_{1,2} = S_1+S_2$.  
(`Compound' does not mean chemical compound; a compound system is just 
a collection of several systems.)  To appreciate this fact it is 
necessary to recognize that the systems comprising a compound system 
can interact with each other in several ways, and therefore the 
possible adiabatic transitions in a compound are far more numerous 
than those allowed for separate, isolated systems.  Nevertheless, the 
increase of the function $S_1+S_2$ continues to describe the adiabatic 
processes exactly---neither allowing more nor allowing less than 
actually occur.  The statement $S_1(X_1)+S_2(X_2)\leq 
S_1(X'_1)+S_2(X'_2)$ does not require $S_1(X_1)\leq S_1(X'_1)$.

The main problem, from our point of view, is this: What properties of 
adiabatic processes permit us to construct such a function?  To what 
extent is it unique?  And what properties of the interactions of 
different systems in a compound system result in additive entropy 
functions?

The existence of an entropy function can be discussed in principle, as 
in Section II, without parametrizing the equilibrium states by 
quantities such as energy, volume, etc..  But it is an additional fact 
that when states are parametrized in the conventional ways then the 
derivatives of $S$ exist and contain all the information about the 
equation of state, e.g., the temperature $T$ is defined by $\partial 
S(U,V)/ \partial U|_V^{\phantom Y} = 1/T$.

In our approach to the second law temperature is never formally 
invoked until the very end when the differentiability of $S$ is 
proved---not even the more primitive relative notions of `hotness' and 
`coldness' are used.  The priority of entropy is common in statistical 
mechanics and in some other approaches to thermodynamics such as in 
(Tisza, 1966) and (Callen, 1985), but the elimination of hotness and 
coldness is not usual in thermodynamics, as the formulations of 
Clausius and Kelvin show.  The laws of thermal equilibrium (Section 
V), in particular the zeroth law of thermodynamics, do play a crucial 
role for us by relating one system to another (and they are ultimately 
responsible for the fact that entropies can be adjusted to be 
additive), but thermal equilibrium is only an equivalence relation 
and, in our form, it is not a statement about hotness.  It seems to us 
that temperature is far from being an `obvious' physical quantity.  It 
emerges, finally, as a derivative of entropy, and unlike quantities in 
mechanics or electromagnetism, such as forces and masses, it is not 
vectorial, i.e., it cannot be added or multiplied by a scalar.  Even 
pressure, while it cannot be `added' in an unambiguous way, can at 
least be multiplied by a scalar.  (Here, we are not speaking about 
changing a temperature scale; we mean that once a scale has been 
fixed, it does not mean very much to multiply a given temperature, 
e.g., the boiling point of water, by the number 17.  Whatever meaning 
one might attach to this is surely not independent of the chosen 
scale.  Indeed, is $T$ the right variable or is it $1/T$?  In 
relativity theory this question has led to an ongoing debate about the 
natural quantity to choose as the fourth component of a four-vector.  
On the other hand, it does mean something unambiguous, to multiply the 
pressure in the boiler by 17.  Mechanics dictates the meaning.)

Another mysterious quantity is `heat'.  No one has ever seen heat, nor 
will it ever be seen, smelled or touched.  Clausius wrote about ``the 
kind of motion we call heat", but thermodynamics---either practical or 
theoretical---does not rely for its validity on the notion of 
molecules jumping around.  There is no way to measure heat flux 
directly (other than by its effect on the source and sink) and, while 
we do not wish to be considered antediluvian, it remains true that 
`caloric' accounts for physics at a macroscopic level just as well as 
`heat' does.  The reader will find no mention of heat in our 
derivation of entropy, except as a mnemonic guide.

To conclude this very brief outline of the main conceptual points, the
concept of {\it convexity} has to be mentioned. It is well known, as
Gibbs (Gibbs 1928), Maxwell and others emphasized, that thermodynamics
without convex functions (e.g., free energy per unit volume as a
function of density) may lead to unstable systems.  (A good discussion
of convexity is in (Wightman, 1979).) Despite this fact, convexity is
almost invisible in most fundamental approaches to the second law.  In
our treatment it is {\it essential} for the description of simple
systems in Section III, which are the building blocks of
thermodynamics.

The concepts and goals we have just enunciated will be discussed in
more detail in the following sections.  The reader who impatiently
wants a quick survey of our results can jump to Section VII where it
can be found in capsule form. We also draw the readers attention to the
article (Lieb-Yngvason 1998), where a summary of this work appeared.
Let us now turn to a brief discussion of other modes of thought about
the questions we have raised.

\bigskip\bigskip\noindent
{\subt B. Other approaches}
\bigskip

The simplest solution to the problem of the foundation of 
thermodynamics is perhaps that of Tisza (1966), and expanded by Callen 
(1985) (see also (Guggenheim, 1933)), who, following the tradition of 
Gibbs (1928), postulate the existence of an additive entropy function 
from which all equilibrium properties of a substance are then to be 
derived.  This approach has the advantage of bringing one quickly to 
the applications of thermodynamics, but it leaves unstated such 
questions as: What physical assumptions are needed in order to insure 
the existence of such a function?  By no means do we wish to minimize 
the importance of this approach, for the manifold implications of 
entropy are well known to be non-trivial and highly important 
theoretically and practically, as Gibbs was one of the first to show 
in detail in his great work (Gibbs, 1928).

Among the many foundational works on the existence of entropy, the 
most relevant for our considerations and aims here are those that we 
might, for want of a better word, call `order theoretical' because the 
emphasis is on the derivation of entropy from postulated properties of 
adiabatic processes.  This line of thought goes back to Carath\'eodory 
(1909 and 1925), although there are some precursors (see Planck, 1926) 
and was particularly advocated by (Born, 1921 and 1964).  This basic 
idea, if not Carath\'eodory's implementation of it with differential 
forms, was developed in various mutations in the works of Landsberg 
(956), Buchdahl (1958, 1960, 1962, 1966), Buchdahl and Greve (1962), 
Falk and Jung (1959), Bernstein (1960), Giles (964), Cooper (1967), 
Boyling, (1968, 1972), Roberts and Luce (1968), Duistermaat (1968), 
Hornix (1968), Rastall (1970), Zeleznik (1975) and Borchers (1981).  
The work of Boyling (1968, 1972), which takes off from the work of 
Bernstein (1960) is perhaps the most direct and rigorous expression of 
the original Carth\'eodory idea of using differential forms.  See also 
the discussion in Landsberg (1970).

Planck (1926) criticized some of Carath\'eodory's work for not
identifying processes that are not adiabatic. He suggested basing
thermodynamics on the fact that `rubbing' is an adiabatic process that
is not reversible, an idea he already had in his 1879 dissertation.
{}From this it follows that while one can undo a rubbing operation by
some means, one cannot do so adiabatically.  We derive  this principle
of Planck from our axioms. It is very convenient because it means that
in an adiabatic process one can effectively add as much `heat'
(colloquially speaking) as one wishes, but the one thing one cannot do
is subtract heat, i.e., use a `refrigerator'.

Most authors introduce the idea of an `empirical temperature', and 
later derive the absolute temperature scale.  In the same vein they 
often also introduce an `empirical entropy' and later derive a 
`metric', or additive, entropy, e.g., (Falk and Jung, 1959) and 
(Buchdahl, 1958, et seq., 1966), (Buchdahl and Greve, 1962), (Cooper, 
1967).  We avoid all this; one of our results, as stated above, is the 
derivation of absolute temperature directly, without ever mentioning 
even `hot' and `cold'.

One of the key concepts that is eventually needed, although it is not 
obvious at first,  is that of the comparison principle (or
hypothesis), (CH). It concerns classes of thermodynamic states and 
asserts that for any two states $X$ and $Y$ within a class one can either
go {\it adiabatically} from $X$ to $Y$, which we write as
$$
X \prec Y,
$$ 
(pronounced ``$X$ precedes $Y$" or ``$Y$ follows $X$") or else one can
go from $Y$ to $X$, i.e., $Y \prec X$.  Obviously, this is not always
possible (we cannot transmute lead into gold, although we {\it can}
transmute hydrogen plus oxygen into water), so we would like to be able
to break up the universe of states into equivalence classes, inside
each of which the hypothesis holds. It turns out that the key
requirement for an equivalence relation is  that if  $X\prec Y$ and
$Z\prec Y$ then  either $X\prec Z$ or  $Z\prec X$.  Likewise, if
$Y\prec X$ and $Y\prec Z$ by then either $X \prec Z$ or $Z\prec X$.
We  find this first clearly stated in Landsberg (1956) and it is also
found in one form or another in many places, see e.g., (Falk and Jung,
1959), (Buchdahl, 1958, 1962), (Giles, 1964).  However, all authors,
except for Duistermaat (1968), seem to take this postulate for granted
and do not feel obliged to obtain it from something else. One of the
central points in our work is to {\it derive} the comparison
hypothesis. This is discussed further below.

The formulation of the second law of thermodynamics that is closest to 
ours is that of Giles (Giles, 1964).  His book is full of deep 
insights and we recommend it highly to the reader.  It is a classic 
that does not appear to be as known and appreciated as it should.  His 
derivation of entropy from a few postulates about adiabatic processes 
is impressive and was the starting point for a number of further 
investigations.  The overlap of our work with Giles's is only partial 
(the foundational parts, mainly those in our section II) and where 
there is overlap there are also differences.

To define the entropy of a state, the starting point in both 
approaches is to let a process that by itself would be adiabatically 
impossible work against another one that is possible, so that the 
total process is adiabatically possible.  The processes used by us and 
by Giles are, however, different; for instance Giles uses a fixed 
external calibrating system, whereas we define the entropy of a state 
by letting a system interact with a copy of itself.  ( According to 
R.\ E.\ Barieau (quoted in (Hornix, 1967-1968)) Giles was unaware of 
the fact that predecessors of the idea of an external entropy meter 
can be discerned in (Lewis and Randall, 1923).)  To be a bit more 
precise, Giles uses a standard process as a reference and counts how 
many times a reference process has to be repeated to counteract some 
multiple of the process whose entropy (or rather `irreversibility') is 
to be determined.  In contrast, we construct the entropy function for 
a single system in terms of the amount of substance in a reference 
state of `high entropy' that can be converted into the state under 
investigation with the help of a reference state of `low entropy'.  
(This is reminiscent of an old definition of heat by Laplace and 
Lavoisier (quoted in (Borchers, 1981)) in terms of the amount of ice 
that a body can melt.)  We give a simple formula for the entropy; 
Giles's definition is less direct, in our view.  However, when we 
calibrate the entropy functions of different systems with each other, 
we do find it convenient to use a third system as a `standard' of 
comparison.

Giles' work and ours use very little of the calculus.  Contrary to 
almost all treatments, and contrary to the assertion 
(Truesdell-Bharata, 1977) that the differential calculus is the 
appropriate tool for thermodynamics, we and he agree that entropy and 
its essential properties can best be described by maximum principles 
instead of equations among derivatives.  To be sure, real analysis 
does eventually come into the discussion, but only at an advanced 
stage (Sections III and V in our treatment).

In Giles, too, temperature appears as a totally derived quantity, but 
Giles's derivation requires some assumptions, such as 
differentiability of the entropy.  We prove the required 
differentiability from natural assumptions about the pressure.

Among the differences, it can be mentioned that the `cancellation 
law', which plays a key role in our proofs, is taken by Giles to be an 
axiom, whereas we derive it from the assumption of `stability', which 
is common to both approaches (see Section II for definitions).

The most important point of contact, however, and at the same time the 
most significant difference, concerns the comparison hypothesis which, 
as we emphasized above, is a concept that plays an essential role, 
although this may not be apparent at first.  This hypothesis serves to 
divide the universe nicely into equivalence classes of mutually 
accessible states.  Giles takes the comparison property as an axiom 
and does not attempt to justify it from physical premises.  The main 
part of our work is devoted to just that justification, and to inquire 
what happens if it is violated.  (There is also a discussion of this 
point in (Giles, 1964, Sect 13.3) in connection with hysteresis.)  To 
get an idea of what is involved, note that we can easily go 
adiabatically from cold hydrogen plus oxygen to hot water and we can 
go from ice to hot water, but can we go either from the cold gases to 
ice or the reverse---as the comparison hypothesis demands?  It would 
appear that the only real possibility, if there is one at all, is to 
invoke hydrolysis to dissociate the ice, but what if hydrolysis did 
not exist?  In other examples the requisite machinery might not be 
available to save the comparison hypothesis.  For this reason we 
prefer to derive it, when needed, from properties of `simple systems' 
and not to invoke it when considering situations involving variable 
composition or particle number, as in Section VI.

Another point of difference is the fact that convexity is central to 
our work.  Giles mentions it, but it is not central in his work 
perhaps because he is considering more general systems than we do.  To 
a large extent convexity eliminates the need for explicit topological 
considerations about state spaces, which otherwise has to be put in 
`by hand'.

Further developments of the Giles' approach are in (Cooper, 1967), 
(Roberts and Luce, 1968) and (Duistermaat, 1968).  Cooper assumes the 
existence of an empirical temperature and introduces topological 
notions which permits certain simplifications.  Roberts and Luce have 
an elegant formulation of the entropy principle, which is 
mathematically appealing and is based on axioms about the order 
relation, $\prec$, (in particular the comparison principle, which they 
call conditional connectedness), but these axioms are not physically 
obvious, especially axiom 6 and the comparison hypothesis.  
Duistermaat is concerned with general statements about morphisms of 
order relations, thermodynamics being but one application.

A line of thought that is entirely different from the above starts 
with Carnot (1824) and was amplified in the classics of Clausius and 
Kelvin (cf.\  (Kestin, 1976)) and many others.  It has dominated most 
textbook presentations of thermodynamics to this day.  The central 
idea concerns cyclic processes and the efficiency of heat engines; 
heat and empirical temperature enter as primitive concepts.  Some of 
the modern developments along these lines go well beyond the study of 
equilibrium states and cyclic processes and use some sophisticated 
mathematical ideas.  A representative list of references is Arens 
(1963), Coleman and Owen (1974, 1977), Coleman, Owen and Serrin 
(1981), Dafermos (1979), Day (1987, 1988), Feinberg and Lavine (1983), 
Green and Naghdi (1978), Gurtin (1975), Man (1989), Owen (1984), 
Pitteri (1982), Serrin (1979, 1983, 1986), Silhavy (1997), Truesdell 
and Bharata (1977), Truesdell (1980, 1984).  Undoubtedly this approach 
is important for the practical analysis of many physical systems, but 
we neither analyze nor take a position on the validity of the claims 
made by its proponents.  Some of these are, quite frankly, highly 
polemical and are of two kinds: claims of mathematical rigor and 
physical exactness on the one hand and assertions that these qualities 
are lacking in other approaches.  See, for example, Truesdell's 
contribution in (Serrin, 1986, Chapter 5).  The chief reason we omit 
discussion of this approach is that it does not directly address the 
questions we have set for ourselves.  Namely, using only the existence 
of equilibrium states and the existence of certain processes that take 
one into another, when can it be said that the list of allowed 
processes is characterized {\it exactly} by the increase of an entropy 
function?

Finally, we  mention an interesting recent paper by Macdonald
(1995) that falls in neither of the two categories described
above. In this paper  \lq heat\rq\ and \lq reversible processes\rq\ are
among the primitive concepts and the existence of reversible processes
linking any two states of a system is taken as a postulate.  Macdonald
gives a simple definition of entropy of a state in terms of the maximal
amount of heat, extracted from an infinite reservoir, that the system
absorbs in processes terminating in the given state. The reservoir thus
plays the role of an entropy meter. The further development of the
theory along these lines, however, relies on unstated assumptions about
differentiability of the so defined entropy that are not entirely
obvious.

%\vfill\eject
\bigskip\noindent
{\subt C. Outline of the paper}
\bigskip

In Section II we formally introduce the relation $\prec$ and explain 
it more fully, but it is to be emphasized, in connection with what was 
said above about an ideal physical theory, that $\prec$ has a well 
defined mathematical meaning independent of the physical context in 
which it may be used.  The concept of an entropy function, which 
characterizes this accessibility relation, is introduced next; at the 
end of the section it will be shown to be unique up to a trivial 
affine transformation of scale.  We show that the existence of such a 
function is {\it equivalent} to certain simple properties of the 
relation $\prec$, which we call axioms A1 to A6 and the `hypothesis' 
CH. Any formulation of thermodynamics must implicitly contain these 
axioms, since they are equivalent to the entropy principle, and it is 
not surprising that they can be found in Giles, for example.  We do 
believe that our presentation has the virtue of directness and 
clarity, however.  We give a simple formula for the entropy, entirely 
in terms of the relation $\prec$ without invoking Carnot cycles or any 
other gedanken experiment.  Axioms A1 to A6 are highly plausible; it 
is CH (the comparison hypothesis) that is not obvious but is {\it 
crucial} for the existence of entropy.  We call it a hypothesis rather 
than an axiom because our ultimate goal is to derive it from some 
additional axioms.  In a certain sense it can be said that the rest of 
the paper is devoted to {\it deriving} the comparison hypothesis from 
plausible assumptions.  The content of Section II, i.e., the 
derivation of an entropy function, stands on its own feet; the 
implementation of it via CH is an independent question and we feel it 
is pedagogically significant to isolate the main input in the 
derivation from the derivation itself.

Section III introduces one of our most novel contributions.  We {\it 
prove } that comparison holds for the states inside certain systems 
which we call {\it simple systems}.  To obtain it we need a few new 
axioms, S1 to S3.  These axioms are mainly about {\it mechanical} 
processes, and not about the entropy.  In short, our most important 
assumptions concern the continuity of the generalized pressure and the 
existence of irreversible processes.  Given the other axioms, the 
latter is equivalent to Carath\'eodory's principle.

The comparison hypothesis, CH, does not concern simple systems alone, 
but also their products, i.e., compound systems composed of possibly 
interacting simple systems.  In order to compare states in different 
simple systems (and, in particular, to calibrate the various entropies 
so that they can be added together) the notion of a {\it thermal join} 
is introduced in Section IV. This concerns states that are usually 
said to be in thermal equilibrium, but we emphasize that temperature 
is not mentioned.  The thermal join is, by assumption, a simple system 
and, using the zeroth law and three other axioms about the thermal 
join, we reduce the comparison hypothesis among states of {\it 
compound systems} to the previously derived result for simple systems.  
This derivation is another novel contribution.  With the aid of the 
thermal join we can prove that the multiplicative constants of the 
entropies of all systems can be chosen so that entropy is additive, 
i.e., the sum of the entropies of simple systems gives a correct 
entropy function for compound systems.  This entropy correctly 
describes all adiabatic processes in which there is no change of the 
constituents of compound systems.  What remains elusive are the 
additive constants, discussed in Section VI. These are important when 
changes (due to mixing and chemical reactions) occur.

Section V establishes the continuous differentiability of the
entropy and defines inverse temperature as the derivative of the entropy
with respect to the energy---in the usual way. No new assumptions are
needed here. The fact that the entropy of a simple system 
is determined uniquely by its adiabats and
isotherms is also proved here.

In Section VI we discuss  the vexed question of comparing
states of systems that differ in constitution or in quantity of matter.
How can the entropy of a bottle of water be compared with  the sum of
the entropies of a container of hydrogen and a container of oxygen? To
do so requires being able to transform one into the other, but this may
not be easy to do reversibly. The usual theoretical underpinning here
is the use of semi-permeable membranes in a `van't Hoff box' but such
membranes are usually far from perfect physical objects, if they exist
at all. We examine in detail just how far one can go in
determining the {\it additive} constants for the entropies of different
systems in the the real world in which perfect semi-permeable 
membranes do not exist.

In Section VII we collect all our axioms together and
summarize our results briefly.
%\vfill\eject

\bigskip\noindent
{\subt D. Acknowledgements}
\bigskip

We are deeply indebted to Jan Philip Solovej for many useful discussions
and important insights, especially in regard to Sections III and VI. 
Our thanks also go to Fredrick Almgren for helping us understand convex
functions, to Roy Jackson, Pierluigi Contucci, Thor Bak and Bernhard
Baumgartner for critically reading our manuscript and to Martin Kruskal
for emphasizing the importance of Giles' book to us.  We thank Robin 
Giles for a thoughtful and detailed review with many helpful
comments. We thank John C. Wheeler for a clarifying correspondence about
the relationship  between adiabatic processes, as usually understood,
and our definition of adiabatic accessibility.  Some of the rough spots
in our story were pointed out to us by various people during various
public lectures we gave, and that is also very much appreciated.

A significant part of this work was carried out at Nordita in 
Copenhagen and at the Erwin Schr\"odinger Institute in Vienna;  we are
grateful for their hospitality and support.

%%%%%%%%%%%%%%%%%%%%%%%%%%%%%%%%%%%%%%%%%%%%
\vfill\eject
%%%%%%%%%%%%%%%%%%%%
\noindent
\leftline {\tit II. ADIABATIC ACCESSIBILITY } \smallskip
\leftline {\tit \phantom{Ix}\enspace AND CONSTRUCTION OF ENTROPY }
\bigskip

Thermodynamics concerns systems, their states and an order relation
among these states.  The order relation is that of {\bf adiabatic
accessibility}, which, physically, is defined by processes whose only
net effect on the surroundings is exchange of energy with a mechanical
source.  The glory of classical thermodynamics is that there always is
an {\it additive}   function, called {\bf entropy}, on the state space
of any system, that {\it exactly} describes the order relation in terms
of the increase of entropy.

Additivity is very important physically and is certainly not obvious;
it tells us that the entropy of a compound system composed of two
systems that can interact and exchange energy with each other is the
sum of the individual entropies. This means that the pairs of states
accessible from a given pair of states, which is a far larger set than
merely the pairs individually accessible by the systems in isolation,
is given by studying the sum of the individual entropy functions. This
is even more surprising when we consider  that the individual entropies
each have undetermined multiplicative constants; there is a way to
adjust, or calibrate the constants in such a way that the sum gives the
correct result for the accessible states---and  this can be done once
and for all so that the same calibration works for all possible pairs
of systems.  Were additivity to fail we would have to rewrite the steam
tables every time a new steam engine is invented.

The other important point about entropy, which is often overlooked, is
that entropy not only increases, but entropy also tells us exactly which
processes are adiabatically possible in any given system; states of high
entropy in a system are {\it always } accessible from  states of lower
entropy. As we shall see this is generally true but it could conceivably
fail when there are chemical reactions or mixing, as discussed in
Section VI.  

In this section we begin by defining these basic concepts more precisely,
and then we present the entropy principle.  Next, we introduce certain
axioms, A1-A6, relating the concepts.  All these axioms are completely
intuitive. However, one other assumption---which we call the {\it
comparison hypothesis}---is needed for the construction of entropy.  It 
is
not at all obvious physically, but it is an essential part of 
conventional
thermodynamics.  Eventually, in Sections III and IV, this hypothesis will
be {\it derived} from some more detailed physical considerations. For the
present, however, this hypothesis will be assumed  and, using it, the
existence of an entropy function will be proved. We  also discuss  the 
extent to which the entropy function is uniquely determined by the order
relation; the comparison hypothesis plays a key role here.

The existence of an entropy function is equivalent to axioms A1-A6 in
conjunction with CH, neither more nor less is required.  The state
space need not have any structure besides the one implied by the order
relation.  However, state spaces parametrized by the energy and work
coordinates have an additional, convex structure, which implies
concavity of the entropy, provided that the formation of convex
combination of states is an adiabatic process.  We add this requirement
as axiom A7 to our list of general axioms about the order relation.

The axioms in this section are so general that they  encompass
situations where {\it all} states in a whole neighborhood of a given
state are adiabatically accessible from it. {\bf Carath\'eodory's
principle} is the statement that this does {\it not} happen for
physical thermodynamic systems. In contrast, ideal mechanical systems
have the property that every state is accessible from every other one
(by mechanical means alone), and thus the world of mechanical systems
will trivially obey the entropy principle in the sense that every state
has the same entropy.  In the last subsection we discuss the connection
between Carath\'eodory's principle and the existence of irreversible
processes starting from a given state.  This principle will again be
invoked when, in Section III, we derive the comparison hypothesis for
simple thermodynamic systems.

Temperature will not be used in this section, not even the notion of
`hot' and `cold'. There will be no cycles, Carnot or otherwise.  The
entropy only depends on, and is defined by the order relation. Thus,
while the approach given here is  not the only path to the second law,
it has the advantage of a certain simplicity and clarity that at least
has pedagogic and conceptual value.  We ask the reader's
patience with our syllogisms, the point being that everything is here
clearly spread out in full view. There are no hidden assumptions, as
often occur in many textbook presentations. 

Finally, we hope that the reader will not be confused by our sometimes
lengthy asides about the motivation and heuristic meaning of our
various definitions and theorems. We also hope these remarks will not
be construed as part of the structure of the second law. The
definitions and theorems are self-contained, as we state them, and the
remarks that surround them are intended only as a helpful guide.

\bigskip\noindent
{\subt A. Basic concepts }
\bigskip

\noindent
{\subsubt 1. Systems and their state spaces}  
\medskip

Physically speaking a thermodynamic {\it system} consists of certain
specified amounts of different kinds of matter; it might be divisible
into parts that can interact with each other in a specified way. A
special class of systems called simple systems will be discussed in the
next chapter.  In any case the possible interaction of the system with
its surroundings is specified.  It is a ``black box" in the sense that
we do not need to know what is in the box, but only its response to
exchanging energy, volume, etc.  with other systems.  The states of a
system to be considered here are {\it always}  equilibrium states, but
the equilibrium may depend upon the existence of internal barriers in
the system.  Intermediate, non-equilibrium states that a system passes
through when changing from one equilibrium state to another will not be
considered.  The entropy of a system not in equilibrium may, like the
temperature of such a system, have a meaning as an approximate and
useful concept, but this is not our concern in this treatment.

Our systems can be quite complicated and the outside world can act on
them in several ways, e.g., by changing the volume and magnetization,
or removing barriers.  Indeed, we are allowed to chop a system into
pieces violently and reassemble them in several ways, each time waiting
for the eventual establishment of equilibrium.

Our systems must be macroscopic, i.e, not too small. Tiny systems
(atoms, molecules, DNA) exist, to be sure, but we cannot  describe
their equilibria thermodynamically, i.e., their equilibrium states
cannot be described in terms of the simple coordinates we use later
on.  There is a gradual shift from tiny systems to macroscopic ones,
and the empirical fact is that  large enough systems conform to the
axioms given below.  At some stage a system becomes `macroscopic'; we
do not attempt to explain  this phenomenon or to give an exact rule
about which systems are `macroscopic'.

On the other hand, systems that are too large are also ruled out
because gravitational forces become important.  Two suns cannot unite
to form one bigger sun with the same properties (the way two glasses of
water can unite to become one large glass of water). A  star with two
solar masses is intrinsically different from a sun of one solar mass.
In principle,  the two suns  could be kept apart and regarded as one
system, but then this would only be a `constrained' equilibrium because
of the gravitational attraction.  In other words the conventional
notions of `extensivity' and `intensivity' fail for cosmic bodies.
Nevertheless, it is possible to define an entropy for such systems by
measuring its effect on some standard body.  Giles' method is
applicable, and our formula (2.20) in Section II.E (which, in the
context of our development, is used only for calibrating the entropies
defined by (2.14) in Section II.D, but which could be taken as an
independent definition) would allow it, too.  (The `nice' systems that
do satisfy size-scaling are called `perfect' by Giles.) The entropy, so
defined, would satisfy additivity but not extensivity, in the `entropy
principle' of Section II.B.  However, to prove this would requires a
significant enhancement of the basic axioms.  In particular, we  would
have to take the comparison hypothesis, CH, for all systems as an axiom
--- as Giles does.  It is left to the interested reader to carry out
such an extension of our scheme.

A basic operation is {\bf composition} of two or more systems to form a
new system.  Physically, this simply means putting the individual
systems side by side and regarding them as one system.  We then speak
of each system in the union as a {\bf subsystem}.  The subsystems may
or may not interact for a while, by exchanging heat or volume for
instance, but the important point is that a state of the total system
(when in equilibrium) is described completely by the states of the
subsystems.

{}From the mathematical point of view a system is just a collection of
points called a {\bf state space}, usually denoted by $\Gamma$. The
individual points of a state space are called {\bf states} and are
denoted here by capital Roman letters, $X, Y, Z, $ etc.  {}From the
next section on we shall build up our collection of states satisfying
our axioms from the states of certain special systems, called {\it
simple systems}.  (To jump ahead for the moment, these are systems with
one or more work coordinates but with only one energy coordinate.)  In
the present section, however, the manner in which states are described
(i.e., the coordinates one uses, such as energy and volume, etc.)  are
of no importance.  Not even topological properties are assumed here
about our systems, as is often done.  In a sense it is amazing that
much of the second law follows from certain abstract properties of the
relation among states, independent of physical details (and hence of
concepts such as Carnot cycles). In approaches like Giles', where it is
taken as an axiom that comparable states fall into equivalence classes,
it is even possible to do without the system concept altogether, or
define it simply as an equivalence class of states. In our approach,
however,  one of the main goals is to derive the property which Giles
takes as an axiom, and systems are basic objects in our axiomatic
scheme.

Mathematically, the composition of two spaces, $\Gamma_1 $ and
$\Gamma_2 $ is simply the Cartesian product of the state spaces 
$\Gamma_1 \times \Gamma_2$.  In other words, the states in $\Gamma_1 
\times 
\Gamma_2 $ are pairs
$(X_1,X_2)$ with $X_1 \in \Gamma_1 $ and $X_2 \in \Gamma_2 $.  {}From 
the physical interpretation of the composition it is clear that the two
spaces $\Gamma_1 \times \Gamma_2 $ and $\Gamma_2 \times \Gamma_1$
are to be identified. Likewise, when forming multiple compositions of 
state spaces, the order and the grouping of the spaces is immaterial. 
Thus $(\Gamma_1 \times \Gamma_2)\times \Gamma_3$,  
$\Gamma_1 \times (\Gamma_2\times \Gamma_3)$ and 
$\Gamma_1 \times \Gamma_2\times \Gamma_3$ are to be identified as far 
as composition of state spaces is concerned. Strictly speaking, a 
symbol like $(X_1,\dots , X_{N})$  with states $X_{i}$ in state 
spaces $\Gamma_{i}$, $i=1,\dots,N$ thus stands for an equivalence 
class of $n$-tuples, corresponding to the different groupings and 
permutations of the state spaces. Identifications of this type are 
not uncommon in mathematics (the formation of direct sums of vector 
spaces is an example).

A further operation we shall assume is the formation of {\bf scaled
copies} of a given system whose state space is $\Gamma$.  
If $t>0$ is some fixed number (the scaling parameter) the state space
$\Gamma^{(t)}$ consists of points denoted $tX$ with $X\in \Gamma$. On the
abstract level $tX$ is merely a symbol, or mnemonic, to define points in
$\Gamma^{(t)}$, but the symbol acquires meaning through the axioms given
later in Sect.\ II.C. In the physical world, and from Sect.\ III onward, 
the state spaces will  always be subsets of some $\R^n$ (parametrized by
energy, volume, etc.). In this case $tX$ has the concrete 
representation as the product of the real number $t$ and the vector 
$X\in\R^n$. Thus in this case $\Gamma^{(t)}$ is simply the image of 
the set 
$\Gamma\subset \R^n$ under scaling by the real parameter $t$.
Hence, we shall sometimes denote $\Gamma^{(t)}$ by $t\Gamma$.

Physically, $\Gamma^{(t)}$ is interpreted as the state space of a system
that has the same  properties as the system with state space $\Gamma$,
except that the amount of each chemical substance in the system has been
scaled by the factor $t$ and the range of extensive variables like 
energy,
volume etc. has been scaled accordingly.  Likewise, $tX$ is obtained from
$X$ by scaling energy, volume etc., but also the matter content of a 
state
$X$  is scaled by the parameter $t$.  
{}From this physical interpretation it is clear
that $s(tX)=(st)X$ and ${(\Gamma^{(t)})}^{(s)}=\Gamma^{(st)}$ and we take
these relations also for granted on the abstract level. The same 
apples to the identifications $\Gamma^{(1)}=\Gamma$ and $1X=X$, and 
also
$(\Gamma_{1}\times\Gamma_{2})^{(t)}=\Gamma_{1}^{(t)}
\times\Gamma_{2}^{(t)}$ and $t(X,Y)=(tX,tY)$.

The operation of forming compound states is thus an associative and 
commutative binary operation on the set of all states, and the group 
of positive real numbers acts by the scaling operation on this set in 
a way compatible with the binary operation and the multiplicative 
structure of the real numbers.  The same is true for the set of all 
state spaces.  {}From an algebraic point of view the simple systems, 
to be discussed in Section III, are a basis for this algebraic 
structure.

While the relation between $\Gamma$ and $\Gamma^{(t)}$ is physically
and intuitively fairly obvious, there can be surprises. Electromagnetic
radiation in a cavity (`photon gas'), which is mentioned after (2.6),
is an interesting case; the two state spaces $\Gamma$ and
$\Gamma^{(t)}$ and the thermodynamic functions on these spaces are
identical in this case! Moreover, the two spaces are physically
indistinguishable. This will be explained in more detail in Section
II.B.

The formation of scaled copies involves a certain physical 
idealization because it ignores the molecular structure of matter.  
Scaling to arbitrarily small sizes brings quantum effects to the fore 
and macroscopic thermodynamics is no longer applicable.  At the other 
extreme, scaling to arbitrarily large sizes brings in unwanted 
gravitational effects as discussed above.  In spite of these well 
known limitations the idealization of continuous scaling is common 
practice in thermodynamics and simplifies things considerably.  (In 
the statistical mechanics literature this goes under the rubric of the 
\lq thermodynamic limit\rq.)  It should be noted that scaling is quite 
compatible with the inclusion of `surface effects' in thermodynamics.  
This will be discussed in Section III. A.

By composing scaled copies of $N$ systems with state spaces $\Gamma_1,
\dots , \Gamma_N$, one can form, for $t_1,\dots,t_N>0$, their {\bf scaled
product} $\Gamma^{(t_1)}_1 \times \cdots \times \Gamma^{(t_N)}_N$ whose 
points are $(t_1 X_1, t_2 X_2, \dots , t_N X_N)$. 
In the particular case that the  $\Gamma_j$'s are identical, i.e.,
$\Gamma_1= \Gamma_2 = \cdots =\Gamma$, we shall call any space  of the
the form $\Gamma^{(t_1)} \times \cdots \times \Gamma^{(t_N)}$ a {\bf
multiple scaled copy} 
of $\Gamma$.
As will be explained later in connection with Eq.\ (2.11), it is 
sometimes 
convenient in calculations to allow $t=0$ as scaling parameter (and even 
negative values). For the moment let us just note that if $\Gamma^{(0)}$ 
occurs 
the reader is asked to regard it as the empty set or 'nosystem'. In other 
words, 
ignore it. 

\smallskip

Some examples may help clarify  the concepts of systems and state
spaces. \smallskip

\item{(a)} $\Gamma_a$: 1 mole of hydrogen, H$_2$. The state space can
be identified with a subset of $\R^2$ with coordinates $U$ ($=$
energy), $V (=$ volume).

\item{(b)} $\Gamma_b$: $\mfr1/2$ mole of H$_2$.  If $\Gamma_a$ and
$\Gamma_b$ are regarded as subsets of $\R^2$ then  $\Gamma_b =
\Gamma_a^{(1/2)} = \{(\mfr1/2 U,\mfr1/2 V) : (U,V) \in \Gamma_a \}$.

\item{(c)} $\Gamma_c$: 1 mole of H$_2$ and $\mfr1/2$ mole of O$_2$
(unmixed). $\Gamma_c = \Gamma_a \times \Gamma_{(\mfr1/2 \ {\rm mole \
O}_2)}$. This is a compound system.

\item{(d)} $\Gamma_d$: 1 mole of H$_2$O.

\item{(e)} $\Gamma_e$: 1 mole of H$_2 + \mfr1/2$ mole of O$_2$ (mixed).
Note that 
$\Gamma_e \not= \Gamma_d$ and $\Gamma_e \not= \Gamma_c$. This system
shows the perils inherent in the concept of equilibrium. The system
$\Gamma_e$ makes sense as long as one does not drop in a piece of
platinum or walk across the laboratory floor  too briskly. Real world
thermodynamics requires that we admit such quasi-equilibrium systems,
although perhaps not quite as dramatic as this one.

\item{(f)} $\Gamma_f$: All the  equilibrium states of one mole of H$_2$
and half a mole of O$_2$ (plus a tiny bit of platinum to speed up the
reactions) in a container.  A typical state will have some fraction of
H$_2$O, some fraction of H$_2$ and some O$_2$. Moreover, these
fractions can exist in several phases.

\bigskip\noindent
{\subsubt 2.  The order relation}  
\medskip

The basic ingredient of thermodynamics is the relation 
$$
\prec
$$ 
of {\bf adiabatic accessibility} among  states of a system--- or even
different systems.  The statement $X\prec Y$, when $X$ and $Y$ are
points in some (possibly different) state spaces, means that there is
an adiabatic transition, in the sense explained below, that takes the
point $X$ into the point $Y$.

Mathematically, we do not have to ask the meaning of \lq adiabatic\rq.
All  that matters is that a list of all possible pairs of states $X$'s
and $Y$'s such that $X \prec Y$ is regarded as given.  This list has to
satisfy certain axioms that we prescribe below in subsection C. Among
other things it must be reflexive, i.e., $X\prec X$, and transitive,
i.e., $X\prec Y$ and $Y\prec Z$ implies $X\prec Z$.  (Technically, in
standard mathematical terminology this is called a {\it pre}order
relation because we can have both $X\prec Y$ and $Y\prec X$ without
$X=Y$.) Of course, in order to have an interesting thermodynamics
result from our  $\prec$ relation it is essential that there are pairs
of points $X,Y$ for which $X\prec Y$ is {\it not} true.

Although the physical interpretation of the relation $\prec$ is
not needed for the mathematical development, for applications it
is essential to have a clear understanding of its meaning. It is
difficult to avoid some circularity when defining the concept of
adiabatic accessibility.
The following version (which is in the spirit of Planck's formulation
of the second law (Planck, 1926)) appears to be sufficiently general
and precise and appeals to us. It has the great virtue (as discovered
by Planck) that it avoids having to distinguish between work and
heat---or even having to define the concept of heat; heat, in the
intuitive sense, can always be generated by rubbing---in accordance
with Count Rumford's famous discovery while boring cannons! We
emphasize, however, that other definitions are certainly possible.  Our
physical definition is the following:  \medskip

{\bf  Adiabatic accessibility:} {\it A state $Y$ is adiabatically
accessible from a state $X$, in symbols $X\prec Y$, if it is possible
to change the state from $X$ to $Y$ by means of an interaction with
some device (which may consist of mechanical and electrical parts as
well as auxiliary thermodynamic systems) and a weight, in such a way
that the device returns to its initial state at the end of the process
whereas the weight may have changed its position in a gravitational
field.}

Let us write
$$
X\prec \prec Y \ \ \  {\rm if}\ \ \  X\prec Y \ \ \ {\rm but}\ \ \
 Y\not\prec X . \eqno(2.1)
$$
In the real world $Y$ is adiabatically accessible from $X$ only
if $X\prec \prec Y$. When $X\prec Y$ and also 
$Y\prec X$ 
then the state change can only be realized in an idealized sense, for
it will take infinitely long time to achieve it in the manner
decribed.  An alternative way is to say that the \lq device\rq\ that
appears in the definition of accessibility has to return to within
\lq$\varepsilon$\rq\ of its original state (whatever that may mean) and
we take the limit $\varepsilon \to 0$. To avoid this kind of discussion
we have taken the definition as given above, but we emphasize that it
is certainly possible to redo the whole theory using only the notion of
$\prec \prec $. An emphasis on $\prec \prec $ appears in Lewis and
Randall's discussion of the second law (Lewis and Randall, 1923, page
116).

{\it Remark:} It should be noted that the operational definition above is 
a 
definition of the concept of
`adiabatic accessibility' and not the concept of an `adiabatic
process'. A state change leading from $X$ to $Y$ can be achieved in
many different ways (usually infinitely many), and not all of them will
be `adiabatic processes' in the usual terminology. Our concern is not
the temporal development of the state change which, in real processes,
always leads out of the space of equilibrium states.
Only the end result for the system and for the rest of the world
interests us. However,  it is important to clarify the relation between
our definition of adiabatic accessiblity and the usual textbook
definition of an adiabatic process. This will be discussed in Section C
after Theorem 2.1 and again in Sec. III; cf. Theorem 3.8. There it will
be shown that our definition indeed coincides with the usual notion
based on processes taking place within an 'adiabatic enclosure'.
A further point to notice is that the word \lq adiabatic\rq\ is
sometimes used to mean ``slow" or quasi-static, but nothing of the sort
is meant here. Indeed, an adiabatic process can be quite violent. The
explosion of a bomb in a closed container is an adiabatic process.

\smallskip

Here are some further examples of adiabatic  processes: \smallskip

\item{1.} Expansion or compression of a gas, with or without the help
of a weight being raised or lowered.

\item{2.} Rubbing or stirring.

\item{3.} Electrical heating. (Note that the concept of `heat' is not
needed here.)

\item{4.} Natural processes that occur within an isolated compound 
system after some barriers
have been removed. This includes  mixing and chemical or nuclear 
processes.

\item{5.} Breaking a system into pieces with a hammer and 
reassembling (Fig. 1).

\item{6.} Combinations of such changes.

In the usual parlance, rubbing would be an adiabatic process, but not
electrical `heating', because the latter requires the introduction of a
pair of wires through the `adiabatic enclosure'. For us, both processes
are adiabatic because what is required is that apart from the
change of the system itself, nothing more than
the displacement of a weight occurs. To achieve electrical heating, one
drills a hole in the container, passes a heater wire through it,
connects the wires to a generator which, in turn, is connected to a
weight. After the heating the generator is removed along with the wires,
the hole is plugged, and the system is observed to be in a new state. The
generator, etc. is in its old state and the weight is lower.

\centerline{\sevenpoint ---- (Insert Figure 1 here) ----}
%\epsfxsize 15truecm
%\epsfysize 7.5truecm 
%\epsffile{figure1.eps}

We shall use the following terminology concerning any two states $X$
and $Y$. These states are said to be {\bf comparable} (with respect to
the relation $\prec$, of course) if either $X \prec Y$ or $Y\prec X$.
If both relations hold we say that $X$ and $Y$ are {\bf adiabatically
equivalent} and write
$$
X\sima Y . \eqno(2.2)
$$
The comparison hypothesis referred to above is the statement that any
two states in the {\it same} state space are comparable.  In  the
examples of systems (a) to (f) above, all satisfy the comparison
hypothesis. Moreover, every point in $\Gamma_c$ is in the relation
$\prec$ to many (but not all) points in $\Gamma_d$. States in different
systems may or may not be comparable.  An example of non-comparable
systems is one mole of H$_2$ and one mole of O$_2$.
Another is one mole of H$_2$ and two moles of H$_2$.

One might think that if the comparison hypothesis, which will be
discussed further in Sects. II.C and II.E, were to fail for some state
space then the situation could easily be remedied by breaking up the
state space into smaller pieces inside each of which the hypothesis
holds.  This, generally, is false. What is needed to accomplish  this
is the extra requirement that {\it comparability is an equivalence
relation;} this, in turn, amounts to saying that the condition $X \prec
Z$ and $Y \prec Z$ implies that $X$ and $Y$ are comparable and,
likewise, the condition $Z \prec X$ and $Z \prec Y$ implies that $X$
and $Y$ are comparable.   (This axiom can be found in (Giles, 1964),
see axiom 2.1.2, and similar requirements were made earlier by Landsberg 
(1956),
Falk and Jung (1959) and Buchdahl (1962, 1966).) While these two 
conditions 
are
logically independent, they can be shown to be equivalent if the axiom
A3 in Section II. C is adopted. In any case, we do not adopt the
comparison hypothesis as an axiom because we find it hard to regard it
as a physical necessity. In the same vein, we do not assume that
comparability is an equivalence relation (which would then lead to the
validity of the comparison hypothesis for suitably defined
subsystems).  Our goal is to prove the comparison hypothesis starting
from axioms that we find more appealing physically.

\bigskip\noindent
{\subt B. The entropy principle}
\bigskip

Given the relation $\prec$ for all possible states of all possible
systems, we can ask whether this relation can be encoded in an entropy
function according to the following principle, which expresses the {\bf 
second 
law of thermodynamics} in a precise and quantitative way:

{\bf Entropy principle:} {\it There is a real-valued function on all
states of all systems (including compound systems), called {\bf entropy}
and denoted by $S$ such that

\item{a)} \underbar{{\tt Monotonicity:}} When $X$ and $Y$ are
comparable states then
$$
X\prec Y \hbox{ \ \ {\rm if and only if} \ \ } S(X) \leq S(Y) . 
\eqno(2.3)
$$
(See (2.6) below.)
\item{b)} \underbar{{\tt Additivity and extensivity:}} If $X$ and $Y$
are states of some (possibly different) systems and if $(X,Y)$ denotes
the corresponding state in the composition of the two systems,  then
the entropy is additive for these states, i.e.,
$$
S((X,Y)) = S(X) + S(Y) . \eqno(2.4)
$$
$S$ is also extensive, i.e.,  for each $t>0$ and each
state $X$ and its scaled copy $tX$,
$$
S(t X)=t S(X) .\eqno(2.5)
$$}

\noindent[Note: {}From now on we shall omit the double parenthesis and
write simply $S(X,Y)$ in place of $S((X,Y))$.]

A logically equivalent formulation of (2.3), that does not use
the word \lq comparable\rq\ is the following pair of statements:
$$
\eqalignno{
X\sima Y &\Longrightarrow S(X) = S(Y) \ \ \ \ {\rm and} \cr
X\prec\prec Y &\Longrightarrow S(X) < S(Y).& (2.6)\cr } 
$$
The last line is especially noteworthy. It says that entropy must
increase in an irreversible process.

Our goal  is to construct an entropy function that satisfies the
criteria (2.3)-(2,5), and to show that it is essentially unique. We
shall proceed in stages, the first being to construct an entropy
function for a  single system, $\Gamma$,  and its multiple scaled
copies (in which comparability is assumed to hold). Having done this,
the problem of relating different systems will then arise, i.e., the
comparison question for compound systems. In the present Section II
(and {\it only} in this section) we shall simply complete the project
by {\it assuming} what we need by way of comparability. In Section IV,
the thermal axioms (the {\it zeroth law of thermodynamics}, in
particular) will be invoked to verify our assumptions about
comparability in compound systems.  In the remainder of this subsection
we discuss he significance of conditions (2.3)-(2.5).

The physical content of (2.3) was already commented on; adiabatic
processes not only increase entropy but an increase of entropy also
dictates which adiabatic processes are possible (between comparable
states, of course).

The content of additivity, (2.4), is considerably more far reaching
than one might think from the simplicity of the notation---as
we mentioned earlier. Consider four
states $X,X',Y,Y'$ and suppose that $X\prec Y$ and $X'\prec Y'$. Then
(and this will be one of our axioms) $(X,X')\prec (Y,Y')$,  and (2.4)
contains nothing new in this case. On the other hand,  the compound
system can well have an adiabatic process in which $(X,X')\prec (Y,Y')$
but $X\not\prec Y$.  In this case, (2.4) conveys much
information. Indeed, by monotonicity, there will be many cases of this
kind because the inequality $S(X) + S(X') \leq S(Y) + S(Y')$ certainly
does not imply that $S(X) \leq S(Y)$. The fact that the inequality
$S(X) + S(X') \leq S(Y) + S(Y')$ tells  us {\it exactly } which
adiabatic processes are allowed in the compound system (assuming
comparability), independent of any detailed knowledge of the manner in
which the two systems interact, is astonishing and is at the 
{\it heart of thermodynamics.}

Extensivity, (2.5), is {\it almost} a consequence of (2.4) alone---but 
logically it is independent.  Indeed, (2.4) implies that (2.5) holds 
for {\it rational} numbers $t$ provided one accepts the notion of
recombination as given in Axiom A5 below, i.e., one can combine
two samples of a system in the same state into a bigger system in a
state with the same intensive properties. (For systems, such as cosmic
bodies, that  do not obey this axiom, extensivity and additivity are
truly independent concepts.) On the other hand, using the
axiom of choice, one may always change a given entropy function
satisfying (2.3) and (2.4) in such a way that (2.5) is violated for
some irrational $t$, but then the function $t\mapsto S(tX)$ would end
up being unbounded in every $t$ interval.  Such pathological cases
could be excluded by supplementing (2.3) and (2.4) with the requirement
that $S(t X)$ should locally be a bounded function of $t$, either from
below or above.  This requirement, plus (2.4), would then imply (2.5).
For a discussion related to this point see (Giles, 1964), who
effectively considers {\it only} rational $t$.  See also (Hardy,
Littlewood, Polya 1934) for a discussion of the concept of Hamel bases
which is relevant in this context.

The extensivity condition can sometimes have surprising results, as in
the case of electromagnetic radiation (the `photon gas').  As is well
known (Landau and Lifschitz, 1969, Sect. 60), the phase space of such a
gas (which we imagine to reside in a box with a piston that can be used
to change the volume) is the quadrant $\Gamma=\{(U, V) \ : \
0<U<\infty,\  0<V<\infty \}$. Thus, 
$$
\Gamma^{(t)} =\Gamma
$$
as {\it sets}, which is not surprising or even exceptional. What is 
exceptional is that $S_{\Gamma}$, which gives the entropy of the states
in $\Gamma$, satisfies
$$
S_\Gamma(U,V) = \hbox{\rm (const.) } V^{1/4} U^{3/4}.
$$
It is homogeneous of first degree in the coordinates and, therefore,
the extensivity law tells us that the entropy function on 
the scaled copy $\Gamma^{(t)}$ is
$$
S_{\Gamma^{(t)}} (U, V) = t S_\Gamma (U/t, V/t)= S_\Gamma (U, V).
$$
Thus, all the thermodynamic functions on the two state spaces are the
same! This unusual situation could, in principle, happen for an ordinary
material system, but we know of no  example besides the photon gas.
Here, the result can be traced to the fact that particle number is not
conserved, as it is for material systems, but it does show that one
should not jump to conclusions. There is, however, a further conceptual
point about the photon gas which is physical rather than mathematical.
If a material system had a homogeneous entropy (e.g., $ S(U,V)=
{\rm (const.) } V^{1/2} U^{1/2}$ )we should still be able to distinguish
$\Gamma^{(t)}$ from $\Gamma$, even though the coordinates and entropy
were indistinguishable. This could be done by weighing the two systems
and finding out that one weighs $t$ times as much as the other.
But the photon gas is different: no experiment can tell the two apart.
However, weight {\it per se} plays no role in thermodynamics, so the 
difference between the material and photon systems is not
thermodynamically significant. 

There are two points of view one could take about this anomalous
situation. One is to continue to use the state spaces $\Gamma^{(t)}$,
even though they happen to represent identical systems.  This is not
really a problem because no one said that $\Gamma^{(t)}$ had to be
different from $\Gamma$. The only concern is to check the axioms, and in
this regard there is no problem. We could even allow the additive entropy
constant to depend on $t$, provided it satisfies the extensivity
condition (2.5). The second point of view is to say that there is only
one $\Gamma$ and no $\Gamma^{(t)}$'s at all. This would cause us to
consider the photon gas as outside our formalism and to require special
handling  from time to time. The first alternative is more attractive to
us for obvious reasons. The photon gas will be mentioned again in
connection with Theorem 2.5.

\bigskip\noindent
{\subt C. Assumptions about the order relation}
\bigskip
We now list our assumptions for the order relation $\prec$. As always,
$X$, $Y$, etc. will denote states (that may belong to different
systems), and if $X$ is a state in some state space $\Gamma$, then $tX$
with $t>0$ is the corresponding state in the scaled state space
$\Gamma^{(t)}$.

\item{{\bf A1)}}  {\bf Reflexivity.} $X \sima X$.

\item{{\bf A2)}}  {\bf Transitivity.} {\it $X \prec Y$ and $Y \prec Z$
implies $X \prec Z$.}

\item{{\bf A3)}} {\bf Consistency.} {\it $X \prec X^\prime$ and $Y 
\prec Y^\prime$ implies $(X,Y) \prec
(X^\prime, Y^\prime)$.}

\item{{\bf A4)}} {\bf Scaling invariance.} {\it If $X\prec Y$, then 
$tX \prec tY$ for all $t>0$.}

\item{{\bf A5)}}  {\bf Splitting and recombination.} {\it  For $0 < 
t < 1$
$$X \sima (t X, (1-t) X). \eqno(2.7)$$}
(If $X \in \Gamma$, then the right side is in the scaled product 
$\Gamma^{(t)}\times \Gamma^{(1-t)}$, of course.)

\item{{\bf A6)}}  {\bf Stability.}  {\it If, for some pair of states, $X$ 
and 
$Y$,
$$(X, \varepsilon Z_0) \prec (Y, \varepsilon Z_1)$$
holds for a sequence of $\varepsilon$'s tending to zero and some states 
$Z_0$, $Z_1$, then 
$$X \prec Y.$$} 

{\it Remark:}  `Stability' means simply that one cannot increase the
set of accessible states with an infinitesimal grain of dust.

Besides these axioms the following property of state spaces, the 
`comparison hypothesis', plays a crucial role in our analysis in this 
section.  It will eventually be established for all state spaces after 
we have introduced some more specific axioms in later sections.

\item{{\bf CH)}} {\bf Definition:}
{\it We say the {\bf comparison hypothesis} (CH) holds for a state 
space if any two states $X$ and $Y$ in the space are comparable, i.e., 
$X\prec Y$ or $Y\prec X$.}

In the next subsection we shall show that, for every state space,
$\Gamma$, assumptions A1-A6, and CH for all two-fold scaled products,
$(1-\lambda) \Gamma \times \lambda \Gamma$, not just $\Gamma$ itself,
are in fact {\it equivalent} to the existence of an additive and
extensive entropy function that characterizes the order relation on the
states in {\it all} scaled products of $\Gamma$. Moreover, for each
$\Gamma$, this function is unique, up to an affine transformation of
scale, $S(X) \rightarrow a S(X)+B$. Before we proceed to the
construction of entropy we derive a simple property of the order
relation from assumptions A1-A6, which is clearly necessary if the
relation is to be characterized by an additive entropy function.
\medskip

{\bf THEOREM 2.1 (Stability implies cancellation law).}  {\it Assume
properties A1-A6, especially A6---the stability law.  Then the {\bf
cancellation law} holds as follows.  If $X,Y$ and $Z$ are states of three
(possibly distinct) systems then
$$(X,Z) \prec (Y,Z) \ \ \ {\rm implies} \ \ \ X \prec Y \qquad {\rm
(Cancellation \ Law)}.
$$}
\medskip

{\it Proof:}  Let $\varepsilon = 1/n$ with $n = 1,2,3, \dots$.  Then we
have
$$
\eqalignii{(X,\varepsilon Z) &\sima ((1-\varepsilon) X, \varepsilon X, 
\varepsilon Z) \quad &\hbox{(by A5)} \cr
&\prec ((1-\varepsilon) X, \varepsilon Y, \varepsilon Z) \quad
&\hbox{(by
A1, A3 and A4)} \cr
&\sima ((1-2 \varepsilon) X, \varepsilon X,
\varepsilon Y,  \varepsilon Z)
\quad &\hbox{(by A5)} \cr
&\prec ((1-2 \varepsilon) X, 
2 \varepsilon Y,  \varepsilon Z) \quad
&\hbox{(by A1, A3, A4 and A5).} \cr}
$$
By doing this $n = 1/\varepsilon$ times we find that $(X, \varepsilon Z)
\prec (Y, \varepsilon Z)$. By the stability axiom A6 we then have 
$X \prec Y $. \hfill\lanbox

{\it Remark:}  Under the additional assumption that $Y$ and $Z$ are
comparable states (e.g., if they are in the same state space for which
CH holds), the cancellation law is logically equivalent to the following
statement (using the consistency axiom A3):
$$
{\sl If} \  X \prec\prec Y 
\ {\sl then}\  (X,Z) \prec\prec (Y,Z) \ {\sl for \  all}\ Z.
$$

The cancellation law looks innocent enough, but it is really rather
strong.  It is a partial converse of the consistency condition A3 and
it says that although the ordering in $\Gamma_1 \times \Gamma_2 $ is
{\it not} determined simply by the order in $\Gamma_1$ and $\Gamma_2$,
there are limits to how much the ordering can vary beyond the minimal
requirements of A3. It should also be noted that the cancellation law
is in accord with our physical interpretation of the order relation in
Subsection II.A.2.; a ``spectator'', namely $Z$, cannot change the
states that are adiabatically accessible from $X$.

\bigskip

{\it Remark about `Adiabatic Processes':\ \ }
With the aid of the cancellation law we can now discuss the connection
between our notion of adiabatic accessibility and the textbook concept
of an `adiabatic process'. One problem we face is that this latter
concept is hard to make precise (this was  our reason for
avoiding it in our operational definition) and therefore the discussion 
must
necssearily be somewhat informal. The general idea of an adiabatic
process, however, is that the system of interest is locked in a
thermally isolating enclosure that prevents `heat' from flowing into or
out of our system. Hence, as far as the system is concerned, all the
interaction it has with the external world during an  adiabatic process
can be thought of as being accomplished by means of some mechanical or
electrical devices. Our operational definition of the relation $\prec$
appears at first sight to be based on more general processes, since we
allow an auxilary thermodynamical system as part of the device. We
shall now show that, despite appearances, our definition coincides with
the conventional one.

Let us temporarily denote by $\prec^*$ the relation between states based
on adiabatic processes, i.e., $X \prec^* Y$ if and only if there is a
mechanical/electrical device that starts in a state $M$ and ends up in a
state $M'$ while the system changes from $X$ to $Y$. We now assume that
the mechanical/electrical device can be restored to the initial state
$M$ from the final state $M'$ by adding or substracting mechanical
energy, and this latter process can be reduced to the raising or
lowering of a weight in a gravitational field.  (This can be taken as a
definition of what we mean by a 'mechanical/electrical device'. Note
that devices with 'dissipation' do not have this property.) Thus,
$X\prec^*Y$ means there is a process in which the mechanical/electrical
device starts in some state $M$ and ends up in the same state, a weight
moves from height $h$ to height $h'$, while the state of our system
changes from $X$ to $Y$.  In symbols, 
$$ 
(X,M,h)\longrightarrow (Y,M,h').\eqno(2.8) 
$$

In our definition of adiabatic accessibility, on the other hand, we have
some {\it arbitrary} device, which interacts with our system and  which
can generate or remove heat if desired.
There is no thermal enclosure. The important constraint is that the
device starts in some state $D$ and ends up in the same state $D$.  As
before a weight moves from height $h$ to height $h'$, while our system
starts in state $X$ and ends up in state $Y$. In symbols,
$$
(X,D,h) \longrightarrow (Y,D,h') \eqno(2.9)  .
$$
It is clear that (2.8) is a
special case of (2.9), so we conclude that $X\prec^*Y$ implies $X\prec
Y$. The device in (2.9) may consist of a thermal part in some state $Z$
and electrical and mechanical parts in some state $M$. Thus $D=(Z,M)$,
and (2.9) clearly implies that $(X,Z)\prec^*(Y,Z)$.

It is natural to assume that $\prec^*$ satisfies axioms A1-A6, just as
$\prec$ does.  In that case we can infer the cancellation law for
$\prec^*$, i.e.,  $(X,Z) \prec^*(Y,Z,)$ implies $X \prec^* Y$.  Hence,
$X\prec Y$ (which is what (2.9) says) implies $X\prec^*Y$. Altogether we
have thus shown that $\prec$ and $\prec^*$ are really the same relation.
In words: {\it adiabatic accessibility can always be achieved by an
adiabatic process applied to the system plus a device and, furthermore,
the adiabatic process can be simplified (although this may not be easy
to do  experimentally) by eliminating all thermodynamic parts of the
device, thus making the process an adiabatic one for the system alone.}

\vfill\eject
\bigskip
\noindent
{\subt D. The construction of entropy for a single system}
\bigskip

Given a state space $\Gamma$ we may, as discussed in Subsection I.A.1.,
construct its {\it multiple scaled copies}, i.e., states of the form $$
Y=(t_1Y_1,\dots,t_NY_N)
$$
with $t_i>0$, $Y_i\in\Gamma$. It 
follows from our assumption A5 that if CH (comparison hypothesis) holds
in the state space $\Gamma^{(t_1)} \times \cdots \times \Gamma^{(t_N)}$
with $t_1,...,t_N$ fixed, then any
other state of the same form, 
$Y'=(t_1'Y_1',\dots,t_M'Y_M')$ with $Y_i'\in\Gamma$ , is comparable to 
$Y$ provided 
$\sum_i t_i=\sum_jt'_j$ 
(but not, in general, if the sums are not equal). This is proved as 
follows for $N=M=2$; the easy extension to the general case is left to
the reader. Since $t_1+t_2 = t_1'+t_2'$ we can assume,  without loss of
generality, that $t_1-t_1' = t_2'-t_2 >0$, because the case 
$t_1-t_1' =0$ is already covered by CH (which was assumed) 
for $\Gamma^{(t_1)} \times  
\Gamma^{(t_2)}$. By the splitting axiom, A5, we have 
$(t_1Y_1,t_2Y_2) \sima (t_1'Y_1, (t_1-t_1')Y_1, t_2Y_2)$ and
$(t_1'Y_1',t_2'Y_2')\sima (t_1'Y_1', (t_1-t_1')Y_2', t_2Y_2')$.
The comparability now follows from CH on the space
$\Gamma^{(t_1')} \times \Gamma^{(t_1-t_1')} \times \Gamma^{(t_2)}$.

The  entropy principle for the states in the multiple scaled
copies of a single system will now be derived. More precisely, we shall
prove the following theorem:  \medskip

{\bf THEOREM 2.2 (Equivalence of entropy and assumptions A1--A6,
CH).} {\it  Let $ \Gamma$ be a state space and let $\prec$ be a
relation on the multiple scaled copies of $\Gamma$. The following
statements are equivalent.\hfill \item{(1)} The  relation $\prec$
satisfies axioms A1--A6, and CH holds for all multiple scaled copies
of $\Gamma$.  
\item{(2)} There is a function, $S_\Gamma$ on $\Gamma$ that
characterizes the relation in the sense that if
\noindent$t_1+\cdots+t_N=t'_1+\cdots
+t_M'$, (for all $N\geq 1$ and $M\geq 1$) then
$$
(t_1Y_1,...,t_NY_N) \ \prec \ (t_1^{\prime}Y_1^{\prime},
...,t_M^{\prime}Y_M^{\prime})
$$
holds if and only if 
$$
\sum_{i=1}^N t_i S_\Gamma(Y_i) \ \leq \ \sum_{j=1}^M t_j^{\prime} 
S_\Gamma(Y_j')\ .   \eqno (2.10)
$$

The function $S_\Gamma$ is uniquely determined on $\Gamma$, up to an 
affine
transformation, i.e., any other function $S_\Gamma^*$ on $\Gamma$ 
satisfying
(2.10) is of the form $S_\Gamma^*(X)=aS_\Gamma(X)+B$ with constants $a>0$ 
and $B$.}
\medskip

{\bf Definition.} A function $S_\Gamma$ on $\Gamma$ that characterizes 
the 
relation $\prec$ on the multiple scaled copies of $\Gamma$ in the sense
stated in the theorem is called an {\bf entropy function on} $\Gamma$.
\smallskip

We shall split the proof of Theorem 2.2 into Lemmas 2.1, 2.2, 2.3 and 
Theorem 
2.3 below.

At this point it is convenient to introduce the following
notion of {\bf generalized ordering}.  While $(a_1 X_1, a_2 X_2,
\dots, a_N X_N)$ has so far only been defined when all $a_i > 0$, we can 
{\it
define} the meaning of the relation
$$
(a_1 X_1, \dots , a_N X_N) \prec (a^\prime_1 X^\prime_1, \dots ,
a^\prime_M X^\prime_M) \eqno(2.11)
$$ for arbitrary $a_i \in \R$, $a^\prime_i \in \R$, $N$ and $M$ positive 
integers
and $X_i \in \Gamma_i$, $X^\prime_i \in \Gamma^\prime_i$ as follows.
If any $a_i$ (or
$a^\prime_i$) is zero we just ignore the corresponding term. 
Example: $(0X_{1},X_{2})\prec (2X_{3},0X_{4})$ means the same thing as
$X_{2}\prec 2X_{3}$. If any $a_i$ (or
$a^\prime_i$) is negative, just move $a_i X_i$ (or $a^\prime_i 
X^\prime_i$)
to the other side and change the sign of $a_i$ (or $a^\prime_i$).  
Example: 
$$
(2 X_1, X_2) \prec (X_3, - 5 X_4, 2X_5, X_6)
$$
means that
$$
(2X_1, 5 X_4, X_2) \prec (X_3, 2 X_5, X_6)
$$
in $\Gamma_1^{(2)} \times \Gamma_4^{(5)} \times \Gamma_2$ and $\Gamma_3
\times \Gamma_5^{(2)} \times \Gamma_6$.  (Recall that $\Gamma_a \times
\Gamma_b = \Gamma_b \times \Gamma_a)$.  It is easy to check, using the
cancellation law, that {\it the splitting and recombination axiom A5
extends to nonpositive scaling parameters}, i.e., axioms A1-A6 imply 
that  $X\sima (aX,bX)$ for all $a,b\in\R$ with $a+b=1$, if the 
relation $\prec$ for nonpositive $a$ and $b$ is understood in the 
sense just decribed.

For the definition of the entropy function we need the following lemma,
which depends crucially on the stability assumption A6 and on the
comparison hypothesis CH for the state spaces
$\Gamma^{(1-\lambda)}\times\Gamma^{(\lambda)}$.  \medskip

{\bf LEMMA 2.1}  {\it Suppose $X_0$ and $X_1$ are two points in 
$\Gamma$ with $X_0\prec\prec X_1$. For $\lambda\in\R$ define
$$
\S_\lambda = \{ X \in \Gamma : ((1 - \lambda) X_0, \lambda X_1) 
\prec X \}.\eqno(2.12)
$$ 
Then 

(i) For every $X \in \Gamma$ there is a $\lambda \in \R$ such that $X
\in \S_\lambda$.

(ii)  For every $X \in \Gamma$, $\sup \{ \lambda : X \in \S_\lambda \}
< \infty$.  } \medskip

{\it Remark.} Since $X\sima ((1-\lambda)X,\lambda X)$ by assumption A5, 
the definition of $\S_\lambda$ really involves the order relation on 
double scaled copies of $\Gamma$ (or on $\Gamma$ itself, if 
$\lambda=0$ or 1.)

{\it Proof of Lemma 2.1.} (i)  If $X_0 \prec X$ then 
obviously $X \in \S_0$ by axiom A2.  
For general $X$ we claim
that 
$$
(1 + \alpha) X_0 \prec (\alpha X_1,X) \eqno(2.13)
$$
for some $\alpha \geq 0$ and hence $((1 -\lambda) X_0, \lambda X_1)
\prec X$ with $\lambda = - \alpha$.  The proof relies on stability, A6,
and the comparison hypothesis CH (which comes into play for the first
time):  If (2.13) were not true, then by CH we would have
$$(\alpha X_1,X) \prec (1 + \alpha) X_0$$
for all $\alpha >0$ and so, by scaling, A4, and A5
$$
\left (X_1,\, {1 \over \alpha}X\right) \prec 
\left( X_0,\, {1 \over \alpha} X_0\right).
$$
By the stability axiom A6 this would imply $X_1 \prec X_0$ in
contradiction to $X_0 \prec\prec X_1$.

(ii)  If $\sup \{ \lambda : X \in \S_\lambda \} = \infty$, then for
some sequence of $\lambda$'s tending to infinity we would have
$((1-\lambda)X_0,\lambda X)\prec X$ and hence  $(X_0, \lambda X_1)
\prec (X, \lambda X_0)$ by A3 and A5. By A4 this implies $\left( {1
\over \lambda} X_0, X_1 \right) \prec \left( {1 \over \lambda} X, X_0
\right)$ and hence $X_1 \prec X_0$ by stability, A6.  \hfill\lanbox

We can now state our {\bf formula for the entropy function}.  If all
points in $\Gamma $ are adiabatically equivalent there is nothing to
prove (the entropy is constant), so we may assume that there are points
$X_0$, $X_1\in\Gamma$ with $X_0\prec\prec X_1$.  We then define for
$X\in\Gamma$ 
$$ 
S_\Gamma(X):=\sup\{\lambda:\ ((1-\lambda)X_0,\lambda X_1)\prec X\}. \eqno 
(2.14)
$$ 
(The symbol $a:=b$ means that $a$ is defined by $b$.) 
This $S_\Gamma$ will be referred to as the {\bf canonical
entropy} on $\Gamma$ with {\bf reference points} $X_0$ and $X_1$. 
This definition is illustrated in Figure 2.

\centerline{\sevenpoint ---- Insert Figure 2 here ----}

By Lemma 2.1 $S_\Gamma(X)$ is well defined and  $S_\Gamma(X)<\infty$ for
all $X$.  (Note that by stability we could replace $\prec$ by 
$\prec\prec$
in (2.14).) We shall now show that this $S_\Gamma$ has all the right
properties. The first step is the following simple lemma, which does not
depend on the comparison hypothesis.  \medskip

{\bf LEMMA 2.2 ($\prec$ is equivalent to $\leq$).}  
{\it Suppose $X_0
\prec\prec X_1$ are states and $a_0, a_1, a^\prime_0, a^\prime_1$
are real numbers
with $a_0 + a_1 = a^\prime_0 + a^\prime_1$.  Then the following are
equivalent.  \item{(i)}  $(a_0 X_0, a_1 X_1) \prec (a^\prime_0 X_0,
a^\prime_1 X_1)$ \item{(ii)}  $a_1 \leq a^\prime_1$ (and hence $a_0
\geq a^\prime_0$).  \smallskip\noindent In particular, $\sima$ holds in
(i) if and only if $a_1 = a^\prime_1$ and $a_0 = a^\prime_0$.}
\smallskip

{\it Proof:} We give the proof assuming that the numbers $a_0, a_1,
a^\prime_0, a^\prime_1$ are all positive and $a_0 + a_1 = a^\prime_0
+ a^\prime_1=1$. The other cases are similar. We write $a_1=\lambda$
and $a_1'=\lambda'$.

(i) $\Rightarrow$ (ii).  If $\lambda > \lambda^\prime$ then, by A5 and
A3, $((1 - \lambda) X_0, \lambda^\prime  X_1, (\lambda -
\lambda^\prime) X_1) \prec ((1 - \lambda) X_0, (\lambda
-\lambda^\prime) X_0, \lambda^\prime X_1)$.  By the cancellation law,
Theorem 2.1, $((\lambda - \lambda^\prime) X_1) \prec ((\lambda -
\lambda^\prime) X_0)$.  By scaling invariance, A5,  $X_1 \prec X_0$,
which contradicts $X_0 \prec\prec X_1$.  \hfill\break (ii)
$\Rightarrow$ (i).  This follows from the following computation.
$$
\eqalignii {((1-\lambda)X_0, \lambda X_1) &\sima
((1-\lambda^\prime)X_0, (\lambda^\prime - \lambda)X_0, \lambda X_1)
\quad &\hbox{(by axioms A3 and A5)} \cr &\prec ((1-\lambda^\prime)X_0,
(\lambda^\prime - \lambda)X_1, \lambda X_1) \quad &\hbox{(by axioms A3
and A4)} \cr &\sima ((1-\lambda^\prime)X_0, \lambda^\prime X_1)  \quad
&\hbox{(by axioms A3 and A5).} \cr}
$$ 
\hfill\lanbox

The next lemma will imply, among other things, that entropy is unique,
up to an affine transformation. \medskip

{\bf LEMMA 2.3 (Characterization of entropy).}  {\it Let $S_\Gamma$ 
denote the canonical entropy (2.14) on $\Gamma$ with respect to the 
reference points 
$X_0\prec\prec X_1$. If $X \in \Gamma$
then the equality
$$
\lambda = S_\Gamma (X)
$$
is equivalent to
$$
X \sima ((1 - \lambda) X_0, \lambda X_1).
$$}
\smallskip

 {\it Proof:}  First, if $\lambda = S_\Gamma(X)$ then, by the
definition of supremum, there is a sequence $\varepsilon_1 \geq
\varepsilon_2 \geq \dots \geq 0$ converging to zero, such that
$$
((1 - (\lambda - \varepsilon_n)) X_0, (\lambda - \varepsilon_n) X_1)
\prec X$$
for each $n$.  Hence, by A5,
$$((1 - \lambda) X_0, \lambda X_1, \varepsilon_n X_0) \sima ((1 -
\lambda + \varepsilon_n) X_0, (\lambda - \varepsilon_n) X_1,
\varepsilon_n X_1) \prec (X, \varepsilon_n X_1),$$ and thus $((1 -
\lambda) X_0, \lambda X_1) \prec X$ by the stability
property A6.  On the other hand, since $\lambda$ is the supremum we have
$$X \prec ((1 - (\lambda + \varepsilon) X_0, (\lambda + \varepsilon)
X_1)
$$
for all $\varepsilon > 0$ by the comparison hypothesis CH.  Thus,
$$
(X, \varepsilon X_0) \prec ((1 - \lambda) X_0, \lambda X_1,
\varepsilon X_1),
$$
so, by A6,  $X \prec ((1 - \lambda) X_0, \lambda X_1)$.  This shows that 
$X 
\sima
((1 - \lambda) X_0, \lambda X_1)$ when $\lambda = S_\Gamma(X)$.

Conversely, if $\lambda^\prime \in [0,1]$ is such that $X \sima ((1 -
\lambda^\prime) X_0, \lambda^\prime X_1)$, then $((1 - \lambda^\prime)
X_0, \lambda^\prime X_1) \sima ((1 - \lambda) X_0, \lambda X_1)$ by
transitivity.  Thus, $\lambda = \lambda^\prime$ by Lemma 2.2.
\hfill\lanbox
\smallskip
%%%%%%%%
{\it Remark 1:}  Without the comparison hypothesis we could find that
$S_\Gamma(X_0)= 0$ and $S_\Gamma(X) = 1$ for all $X$ such that $X_0 \prec 
X$.
\smallskip

{\it Remark 2:} {}From Lemma 2.3 and the cancellation law it
follows that the canonical entropy with reference points $X_0\prec\prec
X_1$ satisfies $0\leq S_\Gamma (X)\leq 1$ if and only if $X$ belongs to 
the
{\bf strip} $\Sigma (X_0, X_1)$ defined by
$$
\Sigma (X_0, X_1) := \{ X \in \Gamma : X_0 \prec X \prec X_1 \} \subset
\Gamma.$$
Let us make the dependence of the canonical entropy on $X_0$ 
and $X_1$ explicit by writing 
$$
S_\Gamma(X)=S_\Gamma(X\vert X_0,X_1) \ . \eqno(2.15)
$$ 
For
$X$ outside the strip we can then write 
$$
S_\Gamma (X \vert X_0, X_1)= S_\Gamma (X_1 \vert X_0, X)^{-1}
\qquad\hbox{if\ }X_1\prec X$$
and
$$
S_\Gamma (X \vert X_0, X_1)= -{S_\Gamma ( X_0\vert X, X_1)\over 
1-S_\Gamma 
(X_0 \vert X, X_1)}
\qquad\hbox{if\ }X\prec X_0.
$$
\smallskip
%\vfill\eject
{\tt Proof of Theorem 2.2:}

{\it (1) $\Longrightarrow$ (2):}   Put $\lambda_i=S_\Gamma(Y_i)$,
$\lambda_i'=S_\Gamma(Y_i')$. By Lemma 2.3  we know that $Y_i\sima
((1-\lambda_i)X_0,\lambda_i X_1)$ and
$Y_i'\sima ((1-\lambda_i')X_0,\lambda_i' X_1)$. By the consistency 
axiom A3   and the recombination axiom A5  it follows that
$$
(t_1Y_1,\dots,t_NY_N)\sima (\sum_i t_i(1-\lambda_i)X_0, \sum_i 
t_i\lambda_i X_1)
$$
and 
$$
(t_1'Y_1',\dots,t_N'Y_N')\sima (\sum_i t_i'(1-\lambda_i')X_0, \sum_i 
t_i'\lambda_i' X_1) \ .
$$
Statement (2) now follows from Lemma 2.2. 
The implication (2) $\Longrightarrow$ (1) is obvious.

The  proof of Theorem 2.2 is now complete except for the uniqueness part.
We formulate this part separately in Theorem 2.3 below, which is slightly 
stronger than
the last assertion in Theorem 2.2.
It implies that an entropy function for the multiple scaled copies of 
$\Gamma$ is 
already uniquely 
determined, up to an affine transformation, by the relation on states of 
the form $((1-\lambda)X,\lambda Y)$, i.e., it requires only the case
$N=M=2$, in the notation of Theorem 2.2.
\medskip

{\bf THEOREM 2.3 (Uniqueness of entropy)} {\it If $S_\Gamma^*$ is a 
function on $\Gamma$ that satisfies
$$
((1-\lambda)X,\lambda Y)\prec((1-\lambda)X',\lambda Y')
$$
if and only if
$$
(1-\lambda)S_\Gamma^*(X)+\lambda 
S_\Gamma^*(Y)\leq(1-\lambda)S_\Gamma^*(X')+\lambda 
S_\Gamma^*(Y'),
$$
for all $\lambda\in\R$ and $X,Y,X',Y'\in\Gamma$, then 
$$
S_\Gamma^*(X)=aS_\Gamma(X)+B
$$ 
with 
$$
a=S_\Gamma^*(X_1)-S_\Gamma^*(X_0)>0,\qquad B=S_\Gamma^*(X_0).
$$
Here $S_\Gamma$ is the canonical entropy on $\Gamma$ 
with reference points $X_0\prec\prec X_1$.}
\medskip

{\it Proof:} This follows immediately from Lemma 2.3, which says that 
for every $X$ there is a unique $\lambda$, namely 
$\lambda=S_\Gamma(X)$, such that 
$$X\sima 
((1-\lambda)X,\lambda X)\sima ((1-\lambda)X_0,\lambda X_1).$$
Hence, by the hypothesis on $S_\Gamma^*$, and $\lambda=S_\Gamma(X)$, we 
have
$$
S_\Gamma^*(X)=(1-\lambda)S_\Gamma^*(X_0)+
\lambda S_\Gamma^*(X_1) = [S_\Gamma^*(X_1)-S_\Gamma^*(X_0)]S_\Gamma(X)
+S_\Gamma^*(X_0).
$$
The hypothesis on $S_\Gamma^*$ also implies that $a:= 
S_\Gamma^*(X_1)-S_\Gamma^*(X_0) >0$, 
because $X_0\prec\prec X_1$.\hfill\lanbox
\medskip

{\it Remark:} Note that $S_\Gamma^*$ is defined on $\Gamma$ and satisfies
$S_\Gamma^*(X) = aS_\Gamma(X)+B$ there. On the space $\Gamma^{(t)}$ 
a corresponding entropy is, {\it by definition}, given by
$S_{\Gamma^{(t)}}^*(tX) = tS_\Gamma^*(X)= atS_\Gamma(X) + tB =
aS_\Gamma^{(t)}(tX) + tB$, where $S_\Gamma^{(t)}(tX)$ is the canonical
entropy on $\Gamma^{(t)}$ with reference points $tX_0, tX_1$.  Thus,
$S_{\Gamma^{(t)}}^*(tX)\neq  aS_\Gamma^{(t)}(tX) +B$ \ (unless $B=0$,
of course).  \bigskip

It is apparent from formula (2.14) that the definition of the canonical
entropy function on $\Gamma$ involves only the relation $\prec$ on the
double scaled products $\Gamma^{(1-\lambda)}\times \Gamma^{(\lambda)}$
besides the reference points $X_0$ and $X_1$. Moreover, the canonical
entropy uniquely characterizes the relation on all multiple scaled 
copies
of $\Gamma$, which implies in particular that CH holds for all multiple
scaled copies. Theorem 2.3 may therefore be rephrased as follows: 
\medskip

{\bf THEOREM 2.4 (The relation on double scaled copies determines the
relation everywhere).} {\it Let $\prec$ and $\prec^*$ be two relations on
the multiple scaled copies of $\Gamma$ satisfying axioms A1-A6, and also
CH for $\Gamma^{(1-\lambda)}\times \Gamma^{(\lambda)}$ for each fixed
$\lambda\in[0,1]$.  If $\prec$ and $\prec^*$ coincide on
$\Gamma^{(1-\lambda)}\times \Gamma^{(\lambda)}$ for each 
$\lambda\in[0,1]$,
then $\prec$ and $\prec^*$ coincide on all multiple scaled copies of
$\Gamma$, and CH holds on all the multiple scaled copies.}
\medskip
 The proof of Theorem 2.2 is now complete.

\bigskip\noindent
{\subt E.  Construction of a universal entropy in the absence of mixing}
\bigskip

In the previous subsection we showed how to construct an entropy 
for a single system, $\Gamma$, that exactly describes the 
relation $\prec$ within the states obtained by forming multiple
scaled copies of $\Gamma$. It is unique up to a multiplicative
constant $a>0$ and an additive constant $B$, i.e., to within an
affine transformation. We remind the reader that this entropy was
constructed by considering just the product of two scaled copies of 
$\Gamma$, but 
our axioms implied that it automatically worked for {\it all} 
multiple scaled copies of $\Gamma$. We shall refer to $a$ and $B$ as {\bf 
entropy constants} for the system $\Gamma$. 

Our goal is to put these entropies together
and show that they behave in the right way on products of
arbitrarily many copies of {\it different} systems. Moreover,
this \lq universal\rq\ entropy will be unique up to {\it one}
multiplicative constant---but still many additive constants.
The central question here is one of {\it \lq calibration\rq\ },
which is to  say that the multiplicative constant in front of 
each elementary entropy has to be chosen in such a way that
the additivity rule (2.4) holds. It is not even obvious yet
that the additivity can be made to hold at all, whatever the choice
of constants.

Let us note that the number of additive constants depends heavily on the
kinds of adiabatic processes available. The system consisting of one
mole of hydrogen mixed with one mole of helium and the system
consisting of one mole of hydrogen mixed with two moles of helium are
different.  The additive constants are independent {\it unless} a
process exists in which both systems can be unmixed, and thereby making
the constants comparable. In nature we expect only 92 constants, one for
each element of the periodic table, unless we allow nuclear processes as
well, in which case there are only two constants (for neutrons and for
hydrogen).  On the other hand, if un-mixing is not allowed uncountably
many constants are undetermined. In Section VI we address the question
of adiabatic processes that unmix mixtures and reverse chemical
reactions. That such  processes exist is not so obvious.

To be precise, the principal goal  of this subsection is the proof of
the following Theorem 2.5, which is a  case of the entropy principle
that is special in that it is restricted to processes that do not
involve  mixing or chemical reactions. It is a generalization of Theorem
2.2. 
\medskip

{\bf THEOREM 2.5 (Consistent entropy scales). } {\it Consider a family of 
systems fulfilling the following requirements:

\item{(i)} 
The state spaces of any two systems in the family are disjoint sets, 
i.e., 
every 
state of a system in the family belongs to exactly one state space.

\item{(ii)} All multiple scaled products of systems in the family
belong also to the family.  

\item{(iii)} Every system in the family satisfies the 
comparison hypothesis. 

For each state space $\Gamma$ of a system in the family let
$S_{\Gamma}$ be some d
efinite entropy function on $\Gamma$. Then 
there are constants $a_{\Gamma}$ and $B_{\Gamma}$ such that the 
function $S$, defined for all states in all $\Gamma$'s by
$$
S(X)= a_{\Gamma} S_{\Gamma} (X)+ B_{\Gamma}
$$
for $X\in \Gamma$, has the following properties:
\smallskip
\item{a).} If $X$ and $Y$ are  in the same state space then
$$
X\prec Y \quad\quad \hbox{\rm if and only if} \quad\quad S(X)\leq 
S(Y).
$$ \smallskip

\item{b).} $S$ is additive and extensive, i.e.,
$$
S(X,Y) = S(X)+S(Y). \eqno (2.4)
$$ 
and, for $t>0$,
$$
S(tX) = tS(X). \eqno (2.5)
$$  }
\medskip
{\it Remark.\/} Note that $\Gamma_1$ and $\Gamma_1\times \Gamma_2$ are 
disjoint 
as sets for any (nonempty) state spaces $\Gamma_1$ and $\Gamma_2$.
\medskip

{\it Proof:} Fix some system $\Gamma_0$ and two points $Z_0\prec 
\prec Z_1$
in  $\Gamma_0$.  In each state space $\Gamma$ choose some fixed point 
$X_{\Gamma} \in \Gamma$ in such a way that the identities
$$\eqalignno{
X_{\Gamma_1  \times \Gamma_2}&= (X_{\Gamma_1}, X_{\Gamma_2}) 
&(2.16)\cr  \noalign{\smallskip} 
X_{t\Gamma}                 &= tX_{\Gamma}&(2.17)\cr }
$$
hold.  With the aid or the axiom of choice this can  be achieved by
considering the formal vector space spanned by all systems and choosing a
Hamel basis of systems $\{\Gamma_{\alpha}\}$ in this space such that 
every
system can be written uniquely as a scaled product of a finite number of
the $\Gamma_{\alpha}$'s. (See Hardy, Littlewood and Polya, 1934). The
choice of an arbitrary state $X_{\Gamma_{\alpha}}$ in each of these
`elementary' systems $\Gamma_{\alpha}$ then defines for each $\Gamma$ a
unique $X_{\Gamma}$ such that (2.17) holds.  (If the reader does not wish
to invoke the axiom of choice then an alternative is to hypothesize that
every system has a unique decomposition into elementary systems; the 
simple
systems considered in the next section obviously qualify as the 
elementary
systems.) 
 
For $X\in \Gamma$ we consider the space $\Gamma \times \Gamma_0$ with 
its
canonical entropy as defined in (2.14), (2.15) relative to the points
$(X_{\Gamma}, Z_0)$ and $(X_{\Gamma}, Z_1)$. Using this function we 
define
$$
S(X)= S_{\Gamma \times \Gamma_0}((X,Z_0) \, \, \vert \, \, (X_{\Gamma}
, Z_0),(X_{\Gamma} , Z_1)). \eqno(2.18)
$$

Note: Equation (2.18) fixes the entropy of $X_{\Gamma}$ to be zero.

Let us denote $S(X) $ by $\lambda$ which, by Lemma 2.3, is 
characterized by
$$
(X,Z_0) \sima ( (1-\lambda ) (X_{\Gamma} , Z_0) , \lambda  
(X_{\Gamma} , Z_1)).
$$
By the cancellation law this is equivalent to
$$
(X,\lambda Z_0)\sima  (X_{\Gamma}, \lambda Z_1)).  \eqno(2.19)
$$

By (2.16) and (2.17) this immediately implies the additivity and
extensivity of $S$.  Moreover, since $X\prec Y$ holds if and only if $(X,
Z_0) \prec (Y,Z_0) $ it is also clear that $S$ is an entropy function on
any $\Gamma$.  Hence $S$ and $S_{\Gamma}$ are related by an affine
transformation, according to Theorem 2.3.  \hfill \lanbox

\medskip

{\bf Definition (Consistent entropies).} A collection of entropy 
functions $S_\Gamma$ on state spaces $\Gamma$ is called {\it 
consistent} if the appropriate linear combination of the functions is 
an entropy function on all multiple scaled products of these state 
spaces.  In other words, the set is consistent if the multiplicative 
constants $a_{\Gamma}$, referred to in Theorem 2.5, can all be chosen 
equal to 1.  \smallskip

\underbar{{\it Important Remark:}}
{}From the definition, (2.14), of the canonical entropy
and (2.19) it follows that  the entropy (2.18) is given by the formula
$$
S(X) = \sup \{ \lambda \, \, : \, \, (X_{\Gamma}, \lambda Z_1)
\prec (X , \lambda Z_0) \}   \eqno (2.20)
$$
for $X\in\Gamma$.  The auxiliary system $\Gamma_0$ can thus be 
regarded as an `entropy meter' in the spirit of (Lewis and Randall, 
1923) and (Giles, 1964).  Since we have chosen to define the entropy 
for each system independently, by equation (2.14), the role of $\, 
\Gamma_0$ in our approach is solely to calibrate the entropy of 
different systems in order to make them consistent.

\medskip

{\it Remark about the photon gas:\/} As we discussed in Section II.B
the photon gas is special and there are two ways to view it.  One way
is to regard the scaled copies $\Gamma^{(t)}$ as distinct systems and
the other is to say that there is only one $\Gamma$ and the scaled
copies are identical to it and, in particular, must have exactly the
same entropy function.  We shall now see how the first point of view
can be reconciled with the latter requirement.  Note, first, that in
our construction above we cannot take the point $(U,V)=(0,0)$ to be the
fiducial point $X_{\Gamma}$ because $(0,0)$ is not in our state space
which, according to the discussion in Section III below, has to be an
open set and hence cannot contain any of its boundary points such as
$(0,0)$. Therefore, we have to make another choice, so let us take
$X_{\Gamma}= (1,1)$. But the construction in the proof above sets
$S_{\Gamma} (1,1)= 0$ and therefore $S_{\Gamma}(U,V) $ will not have
the homogeneous form $S^{\rm hom}(U,V)= V^{1/4}U^{3/4}$.  Nevertheless,
the entropies of the scaled copies will be extensive, as required by
the theorem.  If one feels that all scaled copies should have the same
entropy (because they represent the same physical system) then the
situation can be remedied in the following way: With $S_{\Gamma}(U,V) $
being the entropy constructed as in the proof using $(1,1)$, we note
that $S_{\Gamma}(U,V) = S^{\rm hom}(U,V) + B_{\Gamma}$ with the
constant $B_{\Gamma}$  given by $B_{\Gamma}= -S_{\Gamma}(2,2)$. This
follows from simple algebra and the fact that we know that the entropy
of the photon gas constructed in our proof must equal $S^{\rm hom}$ to
within an additive constant. (The reader might ask how we know this and
the answer is that the entropy of the `gas' is unique up to additive
and multiplicative constants, the latter being determined by the system
of units employed.  Thus, the entropy determined by our construction
must be the `correct entropy', up to an additive constant, and this
`correct entropy' is what it is, as determined by physical measurement.
Hopefully it agrees with the function deduced in (Landau and Lifschitz,
1969).) Let us use our freedom to alter the additive constants as we
please, provided we maintain the extensivity condition (2.5).  It will
not be until Section VI that we have to worry about the additive
constants {\it per se} because it is only there that mixing and
chemical reactions are treated.  Therefore, we redefine the entropy of
the state space $\Gamma$ of the photon gas to be $S^*(U,V) :=
S_{\Gamma}(U,V) + S_{\Gamma}(2,2)$.  which is the same as $S^{\rm
hom}(U,V)$. We also have to alter the entropy of the scaled copies
according to the rule that preserves extensivity, namely
$S_{\Gamma^{(t)}}(U,V) \rightarrow S_{\Gamma^{(t)}}(U,V)
+tS_{\Gamma}(2,2) =S_{\Gamma^{(t)}}(U,V) + S_{\Gamma^{(t)}}(2t,2t) =
S^{\rm hom}(U,V)$.  In this way, all the scaled copies now have the
same (homogeneous) entropy, but we remind the reader that the same
construction could be carried out for any material system with a
homogeneous (or, more exactly an affine) entropy function---if one
existed. {}From the thermodynamic viewpoint, the photon gas is unusual
but not special. \bigskip \bigskip

\bigskip\noindent
{\subt F. Concavity of entropy}
\bigskip                                                        

Up to now we have not used, or assumed, any geometric property of a
state space $\Gamma$.  It is an important stability  property of
thermodynamical systems, however, that the entropy function is a {\it
concave} function of the state variables ---a requirement that was
emphasized by Maxwell, Gibbs, Callen and many others.  Concavity also
plays an important role in the definition of temperature, as in section
V.

In order to have this concavity it is first necessary to make the
state space on which entropy is defined into a convex set, and for
this purpose the choice of coordinates is important. Here, we begin
the discussion of concavity by discussing this geometric property
of the underlying state space and some of the consequences of the
{\it convex combination axiom} A7 for the relation $\prec$, to be given
after the following definition.

{\bf Definition:} By a {\bf state space with  a 
convex structure}, or simply a {\bf convex state space}, we 
mean a
state space $\Gamma$, that is a convex subset of some 
linear space, e.g., $\R^n$. That is, if $X$ and $Y$ are any two 
points in $\Gamma$ and if $0 \leq t \leq 1$,  then
the point $tX + (1-t)Y$ is a well-defined point in $\Gamma$. 
A {\it concave function}, $S$,  on $\Gamma$ is one satisfying the
inequality
$$
S(tX + (1-t)Y) \geq tS(X) + (1-t)S(Y).  \eqno(2.21)
$$

Our basic convex combination axiom for the relation $\prec$ is the
following.
\medskip

\item{\bf A7)}  {\bf Convex combination.} 
Assume $X$ and $Y$ are states in the same {\it convex} state space,
$\Gamma$. For $t \in [0,1]$ let $tX$ and  
$(1-t)Y$ be the corresponding states of their $t$ scaled and
$(1-t)$ scaled copies, respectively. Then the point $(t X, (1-t)
Y)$ in the product space $\Gamma^{(t)}\times
\Gamma^{(1-t)}$ satisfies
$$ 
(t X, (1-t) Y) \prec t X + (1-t)Y\ . \eqno(2.22) 
$$
Note that the right side of (2.22) is in $\Gamma$ and is defined 
by ordinary
convex combination of points in the convex set $\Gamma$.
\medskip

The physical meaning of A7 is more or less evident, but it is essential
to note that the 
convex structure depends heavily on the choice of coordinates for
$\Gamma$.  A7 means that if we take a bottle containing $1/4$ moles of
nitrogen and one containing $3/4$ moles (with possibly different
pressures and densities), and if we mix them together, then among the
states of one mole of nitrogen that can be reached adiabatically there
is one in which the energy is the sum of the two energies and, likewise,
the volume is the sum of the two volumes. Again, we emphasize that the
choice of energy and volume as the (mechanical) variables with which we
can make this statement is an important assumption. If, for example,
temperature and pressure were used instead, the statement would not only
not hold, it would not even make much sense.

The physical example above seems not exceptionable for liquids and 
gases.  On the other hand it is not entirely clear how to ascribe an 
operational meaning to a convex combination in the state space of a 
solid, and the physical meaning of axiom A7 is not as obvious in this 
case.  Note, however, that although convexity is a global property, it 
can often be inferred from a local property of the boundary.  (A 
connected set with a smooth boundary, for instance, is convex if every 
point on the boundary has a neighbourhood, whose intersection with the 
set is convex.)  In such cases it suffices to consider convex 
combinations of points that are close together and close to the 
boundary.  For small deformation of an isotropic solid the six strain 
coordinates, multiplied by the volume, can be taken as work 
coordinates.  Thus, A7 amounts to assuming that a convex combination 
of these coordinates can always be achieved adiabatically.  See, e.g., 
(Callen, 1985).

If $X \in \Gamma$ we denote by $A_X$ the set $\{ Y \in \Gamma : X
\prec Y \}$. $A_X$ is called the {\bf forward sector } of $X$ in
$\Gamma$.  More generally, if $\Gamma^\prime $ is another system,
we call the set
$$
\{Y\in \Gamma': X\prec Y\},
$$
the forward sector of $X$ in $\Gamma^\prime $.  

Usually this concept is applied to the case in which  $\Gamma$ and
$\Gamma^\prime $ are identical, but it can also be useful in cases in
which one system is changed into another; an example is the mixing of
two liquids in two containers (in which case $\Gamma $ is a compound
system) into  a third vessel containing the mixture (in which case
$\Gamma^\prime $ is simple).

The main effect of A7 is that forward sectors are convex sets. 
\medskip

{\bf THEOREM 2.6} {\bf (Forward sectors are convex).} {\it Let
$\Gamma$ and $\Gamma'$ be state spaces of two systems, with
$\Gamma'$ a convex state space. Assume that A1--A5 hold for
$\Gamma$ and $\Gamma'$ and, in addition, A7 holds for
$\Gamma'$. Then the forward sector of $X$ in $\Gamma'$, defined
above, is a {\it convex\/} subset of $\Gamma'$ for each $X\in
\Gamma$.}
\smallskip

{\it Proof:\/} Suppose $X\prec Y_1$ and $X\prec Y_2$ and that $0<t <1$.
We want to show that $X\prec t Y_1+(1-t)Y_2$. (The right side defines,
by ordinary vector addition, a point in the convex set $\Gamma'$.  )
First, $X\prec (t X,(1-t)X)\in \Gamma^{(t)}\times \Gamma^{(1-t)}$, by
axiom A5. Next, $(t X,(1-t)X)\prec (t Y_1,(1-t)Y_2)$ by the
consistency axiom A3 and the scaling invariance axiom
A4. Finally, $(t Y_1,(1-t)Y_2)\prec t Y_1+(1-t)Y_2$ by the convex
combination axiom A7.\nobreak\hfill\lanbox
\medskip

Figure 3 illustrates this theorem in the case $\Gamma = \Gamma'$. 

\centerline{\sevenpoint ---- Insert Figure 3 here ----}

{\bf THEOREM 2.7 (Convexity of $\S_{\lambda}$).}  {\it Let the sets 
$\S_\lambda 
\subset \Gamma$ 
be defined as in (2.12) and assume the state space $\Gamma$ satisfies the 
convex combination axiom A7 
in addition to A1-A5. Then:

(i) $\S_\lambda$ is convex. \hfill 

(ii) If $X\in \S_{\lambda_1}$, $Y\in \S_{\lambda_2}$ and 
$0\leq t\leq 1$,  then $tX+(1-t)Y\in {\cal
S}_{t\lambda_1+(1-t)\lambda_2}$. \hfill 
}
\medskip

{\it Proof.} (i) This follows immediately from the scaling, splitting and
convex combination axioms A4, A5 and A7.

(ii) This is proved by splitting, moving the states of the subsystems 
into 
forward sectors and bringing the subsystems together at the end.  More 
precisely, defining $\lambda=t\lambda_1+(1-t)\lambda_2$ we have to show 
that $((1 - \lambda) X_0, \lambda X_1) \prec tX + (1 - t)Y$.  Starting 
with 
$((1-\lambda)X_0,\lambda X_1)$ we split $(1-\lambda)X_0$ into 
$(t(1-\lambda_1)X_0, (1-t)(1-\lambda_2)X_0)$ and $\lambda X_1$ into 
$(t\lambda_1X_1,(1-t)\lambda_2X_1)$.  Next we consider the states 
$(t(1-\lambda_1)X_0,t\lambda_1X_1)$ and 
$((1-t)(1-\lambda_2)X_0,(1-t)\lambda_2X_1)$.  By scaling
invariance A4 and the splitting property A5 we 
can pass from the former to $(t(1-\lambda_1)X,t\lambda_1X)$ and from the 
latter to $((1-t)(1-\lambda_2)Y,(1-t)\lambda_2Y)$.  Now we combine the 
parts 
of $(t(1-\lambda_1)X,t\lambda_1X)$ to obtain $tX$ and the parts of 
$((1-t)(1-\lambda_2)Y,(1-t)\lambda_2Y)$ to obtain $(1-t)Y$, and finally 
we 
use the convex combination property A7 to reach 
$tX+(1-t)Y$.\nobreak\hfill\lanbox

\bigskip
\medskip

{\bf THEOREM 2.8 (Concavity of entropy).}  
{\it Let $\Gamma$ be a convex state space. Assume 
axiom A7 in addition to A1-A6, and CH for multiple scaled copies of 
$\Gamma$. 
Then the entropy  $S_{\Gamma}$ defined by (2.14) is a  concave function 
on 
$\Gamma$.
Conversely, if $S_{\Gamma}$ is concave,
then axiom A7 necessarily holds a-fortiori.
}

{\it Proof:}  If $X \in \S_{\lambda_1}, Y \in \S_{\lambda_2}$, then by 
Theorem
2.7, (ii), $t X + (1 - t)Y \in \S_{t \lambda_1 + (1-t)\lambda_2}$, for 
$t,
\lambda_1, \lambda_2 \in [0,1]$.  By definition, this implies $S_{\Gamma} 
(t X +
(1 - t)Y) \geq t \lambda_1 + (1 - t)\lambda_2$.  Taking the supremum over 
all
$\lambda_1$ and $\lambda_2$ such that $X \in \S_{\lambda_1}, Y \in
\S_{\lambda_2}$, then gives $S_{\Gamma}(t X + (1 - t)Y) \geq t S_{\Gamma} 
(X) +
(1 - t) S_{\Gamma}(Y)$.  The converse is obvious.\hfill\lanbox

\bigskip\noindent
{\subt G. Irreversibility and Carath\'eodory's principle}
\bigskip

One of the milestones in the history of the second law is
Carath\'eodory's attempt to formulate the second law in terms of purely
local properties of the equivalence relation $\sima$.  The
disadvantage of the purely local formulation is, as we said earlier, the
difficulty of deriving a globally defined concave entropy function.
Additionally, Carath\'eodory relies on differentiability (differential
forms), and we would like to avoid this, if possible, because physical
systems do have points (e.g., phase transitions) in their state spaces
where differentiability fails. Nevertheless, Carath\'eodory's idea
remains a powerful one and it does play  an important role in the story.
We shall replace it by a seemingly more natural idea, namely the
existence of irreversible processes.  {\it The existence of many such
processes lies at the heart of thermodynamics.\/} If they did not exist,
it would mean that nothing is forbidden, and hence there would be no
second law. We now  show the relation between the two concepts. There
will be no mention of differentiability, however. 

Carath\'eodory's principle has been criticized (see, for example, the 
remark attributed to Walter in Truesdell's paper in (Serrin, 1986, 
Chapter 5)) on the ground that this principle does not tell us where 
to look for a non adiabatic process that is supposed, by the 
principle, to exist in every neighborhood of every state.  In Sect.  
III and V we show that this criticism is too severe because the 
principle, when properly interpreted, shows exactly where to look and, 
in conjunction with the other axioms, it leads to the Kelvin-Planck 
version of the second law.

\medskip

{\bf THEOREM 2.9 (Carath\'eodory's principle and irreversible 
processes).}
{\it Let $ \Gamma$ be a state space that is  a convex subset of ${\bf 
R}^n$
and assume that axioms A1--A7 hold on $ \Gamma$. Consider the following
two statements. 
\item{(1)}{\bf Existence of irreversible processes:}
For every point $X \in   \Gamma$ there is a $Y \in  \Gamma$ such that
$X \prec \prec Y$.
\item{(2)}{\bf Carath\'eodory's principle:} In every
neighborhood of every $X \in   \Gamma$ there is a point $Z\in  \Gamma$
such that $X \sima Z$ is false. 
\smallskip
Then (1) always implies (2). Indeed,  (1) implies the stronger
statement that there is a $Z$ such that $X \prec Z$ is false. 
On the other hand, if all the forward sectors in $ \Gamma$ 
have non-empty interiors (i.e., they are not contained in lower 
dimensional
hyperplanes) then (2) implies (1). }
\medskip

{\it Proof: \/}  
Suppose that for some $X \in   \Gamma$ there is  a neighborhood, ${\cal
N}_X$ of $X$ such that ${\cal N}_X$ is contained in $A_X$, the forward
sector of $X$. (This is the negation of the statement that in every
neighbourhood of every $X$ there is a $Z$ such that $X\prec Z$ is
false.) Let $Y\in A_X$ be arbitrary.  By the convexity of $A_X$ (which
is implied by the axioms), $X$ is an interior point of a line segment
joining $Y$ and some point $Z\in {\cal N}_X$.
By axiom A7, we thus have 
$$
((1-\lambda )Z,\lambda Y) \prec X \sima ((1-\lambda )X,\lambda X)
$$
for some $\lambda \in (0,1)$. But we also have that $((1-\lambda)X,
\lambda Y)\prec ((1-\lambda )Z,\lambda Y) $ since $Z\in A_X$. This
implies, by the cancellation law, that $Y\prec X$. 
Thus we conclude that for some $X$, we have that $X\prec Y$ implies 
$X\sima 
Y$. This contradicts (1). In particular, we have shown that (1)
$\, \Rightarrow$(2).
\smallskip

Conversely, assuming that (1) is false, there is a point 
$X_0$
whose forward sector is given by $A_{X_0} = \{ Y:Y\sima X_0 \}$. Let 
$X$ be an interior point of $A_{X_0}$, i.e., there is a neighborhood
of $X$, ${\cal N}_{X}$, which is entirely contained in  $A_{X_0}$.
All points in ${\cal N}_{X}$ are adiabatically equivalent to $X_0$, 
however, 
and hence to $X$, since $X\in {\cal N}_{X}$. Thus, (2) is false. 
\hfill\lanbox

\bigskip\noindent
{\subt H. Some further results on uniqueness}
\bigskip

As stated in Theorem 2.2, the existence of an entropy function on a state 
space $\Gamma$ is equivalent to the axioms A1-A6 and CH for the multiple 
scaled copies of $\Gamma$.  The entropy function is unique, up to an 
affine change of scale, and according to  formula (2.14) it is even 
sufficient to know the relation on the double scaled copies 
$\Gamma^{(1-\lambda)}\times\Gamma^{(\lambda)}$ in order to compute the 
entropy.  This was the observation behind the uniqueness Theorem 2.4
which stated 
that the restriction of the relation $\prec$ to the double scaled copies 
determines the relation everywhere.

The following very general result shows that it is in fact not 
necessary to know $\prec$ on all 
$\Gamma^{(1-\lambda)}\times\Gamma^{(\lambda)}$ to determine the 
entropy, provided the relation is such that the range of the entropy 
is connected.  In this case $\lambda=1/2$ suffices.  By Theorem 2.8 
the range of the entropy is necessarily connected if the convex 
combination axiom A7 holds.  \medskip

{\bf THEOREM 2.10 (The relation on $\Gamma\times\Gamma$ determines 
entropy).}  
{\it Let $\Gamma$ be a set
and $\prec$ a relation on $\Gamma\times \Gamma$.  Let
$S$ be a real valued function on $\Gamma$ satisfying the following 
conditions:
\item{(i)} $S$ characterizes the relation on $\Gamma\times \Gamma$ in the 
sense that 
$$(X,Y)\prec (X',Y')\qquad\hbox{\sl if and only if}\qquad S(X)+S(Y)\leq
S(X')+S(Y')$$
\item{(ii)} The range of $S$ is an interval (bounded or unbounded and 
which
could even be a point).

Let $S^*$ be another function on 
$\Gamma$ satisfying condition (i). Then $S$ and 
$S^*$ are affinely related, i.e., there are numbers 
$a > 0$ and $B$ such that $S^*(X) = a S(X) + B$ 
for all $X \in \Gamma$. In particular, $S^*$ must satisfy condition 
(ii). }
\medskip

{\it Proof:}  In general, if $F$ and $G$ are any two real valued 
functions 
on $\Gamma
\times \Gamma$, 
such that $F(X,Y)\leq F(X',Y')$ if and only if $G(X,Y)\leq G(X',Y')$,
it is an easy logical exercise to show that there is a
monotone increasing function $K$ (i.e., $x\leq y$ implies $K(x)\leq 
K(y)$)
defined on the range of $F$, 
so that $G = K \circ F$.  In our
case $F(X,Y)=S(X) + S(Y)$. If the range of $S$ is the interval $L$ then 
the
range of $F$ is $2L$. Thus $K$, which is
defined on $2L$, satisfies
$$
K(S(X) + S(Y)) = S^* (X) + S^* (Y) \eqno(2.23)
$$
for all $X$ and $Y$ in $\Gamma$ because both $S$ and $S^*$ satisfy 
condition (i).  For convenience, define $M$ on $L$ by
$M(t) = \mfr1/2 K (2t)$.  If we now set $Y = X$ in (1) we obtain 
$$
S^* (X) = M (S(X)), \quad X \in \Gamma \eqno(2.24)
$$
and (2.23) becomes, in general,
$$ M \left( {x+y \over 2} \right) = \mfr1/2 M(x) + \mfr1/2 
M(y)\eqno(2.25)
$$
for all $x$ and $y$ in $L$.  Since $M$ is monotone, it is bounded on all 
finite subintervals of $L$. Hence (Hardy, Littlewood, Polya 1934)
$M$ is both concave and convex in the usual sense, i.e.,
$$
M (t x + (1- t) y) = t M(x) + (1 - t) M(y)
$$
for all $0 \leq t \leq 1$ and $x,y \in L$.  {}From this it follows
that $M(x) = a x + B$ with $a\geq 0$.  If $a$ were
zero then $S^*$ would be constant on $\Gamma$ which would imply that
$S$ is constant as well.  In that case we could always replace $a$
by 1 and replace $B$ by $B-S(X)$.
 \hfill\lanbox

{\it Remark:}  It should be noted that Theorem 2.10 does not rely on
any structural property of $\Gamma$, which could be any abstract set.
In particular, continuity plays no role; indeed it cannot be defined
because no topology on $\Gamma$ is assumed. The only residue of
``continuity" is the requirement that the range of $S$ be an interval.

That condition (ii) is not superfluous for the uniqueness
theorem may be seen from the following simple counterexample.

{\bf EXAMPLE:} Suppose the state space $\Gamma$ consists of 3 points,
$X_0$, $X_1$ and $X_2$, and let $S$ and $S^*$ be defined by $S(X_0)=
S^*(X_0)=0$, $S(X_1)=S^*(X_1)=1$, $S(X_2)$=3, $S^*(X_2)$=4.  These
functions correspond to the same order relation on $\Gamma\times
\Gamma$, but they are not related by an affine transformation.

The following sharpening of Theorem 2.4 is an immediate corollary of 
Theorem 2.10 in the 
case that the convexity axiom A7 holds, so that the range of the 
entropy is connected. 

\medskip
{\bf THEOREM 2.11 (The relation on $\Gamma\times\Gamma$ determines the
relation everywhere)} {\it Let $\prec$ and $\prec^*$ be two relations
on the multiple scaled copies of $\Gamma$ satisfying axioms A1-A7,
and CH for $\Gamma^{(1-\lambda)}\times \Gamma^{(\lambda)}$ for each
fixed $\lambda\in[0,1]$.  If $\prec$ and $\prec^*$ coincide on
$\Gamma\times \Gamma$, i.e., $$ (X,Y) \prec (X^{\prime}, Y^{\prime})
\ \ \ {\it if \ and \  only \ if}\ \ \ (X,Y) \prec^* (X^{\prime},
Y^{\prime}) $$ for $X,X',Y,Y'\in\Gamma$, then $\prec$ and $\prec^*$
coincide on all multiple scaled copies of $\Gamma$.} \bigskip

%%%%%
As a last variation on the theme of this subsection let us note that
uniqueness of entropy does even not require knowledge of the order
relation $\prec$ on all of $\Gamma \times \Gamma$. The knowledge of
$\prec$  on a relatively thin ``diagonal" set will suffice, as Theorem
2.12 shows.  \medskip

{\bf THEOREM 2.12 (Diagonal sets determine entropy).} {\it Let
$\prec$ be an order relation on $ \Gamma \times \Gamma$ and let $S$ be
a function on  $\Gamma$ satisfying conditions (i) and (ii) of Theorem
2.10. Let ${\cal D}$ be a subset of $ \Gamma \times \Gamma$ with the
following properties:  

\item{(i)} $(X,X) \in {\cal D}$ for every $X \in  \Gamma$.  
\item{(ii)} The set 
$D= \{(S(X),S(Y))\in {\bf R}^2 \, : \, (X,Y)\in {\cal D} \}$
contains an open subset of ${\bf R}^2$ (which necessarily contains the
set $\{(x,x) : x\in {\rm Range}\, S\}$).

\medskip

Suppose now that $ \prec^*$ is another order relation on $ \ \Gamma 
\times
\Gamma$ and that $S^*$ is a function on $ \Gamma$ satisfying condition 
(i) of Theorem 2.10 with respect to $ \prec^*$ on $ \Gamma \times 
\Gamma$. 
Suppose
further, that $ \prec$ and $ \prec^*$ agree on ${\cal D}$, i.e., 
$$
(X,Y) \prec (X^{\prime}, Y^{\prime}) \ \ \ {\it if \ and \  only \ if}\ \ 
\
(X,Y) \prec^* (X^{\prime}, Y^{\prime})
$$
whenever $(X,Y)$ and $(X^{\prime}, Y^{\prime})$ are both in ${\cal D}$. 
Then $ \prec$ and $ \prec^*$ agree on all of  $\Gamma \times \Gamma$ and
hence, by Theorem 2.10, $S$ and $S^*$ are related by an affine
transformation. }

{\it Proof:} By considering points $(X,X) \in {\cal D}$, the consistency 
of
$S$ and $S^*$ implies that $S^*(X) = M(S(X))$ for all $X \in  \Gamma$, 
where
$M$ is some monotone increasing function on $L \subset {\bf R}$. Again, 
as
in the proof of Theorem 2.10, 
$$
\mfr1/2 M(S(X)) + \mfr1/2 M(S(Y)) = M\Bigl({S(X)) + S(Y)\over 2}\Bigr) 
\eqno(2.26)
$$
for all $(X,Y)\in {\cal D}$. (Note: In deriving Eq.\ (2.25) we did 
not use the fact that $  \Gamma \times \Gamma$ was the Cartesian
product of two spaces; the only thing that was used was the fact that 
$S(X) + S(Y)$ characterized the level sets of $ \Gamma \times \Gamma$.
Thus, the same argument holds with $ \Gamma \times \Gamma$ replaced by
${\cal D}$.)

Now fix $X\in  \Gamma$ and let $x=S(X)$. Since $D$ contains an open set
that contains the point $(x,x) \in {\bf R}^2$, there is an open square
$$
Q=(x-\epsilon , x+\epsilon)\times (x-\epsilon , x+\epsilon)
$$
in $D$. Eqn.~(1) holds on $Q$ and so we conclude, as in the proof of
Theorem 2.10, that, for $y\in (x-\epsilon , x+\epsilon)$ $M(y) = a
y + B$ for some $a, B$, which could depend on $Q$,
a-priori.

The diagonal $\{(x,x):x\in L\}$ is covered by these open squares and,
by the Heine-Borel theorem, any closed, finite section of the diagonal
can be covered by finitely many squares $Q_1, Q_2, ..., Q_N$, which we
order according to their ``diagonal point" $(x_i, x_i)$. They are not
disjoint and, in fact, we can assume that $T_i := Q_i\cap Q_{i+1}$ is
never empty. In each interval $(x_i-\epsilon , x_i+\epsilon)$,
$M(x)=a _i x + B _i$ but agreement in the overlap region $T_i$
requires that $a _1$ and $B_i$ be independent of $i$. Thus,
$S^*(X) = a S(X) +B$ for all $X\in \Gamma$, as claimed.
\hfill\lanbox
%%%%%%%%%%%%%%%%%%%%%%%%%%%%%%
\vfill\eject
%%%%%%%%%%%%%%%%%%%%
%%%%%%%%%%%%%%%%%%%%%%%%%%%%%%%%%%
\noindent
{\tit III.  SIMPLE SYSTEMS }
\bigskip

Simple systems are the building blocks of thermodynamics.  In general, 
the {\it equilibrium} state of a (simple or complex) system is 
described by certain coordinates called {\it work coordinates} and 
certain coordinates called {\it energy coordinates}.  Physically, the 
work coordinates are the parameters one can adjust by mechanical (or 
electric or magnetic) actions.  We denote work coordinates 
collectively by $V$ because the volume is a typical one.  A simple 
system is characterized by the fact that it has exactly one energy 
coordinate, denoted by $U$.

The meaning of these words will be made precise; as always there is a 
physical interpretation and a mathematical one.  The remark we made in 
the beginning of Section II is especially apt here; the mathematical 
axioms and theorems should be regarded as independent of the numerous 
asides and physical discussions that surround them and which are not 
intrinsic to the logical structure, even though they are very 
important for the physical interpretation.  The mathematical 
description of simple systems will require three new assumptions, 
S1--S3.  {\it In our axiomatics simple systems with their energy and 
work coordinates are basic (primitive) concepts that are related to 
the other concepts by the axioms.} The statement that they are the 
building blocks of thermodynamics has in our approach the precise 
meaning that from this section on, {\it all systems under 
consideration are assumed to be scaled products of simple systems}.

{}From the physical point of view, a simple system is a {\it fixed} 
quantity of matter with a {\it fixed} amount of each element of the 
periodic table.  The content of a simple system can be quite 
complicated.  It can consist of a mixture of several chemical species, 
even reactive ones, in which case the amount of the different 
components might change as the external parameters (e.g., the volume) 
change.  A simple system need not be spatially homogeneous.  For 
example a system consisting of two vessels, each with a piston, but 
joined by a heat conducting thread, is simple; it has two work 
coordinates (the volumes of the two vessels), but only one energy 
coordinate since the two vessels are always in thermal equilibrium 
when the total system is in equilibrium.  This example is meant to be 
informal and there is no need to define the words `piston`, `thread' 
and `heat conducting'.  It is placed here as an attempt at 
clarification and also to emphasize that our definition of `simple 
system' is not necessarily the same as that used by other authors.

An example of a compound, i.e., non-simple system is provided by two 
simple systems placed side by side and not interacting with each 
other.  In this case the state space is just the Cartesian product of 
the individual state spaces.  In particular, two energies are needed 
to describe the state of the system, one for each subsystem.

Some examples of simple systems are:  
\item{(a)}  One mole of water in
a container with a piston (one work coordinate).  
\item{(b)}  A half
mole of oxygen in a container with a  piston and in a magnetic field
(two work coordinates, the volume and the magnetization). 
\item{(c)} Systems (a) and (b) joined by a copper thread (three work 
coordinates). 
\item{(d)} A mixture consisting of 7 moles of hydrogen and one mole of
oxygen (one work coordinate). Such a mixture is capable of explosively 
reacting to form water, 
of course, but for certain purposes (e.g., in chemistry, material 
science  and 
in astrophysics) we can regard a 
non-reacting, metastable mixture as
capable of being in an equilibrium state, as long as one is careful not
to bump the container with one's elbow. \smallskip

To a certain extent, the question of which physical states are to be
regarded as equilibrium states is a matter of practical convention.
The introduction of a small piece of platinum in (d) will soon show us
that this system is not truly in equilibrium, although it can be
considered to be in equilibrium for practical purposes if no catalyst is
present. 
 
A few more remarks will be made in the following about the physics of
simple systems, especially the meaning of the distinguished energy
coordinate.  In the real world, it is up to the experimenter to decide
when a system is in equilibrium and when it is simple. If the system
satisfies the mathematical assumptions of a simple system---which we
present next---then our analysis applies and the second law holds for it. 
Otherwise, we cannot be sure.

Our main goal in this section is to show that the forward sectors in 
the state space $\Gamma$ of a simple system form a {\it nested} family 
of closed sets, i.e., two sectors are either identical or one is 
contained in the interior of the other (Theorem 3.7).  Fig.\ 5 
illustrates this true state of affairs, and also what could go wrong 
{\it a priori} in the arrangement of the forward sectors, but is 
excluded by our additional axioms S1-S3.  Nestedness of forward 
sectors means that the comparison principle holds within the state 
space $\Gamma$.  The comparison principle for multiple scaled copies 
of $\Gamma$, which is needed for the definition of an entropy function 
on $\Gamma$, will be derived in the next section from additional 
assumptions about thermal equilibrium.

\bigskip\noindent
{\subt A. Coordinates for simple systems}
\bigskip

A (equilibrium) state of a simple system is parametrized uniquely (for
thermodynamic purposes) by a point in $\R^{n+1}$, for some $n > 0$
depending on the system (but not on the state).

A point in $\R^{n+1}$ is written as $X = (U,V)$ with $U$ a
distinguished coordinate called the {\bf internal energy} and with $V =
(V_1, \dots , V_n) \in \R^n$.  The coordinates $V_i$ are called the
{\bf work coordinates}.

We could, if we wished,  consider the case $n=0$, in which case we
would have a system whose states are parametrized by the energy alone.
Such a system is called a {\bf thermometer} or a {\bf degenerate simple
system.} These systems must be (and will be in Section IV) treated
separately because they will fail to satisfy the transversality axiom
T4, introduced in Section IV. {}From the point of view of the convexity
analysis in the present section, degenerate simple systems can be
regarded as trivial.

The energy is special, both mathematically and physically.  The fact
that it can be defined as a physical coordinate  really goes back to the
{\bf first law of thermodynamics}, which says that the amount of work 
done 
by 
the outside world
in going adiabatically from one state of the system to another is
independent of the manner in which this transition is carried out. This
amount of work is the amount by which a weight was raised or lowered in
the physical definition given earlier of an adiabatic process. (At the
risk of being tiresomely repetitive, we remind the reader that
`adiabatic, means neither `slow' nor `isolated' nor any restriction
other than the requirement that the external machinery returns to its
original state while a weight may have risen or fallen.) Repeatedly,
authors have discussed the question of exactly what has to be assumed in
order that this fact lead to a {\it unique} (up to an additive constant)
energy coordinate for all states in a system with the property that the
difference in the value of the parameter at two points equals the work
done by the outside world in going adiabatically from one point to the
other.  See e.g., (Buchdahl, 1966), (Rastall, 1970), and (Boyling,
1972). These discussions are interesting, but for us the question lies
outside the scope of our inquiry, namely the second law. We simply take
it for granted that the state space of a simple system can be
parametrized by a subset of some $\R^{n+1}$ and  that there is one
special coordinate, which we call `energy' and which we label by $U$. 
Whether or not this parametrization is unique is of no particular
importance for us.  The way in which $U$ is special will become clear
presently when we discuss the tangent planes that define the pressure
function.

Mathematically, we just have coordinates. The question of which physical
variables to attach to them is important in making the transition from
physics to mathematics and back again. Certainly, the coordinates have to
be chosen so that we are capable of specifying states in a one-to-one
manner.  Thus, $U=$ energy and $V=$ volume are better coordinates for
water than, e.g., $H=U+PV$ and $P$, because $U$ and $V$ are  capable of
uniquely specifying the division of a multi-phase system into phases,
while $H$ and $P$ do not have this property. For example, the triple 
point
of water corresponds to a triangle in the $U$, $V$ plane (see Fig.  8),
but in the $H$, $P$ plane the triple point corresponds to a line, in 
which
case one cannot know the amount of the three phases merely by specifying 
a
point on the line.  The fundamental nature of energy and volume as
coordinates was well understood by Gibbs and others, but seems to have
gotten lost in many textbooks. Not only do these coordinates have the
property of uniquely specifying a state but they also have the advantage
of being directly tied to the fundamental classical mechanical variables,
energy and length. We do not mean to imply that energy and volume always
suffice. Additional work coordinates, such as magnetization, components 
of
the strain tensor, etc., might be needed.  

Associated with a simple system is its {\bf state space}, which is a 
non-empty 
{\it convex} and {\it open}
subset $\Gamma \subset \R^{n+1}$.  This
$\Gamma$ constitutes all values of the coordinates that the system can
reach. $\Gamma $ is open because points on the boundary of $\Gamma$ are
regarded as not reachable physically in a finite time, but there could be
exceptions.  

The reason that   $\Gamma $ is convex was discussed at length in
Section II.F.  We  assume axioms A1--A7. In particular, a state space
$\Gamma$, scaled by $t>0$, is the convex set
$$
\Gamma^{(t)} = t \Gamma:=\{t X: X\in\Gamma\} \ .  \eqno(3.1)
$$
Thus, what was formerly
the abstract symbol $tX$ is now concretely realized as the point $(tU,
tV) \in \R^{n+1}$ when $X=(U,V) \in \R^{n+1}$.

{\it Remark. \/} Even if $\Gamma^{(t)}$ happens to coincide with $\Gamma$
as a subset of $\R^{n+1}$ (as it does, e.g.,\ if $\Gamma$ is the orthant
$\Gamma=\R_{+}^n$) it is important to keep in mind that 
the mole numbers that specify the material content of 
the states in $\Gamma^{(t)}$ are $t$-times the mole numbers for the
states in $\Gamma$.  Hence the state spaces must be regarded as 
different.
The photon gas, mentioned in Sect.\ II.B.
is an exception: Particle number is not conserved, and  `material 
content'
is not an independent variable.  Hence the state spaces $\Gamma^{(t)}$ 
are
all {\it physically} identical in this case, i.e., no physical 
measurement
can tell them apart. Nevertheless it is a convenient fiction to regard 
them 
as
mathematically distinguishable; in the end, of course, they must all have 
the 
same properties, e.g., entropy, as a function of the coordinates---up to 
an
additive constant, which can always be adjusted to be zero, as discussed
after Theorem 2.5.

Usually, a forward sector, $ A_X$, with $X = (U^0, V^0)$, contains the 
`half-lines' $\{ (U, V^0) : U \geq U^0 \}$ and $\{ (U^0, V) : V 
_{i}\geq V^0_{i},i=1,\dots,n \}$ but, theoretically, at least, it 
might not do so.  In other words, $\Gamma$ might be a bounded subset 
of $\R^n$.  This happens, e.g., for a quantum spin system.  Such a 
system is a theoretical abstraction from the real world because real 
systems always contain modes, other than spin modes, capable of having 
arbitrarily high energy.  We can include such systems
with bounded state spaces in our theory, however, but then we have to be 
a
bit careful about our definitions of state spaces and the forward sectors
that lie in them.  This partially accounts for what might appear to be 
the
complicated nature of the theorems in this section. 

Scaling and convexity might at first sight appear to be requirements
that exclude from the outset the treatment of `surface effects' in our
framework.  In fact, a system like a drop of a liquid, where volume and
surface effects are coupled, is not a simple system.  But as we shall
now argue, the state space of such a system can be regarded as a subset
of the convex state space of a simple system that contains all the
relevant thermodynamic information.  The independent work coordinates
of this system are the volume $V$ and the surface area $A$.  Such a
system could, at least in principle,  be realized by putting the liquid
in a rectangular pan made out of such a material that the adhesive
energy between the walls of the pan and the liquid exactly matches the
cohesive energy of the liquid.  I.e., there is no surface energy
associated with the boundary beween liquid and walls, only between
liquid and air.  (Alternatively, one can think of an `ocean' of liquid
and separate a fixed amount of it (a `system') from the rest by a
purely fictitious boundary.)  By making the pan (or the fictuous
boundary) longer at fixed breadth and depth and, by pouring in the
necessary amount of liquid, one can scale the system as one pleases.
Convex combination of states also has an obvious operational meaning.
By varying breadth and depth at fixed length the surface area $A$ can
be varied independently of the volume $V$.  Lack of scaling and
convexity enter only when we restrict ourselves to non-convex
submanifolds of the state space, defined by subsidiary conditions like
$A=(4\pi)^{1/3}3^{2/3}V^{2/3}$ that are appropriate for a drop of
liquid.  But such coupling of work coordinates is not special to
surface effects; by suitable devices one can do similar things for any
system with more than one work coordinate.  {\it The important point is
that the thermodynamic properties of the constrained system are
derivable from those of the unconstrained one, for which our axioms
hold.}

It should be remarked that the experimental realization of the simple 
system with volume and surface as independent work coordinates 
described above might not be easy in practice. In fact, the usual 
procedure would 
be to compare measurments on the liquid in bulk and on 
drops of liquid, and then,  by inverting 
the data, infer the properties of the system where volume and surface 
are independent variables. The claim that scaling and convexity are 
compatible with the inclusion of surface effects amounts to saying 
that these properties hold after such a
`disentanglement' of the coordinates.
%%%%%%%%%%%% 

\bigskip
%\vfill\eject
\noindent {\subt B.
Assumptions about simple systems} \bigskip
 
As was already stated, we assume the general axioms A1--A7 of Section II. 
Since the state space $\Gamma$ of a simple system has a convex structure, 
we
recall from Theorem 2.6 that the forward sector of a point $X \in 
\Gamma$,
namely $A_X = \{Y\in \Gamma : X \prec Y\}$ is a convex subset of $\Gamma
\subset \R^{n+1}$. We now introduce three new axioms.  It is also to be 
noted
that the comparison hypothesis, CH, is {\it not} used here---indeed, {\it 
our
chief goal in this section and the next is to derive CH from the other 
axioms.}

The new axioms are:
\smallskip

\item{{\bf S1)}} {\bf Irreversibility.} For each $X \in \Gamma$ there
is a point $Y \in \Gamma$ such that $X \prec\prec Y$. In other words,
each forward sector, $A_X$, consists of {\it more} than merely points
that, like $X$ itself, are adiabatically equivalent to $X$.

\medskip

We remark that  axiom S1 is implied by the thermal transversality axiom 
T4 
in
Section IV. This fact deserves to be noted in any count of the total 
number 
of
axioms in our formulation of the second law, 
and it explains why we gave the number of our axioms as 15 in Section I.  
Axiom S1 is listed here as a separate axiom because it is basic to the
analysis of simple systems and is conceptually independent of the notion 
of
thermal equilibrium presented in Section IV.

By Theorem 2.9 Carath\'eodory's principle holds. This principle
implies that 
$$
X \in \partial  A_X \ , \eqno(3.2)
$$ 
where  $\partial  A_X$ denotes the {\bf boundary} of
$A_X$. By `boundary' we mean, of course, the 
{\it relative} boundary, i.e.,
the part of the usual boundary of $A_X$, (considered as a subset of
$\R^{n+1}$) that lies in $\Gamma$.

Since $X$ lies on the boundary of the convex set $A_X$ we can draw at
least one support plane to $A_X$ that passes through $X$, i. e., a
plane with the property that $A_X$ lies entirely on one side of the
plane.  Convexity alone does not imply that this plane is unique, or
that  this plane intersects the energy axis of $\Gamma$. The next axiom
deals with these matters.
\medskip

\item{{\bf S2)}} {\bf Lipschitz tangent planes.} For each $X\in \Gamma$ 
the
forward sector $A_X$ has a {\it unique} support plane at $X$ (i.e., $ 
A_X$
has a {\it tangent plane} at $X$), denoted by $\Pi_X$ .  The tangent
plane $\Pi_X$ is assumed to have  a finite slope with respect to the work
coordinates and the slope is moreover assumed to be a {\it locally 
Lipschitz 
continuous} function of $X$.
\smallskip

We emphasize that this tangent plane to $A_X$ is initially assumed to 
exist
only at $X$ itself. In principle, $\partial A_X$ could have `cusps' at
points other than $X$, but Theorem 3.5 will state that this does not 
occur.

\medskip

The precise meaning of the statements in axiom S2 is the following: 
The tangent plane at $X = (U^0, V^0)$ is, like any plane in
$\R^{n+1}$, defined  by a linear equation. The finiteness of 
the slope with respect to the work coordinates means that this equation 
can 
be 
written as
$$
U - U^0 + \sum \limits^n_{i=1} P_i (X) (V_i - V^0_i) = 0, \eqno(3.3)
$$
in which the $X$ dependent numbers $P_i (X)$ are the parameters that
define the slope of the plane passing through $X$. (The slope is thus in 
general 
a vector.)
The assumption that $P_i (X)$ is {\it finite} 
means that the plane is never `vertical', i.e., 
it never contains the line $\{(U,V^0): U\in \R\}$. 

The assumption that $\Pi_X$ is the unique  supporting hyperplane  of 
$A_{X}$ at $X$ means that the linear expression, with coefficients $g_i$,
$$
U - U^0 + \sum \limits^n_{i=1} g_i (V_i - V^0_i)\eqno(3.4)
$$
has one  sign for all $(U,V) \in A_X$ (i.e., it is $\geq 0$ or $\leq 0$
for all points in $A_{X}$) if and only if $g_i = P_i (X)$ for all $i = 1,
\dots , n$.  The assumption that the slope of the tangent plane is 
locally 
Lipschitz
continuous means that each $P_i$ is a locally Lipschitz continuous
function on $\Gamma$.  This,
in turn, means that for any closed ball $B \subset \Gamma$ with finite
radius there is a constant $c = c(B)$ such that for all $X$ and $Y \in B$
$$
\vert P_i(X) - P_i(Y) \vert \leq c \vert X-Y \vert_{\R^{n+1}}. \eqno(3.5)
$$

The function $X\mapsto P(X)=(P_1(X), \dots ,P_n(X))$ from $\Gamma$ 
to $\R^n$ is  called the
{\bf pressure}.  {\it Note:}  We do {\it not} need to assume that $P_i
\geq 0$.

{\it Physical motivation: \/}  The uniqueness of the support plane comes
from the following physical consideration.  We interpret the pressure as
realized by a force on a spring that is so adjusted that the system is in
equilibrium at some point $(U^0, V^0)$. By turning the screw on the 
spring
we can change the volume infinitesimally to $V^0 +\delta V$, all the 
while
remaining in equilibrium. In so doing   we change  $U^0$ to $U^0 + \delta
U$.   The physical idea is that a slow reversal of the screw  can take 
the
system  to $(U^0 - \delta U, V^0 - \delta V)$, infinitesimally. The 
energy
change is the same, apart from a sign, in both directions.

The Lipschitz continuity  assumption is weaker than, and is implied by,
the assumption that $P_i$ is continuously differentiable.  By
Rademacher's theorem, however, a locally Lipschitz continuous function
is differentiable almost everywhere, but the relatively rare points of
discontinuity of a derivative are particularly interesting.

The fact that we do {\it not} require the pressure to be a
differentiable function of $X$ is important for real physics because
phase transitions occur in the real world, and the pressure need not be
differentiable at such transition points.  Some kind of continuity
seems to be needed, however, and local Lipschitz continuity does accord
with physical reality, as far as we know.  It plays an important role
here because it guarantees the uniqueness of the solution of the
differential equation given in Theorem 3.5 below.  It is also important
in Section V when we prove the differentiability of the entropy, and
hence the uniqueness of temperature.  This is really the only reason we
invoke continuity of the pressure and this assumption could, in
principle,  be dropped if we could be sure about the uniqueness and
differentiablity just mentioned. There are, in fact statistical
mechanical models with special forces that display discontinuous
pressures (see e.g., (Fisher and Milton, 1983)) and temperatures (which
then makes temperature into an `interval-valued' function, as we
explain in Section V) (see e.g., (Thirring, 1983)). These models are not
claimed to be realistic; indeed, there are some theorems in statistical
mechanics that prove the Lipschitz continuity of the pressure under
some assumptions on the interaction potentials, e.g., (Dobrushin and
Minlos, 1967). See (Griffiths, 1972).

There is another crucial fact about the pressure functions that will
finally be proved in Section V, Theorem 5.4. The surfaces $\partial
A_X$ will turn out to be the surfaces of constant entropy, $S(U,V)$, 
and evidently,
from the definition of the tangent plane (3.3), the functions $P_i(X)$
are truly the pressures in the sense that
$$
P_i(X)= {\partial U \over \partial V_i}(X)     \eqno (3.6)
$$
along the (constant entropy) surface $\partial A_X$.  However, one
would also like to know the following two facts, which are at the 
basis of
Maxwell's relations, and which are the fundamental defining relations
in many treatments.
$$
{1 \over T (X)} := {\partial S \over \partial U} (X) \eqno(3.7) 
$$
and
$$
{P_i(X) \over T(X)} = {\partial S \over \partial V_i}(X),  \eqno (3.8)
$$
where $T(X)$ is the temperature  in the state $X$. Equation (3.7) 
constitutes, for us, the {\it definition} of temperature, but 
we must first  prove that $S(U,V)$
is sufficiently smooth in order to make sense of (3.7). Basically,
this is what Section V is all about.

In Theorems 3.1 and 3.2 we shall show that $A_X$ is closed and
has a non-empty interior, ${\rm Interior}(A_X)$. Physically, the points 
in
${\rm Interior}(A_X)$ represent the states that can be reached from $X$, 
by
some adiabatic means, in a finite time. (Of course, the re-establishment 
of
equilibrium usually requires an infinite time but, practically
speaking, a finite time suffices.) On the other hand, the points in
$\partial A_X$ require a truly infinite time to reach from $X$. In the
usual parlance they are reached from $X$ only by `quasi-static
reversible processes'. However, these boundary points can be reached in
a finite time with the aid of a tiny bit of cold matter---according to
the stability assumption. If we wish to be pedantically `physical' we
should exclude $\partial A_X$ from $A_X$. This amounts to replacing
$\prec$ by $\prec \prec$, and we would still be able to carry out our
theory,  with the help of the stability assumption and some
unilluminating epsilons and deltas.  Thus, the seemingly innocuous, but
important stability axiom permits us to regard certain infinitely slow
processes as physically valid processes.
\medskip

Our third axiom about simple systems is technical but important.
\medskip

\item{{\bf S3)}} {\bf Connectedness of the boundary.} We assume that
$\partial A_X$ is arcwise connected. 
\medskip

Without this assumption counterexamples to
the comparison hypothesis, CH, can be constructed, even ones
satisfying all the other axioms.

\medskip

{\it Physical motivation:}  If $Y \in \partial A_X$, we think of $Y$ as
physically and adiabatically reachable from $X$ by a continuous curve
in $\partial A_X$ whose endpoints are $X$ and $Y$. (It is not possible
to go from $X$ to $Y$ by a curve that traverses the interior of $A_X$
because such a process could not be adiabatic.) Given this conventional
interpretation, it follows trivially that $Y,Z \in \partial A_X$
implies the existence of a continuous curve in $\partial A_X$  from $Y$
to $Z$. Therefore $\partial A_X$ must be a connected set.

We call the family of relatively closed sets $\{ \partial A_X \}_{X \in
\Gamma}$ the {\bf adiabats} of our system. As we shall see later in
Theorem 3.6, $Y\in \partial A_X $ implies that $X\in \partial A_Y $.
Thus, all the points on any given adiabat are equivalent and it is
immaterial which one is chosen to specify the adiabat.

\bigskip
{\subt  C. The geometry of forward sectors}
\bigskip
In this subsection all points are
in the state space of the same fixed, simple system $\Gamma$, if not 
otherwise 
stated. $\Gamma$ is, of course, regarded here as a subset of some 
$\R^{n+1}$.

%%%%%%
We begin with an interesting geometric fact that complements convexity,
in some sense. Suppose that $X, Y,Z$ are three collinear points, with $Y$ 
in
the middle, i.e.,  $Y=tX + (1-t)Z$ with $0<t<1$.  
The convexity  axiom A7 tells us that 
$$
X\prec Z  \quad \quad {\rm implies \ that } \quad \quad X\prec Y 
\eqno(3.9)
$$
because $X\prec ((1-t)X, tX)\prec (1-t)Z,tX) \prec Y$. The next lemma 
is geometrically related to this, but its origins are different. We shall 
use 
this lemma in the proof of Theorems 3.3 and 3.7 below.

{\bf LEMMA 3.1 (Collinear points).} {\it Let  $Y=tX + (1-t)Z$ with 
$0<t<1$
as above and suppose that $Y\prec Z$. Then $X\prec Y$ (and hence $X\prec
Z$).}

\smallskip
{\it Remark:} Equation (3.9) and Lemma 3.1 rely only on the convexity of
$\Gamma $ and on  axioms A1-A7. The same properties hold for compounds
of simple systems (note that the Cartesian product of two convex sets is
convex) and hence (3.9) and Lemma 3.1 hold for compounds as well.
 
\smallskip
{\it Proof:\/} By A7, A5, our hypothesis, and A3
$$
(tX, (1-t)Z)) \prec Y  \sima (tY,  (1-t)Y) \prec (tY, (1-t)Z).
$$
By transitivity, A2, and the cancellation law, Theorem 2.1, $tX\prec 
tY$. By scaling, A4, $X\prec Y$. \hfill\lanbox
\smallskip

%%%%%%%
Our first theorem in this section, about closedness,  is crucial because
it lies behind many of the more complex theorems. Once again,
the seemingly innocuous stability axiom A6 plays a central role. As we
said in Section II, this axiom amounts to some kind of continuity in a
setting in which, at first, there is not even any topology on the 
state spaces. Now that we are in $\R^{n+1}$, the topology is evident and
stability reveals its true character in the statement of closedness in 
the 
usual topological sense. The following proof has some of the spirit
of the proof of Lemma 3.1.
\medskip

{\bf THEOREM 3.1 (Forward sectors are closed).} {\it The forward sector,
$A_X$, of each point $X \in \Gamma$ is a relatively closed subset of
$\Gamma$, i.e., $Closure(A_X) \cap \Gamma = A_X$.}
\medskip

{\it Proof:\/} The proof uses only axioms A1-A7, in 
particular stability, A6, and convexity, A7, but not 
S1-S3. What we have to prove is
that if $Y \in \Gamma$ is on the boundary of $A_X$ then $Y$  is in
$A_X$.  For this purpose we can assume that the set $A_X$ has full
dimension, i.e., the interior of $A_X$ is not empty.  If, on the
contrary, $A_X$ lay in some lower dimensional hyperplane then the
following proof would work, without any changes, simply by replacing
$\Gamma $ by the intersection of $\Gamma$ with this hyperplane.
 
Let $W$ be any point in the interior of
$A_X$. Since $A_X$ is convex, and $Y$ is on the boundary of $A_X$, the
half-open line segment joining $W$ to  $Y$ (call it $[W,Y)$, bearing in
mind that $Y\not\in [W,Y)$) lies in $A_X$.  The prolongation of this
line beyond $Y$
lies in the complement of $A_X$ and has at least one point (call it
$Z$) in $\Gamma$. (This follows from the fact that $\Gamma$ is open and
$Y\in \Gamma$.) For all sufficiently large integers $n$ the point $Y_n$
defined by
$$
{n\over (n+1)} Y_n +{1\over (n+1)} Z = Y \eqno (3.10)
$$
belongs to $[W,Y)$.  We claim that $(X, {1\over n} Z) \prec (Y, {1\over
n} Y)$. If this is so then we are done because, by the stability axiom,
$X\prec Y$.

To prove the last claim, first note that  $(X, {1\over n} Z )\prec  (Y_n,
{1\over n} Z)$ because $X\prec Y_n $ and by axiom A3. By scaling,
A4, the convex combination axiom A7, and (3.10) 
$$
\left(Y_n, {1\over n} Z \right) \ = \ {n+1 \over n} \left({n\over
(n+1)} Y_n, {1\over (n+1)} Z \right) \ \prec \ {n+1 \over n}Y \ .
\eqno(3.11)
$$
But this last equals $ (Y, {1\over n} Y)$ by the splitting axiom, A5. 
Hence $(X, {1\over n} Z) \prec (Y, {1\over
n} Y)$.
\hfill\lanbox
\medskip

The following theorem uses Theorem 3.1 in an essential way.
\medskip

{\bf THEOREM 3.2 (Forward sectors have interiors).} {\it For all
$X$, the forward sector $A_X$ has a non empty interior.}
\medskip

{\it Proof.} The proof uses the transitivity axiom, A2, convexity, A7,
the existence
of irreversible processes, S1, and the tangent plane axiom S2, but
neither local Lipschitz continuity of the pressure nor the
connectedness of the boundary, S3, are required for our proof here.

We start by remarking that a convex set in ${\bf R}^{n+1}$ either has a
non empty interior, or it is contained in a hyperplane. We therefore
assume that $A_X$ is contained in a hyperplane and show that this
contradicts the axioms. [An illustrative  picture to keep in mind here
is  that $A_X$ is a closed, (two-dimensional) disc in $\R^3$ and
$X$ is some point inside this disc and not on its perimeter.
This disc is a closed subset
of $\R^3$ and
$X$ is on its boundary (when the disc is viewed as a subset of $\R^3$).  
The hyperplane is the plane in $\R^3$ that contains the disc.]

Any hyperplane containing $A_X$ is a support plane to $A_X$ at $X$, and
by axiom S2 the support plane is unique, so $A_X\subset \Pi_X$. If
$Y\in A_X$, then $A_Y\subset A_X\subset \Pi_X$ by transitivity, A2. By
the irreversibility axiom S1, there exists a $Y\in A_X$ such that
$A_Y\neq A_X$, which implies that the convex set $A_Y\subset \Pi_X$,
regarded as a subset of $\Pi_X$, has a boundary point in $\Pi_X$. If
$Z\in \Pi_X$ is such a boundary point of $A_Y$, then $Z\in A_Y$ because
$A_Y$ is closed. By transitivity, $A_Z\subset A_Y\subset \Pi_X$, and
$A_Z\neq \Pi_X$ because $A_Y\neq A_X$.

Now $A_Y$, considered as a subset of $\Pi_X$,  has an
$(n-1)$-dimensional supporting hyperplane at $Z$ (because $Z$ is a
boundary point).  Call this hyperplane $\Pi'_Z$. Since $A_Z\subset
A_Y$, $\Pi_Z'$ is a supporting hyperplane for $A_Z$,  regarded as a
subset of $\Pi_X$.  Any $n$-dimensional hyperplane in ${\bf R}^{n+1}$
that contains the $(n-1)$-dimensional hyperplane $\Pi'_Z\subset \Pi_X$
clearly supports $A_Z$ at $Z$, where $A_Z$ is now considered as a
convex subset of ${\bf R}^{n+1}$. Since there are infinitely many such
$n$-dimensional hyperplanes in ${\bf R}^{n+1}$, we have a contradiction
to the uniqueness axiom S2.  \hfill \lanbox
\medskip

Thanks to this last theorem it makes sense to talk about the direction
of the normal to the tangent plane $\Pi_X$ (with respect to the canonical 
scalar product on $\R^{(n+1)}$) pointing to the interior 
of $A_{X}$. The part of axiom S2, that requires the tangent plane to 
have finite slope with respect to the work coordinates, means that the 
normal is never orthogonal to the energy axis. It appears natural to 
extend 
the 
continuity requirement of axiom S2 by requiring not 
only that the slope but also the direction of the normal depends 
continuously on $X$. Since $\Gamma$ is connected it then follows 
immediately that forward
sectors are on the `same side' of the tangent plane, i.e., the 
projection of the normal on the energy axis is either positive for 
all sectors or negative for all sectors. 

In fact, it is not necessary to invoke 
this strengthened continuity requirement to prove that
forward sectors all point the same way. It is already a consequence of 
axioms 
A1-A7, S1 and the finite 
slope part of axiom S2. We shall prove this below as Theorem 3.3, but 
leave the reader the option to accept it simply as a part of the 
continuity 
requirement for tangent planes if preferred. 

As far as our axiomatic framework is concerned the direction of the
energy coordinate and hence of the forward sectors is purely
conventional, except for the proviso that once it  has been set for one
system it is set for all systems. (This follows from Theorem 4.2 in 
the next section.) {\it We shall adopt the convention that they
are on the positive energy side}.  {}From a physical point of view
there is more at stake, however. In fact, our operational interpretation 
of
adiabatic processes in Sect. II involves either the raising or lowering
of a weight in a gravitational field and these two cases are physically
distinct.  Our convention, together with the usual convention for the
sign of energy for mechanical systems and energy conservation, means that
we are concerned with a world where adiabatic process at fixed work
coordinate can never result in the raising of a weight, only in the
lowering of a weight.  The opposite possibility differs from the
former in a mathematically trivial way, namely by an overall sign of the
energy, but given the physical interpretation of the energy direction in
terms of raising and lowering of weights, such a world would be different
from the one we are used to. 

Note that (3.7) tells us that the fact that forward sectors point 
upward is equivalent to the temperature being everywhere positive.  To 
illustrate what is involved here, let us consider a system of $N$ 
independent spins in a magnetic field, so that each spin has energy 
either $0$ or $e$.  In the thermodynamic limit $N,\ U\to\infty$ with 
$X =U/(Ne)$ fixed, the entropy per spin is easily calculated according 
to the rules of statistical mechanics to be $S/N= -X \ln X -(1-X)\ln 
(1-X)$.  The first half of the energy range, $0 < U/(Ne) < 1/2$ has 
positive 
temperature 
while the second half $1/2 < U/(Ne)<1$ has negative temperature, 
according to (3.7).  How can we reconcile this with our formulation of 
simple systems?  That is to say, we insist that the state space 
$\Gamma$ of our spin system consists only of the region $0 < U/(Ne) < 
1/2$, and we ask what feature of our axioms has ruled out the 
complementary region. The answer is that if we included the second 
half then convexity would require that we also include the maximum 
entropy point $X=1/2$.  But the forward sector of $X$ contains only 
$X$ itself and this violates axiom S1.  

This example captures the essential feature that lies  behind the 
following general fact.
\medskip

{\bf LEMMA 3.2 (Range of energy in forward sectors).} {\it Let 
$X=(U^0,V^0)\in\Gamma$ and assume that its forward sector $A_{X}$ is on 
the 
positive energy 
side of $\Pi_{X}$. Then
$$A_{X}\cap \{(U,V^0):U\in\R\}=\{(U,V^0):U\geq U^0 \}
\cap \Gamma.\eqno(3.12)$$}
(If $A_{X}$ is on the negative energy side, then (3.12) holds with 
`$\geq$' replaced by `$\leq$'.)
\medskip

{\it Proof:} The left side of (3.12), denoted $J_{X}$,
is convex and 
relatively closed in $\Gamma$ by Theorem 3.1. It is not larger than the 
right 
side because $A_{X}$ lies above the tangent plane that cuts the line 
$L=\{(U,V^0):U\in\R\}$ at $X$.
If it is strictly smaller than the right side of (3.12), then $J_{X}$ is 
a compact interval. Let $X_{1}$ denote its mid point. 
Then $J_{X_{1}}$, the intersection of $A_{X_{1}}$ with the line $L$,
is a closed subinterval of $J_{X}$ and its 
length is at most half the length of $J_{X}$. (Here we have used 
transitivity, closedness, and that $X_{1}$ is on the boundary of 
$J_{X_{1}}$.)  Repeating this 
procedure we obtain a convergent sequence, $X_{n}$, $n=1,2,\dots$ of 
points in $J_{X}$, 
such that the forward sector of its limit point $X_{\infty}$ contains 
only $X_{\infty}$ itself in violation of S1.\hfill\lanbox
\medskip

The `same sidedness' of forward sectors follows from 
Lemmas 3.1 and 3.2 together with the finite slope of tangent planes. 
\medskip

{\bf THEOREM 3.3 (Forward sectors point the same way).} {\it If $\Gamma$
is the state space of a simple system, and if the forward sector $A_X$
for one $X\in\Gamma$ is on the positive energy side of the tangent plane
$\Pi_X$, then the same holds for all states in $\Gamma$.} \medskip

{\it Proof:\/} For brevity, let us say that a state $X\in \Gamma$ is 
`positive' if $A_X$ is on the positive energy side of $\Pi_X$, and 
that $X$ is `negative' otherwise.  
Let $I$ be the intersection of $\Gamma$ with a line parallel to the 
$U$-axis, i.e., $I=\{(U,V)\in\Gamma, U\in\R\}$ for some $V\in\R^{n}$. 
If $I$ contains a positive point, $Y$, then it follows immediately from 
Lemma 3.2 that all points, $Z$, that lie above it on 
$I$ (i.e., have higher energy)  are also positive. In fact, one can pass 
from 
$Y$ to 
$Z$, and if $Z$ were negative, then, using Lemma 3.2 again, one could 
pass 
from 
$Z$ to a state $X$ below $Y$, violating the positivity of $Y$. Lemma 
3.1, on the other hand, immediately implies that all points $X$ below $Y$ 
are positive, for $Y\prec Z$ for some $Z$ strictly above $Y$, by S1. 
By the analogous argument for negative $Y$ we conclude that all points on 
$I$ have the same `sign'.

Since $\Gamma$ is convex, and therefore connected, the coexistence of 
positive and negative points would mean that there are pairs of 
points of different sign, arbitrarily close together. Now if $X$ and $Y$ 
are sufficiently close, then the line $I_{Y}$ through $Y$ parallel to the 
$U$ axis intersects both $A_{X}$ and its complement. (This follows 
easily from the finite slope of the tangent plane, cf. the proof of 
Theorem 3.5 (ii) below.) Transitivity and Lemma 
3.2 imply that any point in $\partial A_{X}\cap I_{Y}$ has the same sign 
as 
$X$, and since all points on $I_{Y}$ have the same sign, this applies 
also to $Y$. 
\hfill\lanbox
%%%%%%%%%%%%%%%%%%

{}From now on we adopt the convention that the forward sectors in 
$\Gamma$ are on the {\it positive energy side} of all the tangent 
planes.  The mathematical and physical aspects of this choice were 
already discussed above.
%%%%

Since negative states are thus excluded (the possibility to do so is 
the content of Theorem 3.3), we may restate Lemma 3.2 in 
the following way, which we call {\it Planck's principle}
because Planck emphasized the importance for thermodynamics of the fact
that `rubbing' (i.e., increasing the energy at fixed work coordinate)
is an irreversible process (Planck, 1926, 1954).
\medskip

{\bf THEOREM 3.4 (Planck's principle).} {\it If two states, $X$ and $Y$,  
of a simple system have the 
same work coordinates, then $X\prec Y$ if and only if the energy of $Y$ 
is 
no less than the energy of $X$.}
\medskip

Taking  our operational definition of the relation $\prec$ in Sect.  II
into account, the `only if' part of this theorem is essentially a
paraphrasing of the  Kelvin-Planck statement in Section I.A., but
avoiding the concept of `cooling': 

\medskip
{\it `No process is possible, the sole result of which is a change in the
energy of a simple system (without changing the work coordinates) and the
raising of a weight.'}\medskip

This statement is clearly stronger than Carath\'eodory's principle, for 
it explicitly identifies  states 
that are arbitrarily close to a given state, but not adiabatically 
accessible from it. 

It is worth remarking that Planck's principle, and hence this version 
of the Kelvin-Planck statement, already follows from axioms A1-A7, S1 and 
a 
part of S2, namely the requirement that the tangent planes to the 
forward sectors have finite slope with respect to the work coordinates. 
Neither Lipschitz continuity of the slope, nor the connectedness 
axiom S3, are needed for this. However, although Planck's principle 
puts severe restrictions on the geometry of forward 
sectors, it alone does not suffice to establish the comparison principle. 
For instance,  the forward sector 
$A_{Y}$ of a point $Y$ on the boundary $\partial A_{X}$ of another 
forward sector could be properly contained in $A_{X}$. In such a 
situation 
the relation $\prec$ could 
not be characterized by an entropy function. In order to exclude 
pathological cases like this we shall now study the boundary 
$\partial A_{X}$ of a forward sectors in more detail, making full use of 
the axioms S2 and S3.

We denote by $\uprho_X$ the projection of $\partial A_X$ on $\R^n$, i.e.,
$$
\uprho_X = \{ V \in \R^n : (U,V) \in \partial A_X \ \hbox{for some} 
\ U \in \R \}. \eqno(3.13)
$$

Clearly, $\uprho_X$ is a {\it connected subset} of $\R^n$ because of
assumption S3. Note that $\uprho_X$ might be strictly smaller than
the projection of $A_X$. See Figure 4. 
\medskip

\centerline{\sevenpoint ---- Insert Figure 4 here ----}
\bigskip
\bigskip
{\bf THEOREM 3.5 (Definition and properties of the function $u_X$).}
{\it Fix $X = (U^0, V^0)$ in $\Gamma$.

(i).  Let $Y \in \partial  A_X$.  Then $A_X$ has a tangent
plane at $Y$ and it is $\Pi_Y$.

(ii).  $\uprho_X$ is an open, connected subset of $\R^n$.

(iii).  For each $V \in \uprho_X$ there is exactly one number, $u_X
 (V)$, such that $(u_X (V), V) \in \partial A_X$.  I.e., $$\partial  A_X
= \{ (u_X (V), V) : V \in \uprho_X \}. \eqno(3.14)$$ This $u_X (V)$ is
given by
$$
u_X (V) = \inf \{ u: (u,V) \in  A_X \}. \eqno(3.15)$$
The function $u_X$ is continuous on $\uprho_X$ and locally convex,
i.e., $u_X$ is convex on any convex subset of $\uprho_X$. (Note that
$\uprho_X$ need not be convex---or even contractible to a point.)
Moreover, 
$$
A_X \supset\{ (U,V): U \geq u_X (V), \quad V \in \uprho_X
\} \bigcap
\Gamma. \eqno(3.16)$$

(iv).  The function $u_X$ is a differentiable function on $\uprho_X$ with 
a 
locally
Lipschitz continuous derivative and satisfies the system of partial
differential equations 
$$
{{\partial} u_X \over {\partial}V_j} (V) = - P_j (u_X (V), V)
\quad {\rm for} \ j=1,\dots ,n \ . \eqno(3.17)
$$
 
(v).  The function $u_X$ is the only continuous function defined on 
$\uprho_X$ that
satisfies the differential equation, {\rm (3.17)}, in the sense of 
distributions, and that satisfies $u_X(V^0) = U^0$. }
%[Indeed, it was
%precisely our desire to have a unique solution to {\rm (3.17)} 
%that motivated  the axiom S2.] 

\medskip

{\it Remark:}  A solution to (3.17) is not guaranteed {\it a priori}; an
integrability condition on $P$ is needed. However, our assumption S2
implies that $P$ describes the boundary of $A_X$ (cf.\ (i) above), so 
the integrability condition is automatically fulfilled. Thus, a solution
exists. It is the Lipschitz continuity that yields uniqueness; indeed, it
was precisely our desire to have a unique solution to (3.17) that
motivated  axiom S2.  
\medskip

{\it Proof:}  (i). Since $Y \in \partial  A_X$, $ A_X$ has some support
plane, $\Pi$, at $Y$.  Since $A_{X}$ is closed by Theorem 3.1 we have
$Y\in A_X$ and hence $A_Y \subset  A_X$ by transitivity, A2. Thus $\Pi$
also supports $A_Y$ at $Y$.  By assumption S2, $A_Y$ has a {\it unique}
support plane at $Y$, namely $\Pi_Y$.  Therefore, $\Pi = \Pi_Y$.

(ii).  Connectedness of
$\uprho_X$ follows immediately from assumption S3,
i.e., $\partial A_X$  is connected. The following proof that $\uprho_X$ 
is 
open 
does not use assumption S3. The key fact is that by (i) and S2 the 
tangent plane to the convex set $A_{X}$ has finite slope at any 
$Y\in\partial A_{X}$ . Pick a $Y=(U,V)\in \partial A_{X}$. Since  
$\Gamma$ 
is 
open, the closed 
cylinder $C = 
\{ (U^\prime, V^\prime): \vert
V^\prime - V \vert \leq \varepsilon, \ \vert U^\prime - U \vert \leq
\sqrt{\varepsilon} \,\}$ with $Y$ at its center lies in $\Gamma$ for 
$\varepsilon > 0$ small enough.  Since the tangent plane through 
$Y$ has finite slope, the bottom `disc' 
$D_{-}=\{ (U -\sqrt{\varepsilon},
V^\prime):  \vert V^\prime - V \vert < \varepsilon \}$ lies below the 
tangent plane for $\varepsilon$ small enough and thus belongs to the 
complement of $A_{X}$. Consider the intersection of $A_{X}$  with the top 
disc,
$D_{+}=\{ (U +\sqrt{\varepsilon},
V^\prime):  \vert V^\prime - V \vert < \varepsilon \}$.
This intersection  is compact, convex and  contains the 
point $(U+\sqrt{\varepsilon},V)$ by Lemma 3.2 and A2 (the latter 
implies that $A_{Y}\subset A_{X}$). Its boundary is also compact and 
thus
contains a point with minimal distance $\delta$ from the cylinder 
axis (i.e, from the point $(U+\sqrt{\varepsilon},V)$ ).
We are obviously done if we show that $\delta>0$, for then all lines 
parallel to the cylinder axis with distance $<\delta$ from the axis 
intersect both $A_{X}$ and its complement, and hence the boundary
$\partial A_{X}$. Now, if $\delta=0$, it follows from Lemma 3.2 and 
transitivity
that the vertical line joining $(U+\sqrt{\varepsilon},V)$ 
and $(U,V)$ has an empty intersection with the interior of $A_{X}$. 
But then $A_{X}$ has a vertical support plane (because it is a convex 
set), contradicting S2.

(iii).  The proof of (3.14)-(3.16) is already contained in Lemma 3.2, 
bearing in mind that $A_{Y}\subset A_{X}$ for all $Y\in\partial A_{X}$.
The local convexity of $u_X$ follows from its definition: Let $C
\subset \uprho_X$ be convex, let $V^1$ and $V^2$ be in $C$ and let $0
\leq \lambda \leq 1$.  Then the point $V := \lambda V^1 + (1- \lambda)
V^2$ is in $C$ (by definition) and, by axiom A7, $(\lambda u_X (V^1) +
(1 - \lambda) u_X (V^2), V)$ is in $ A_X$.  Hence, by (3.15), $u_X (V)
\leq \lambda u_X (V^1) + (1-\lambda) u_X (V^2)$.  Finally, every convex
function defined on an open, convex subset of $\R^n$ is continuous.

(iv).  Fix $V \in \uprho_X$, let $B \subset \uprho_X$ be an open ball
centered at $V$ and let $Y := (u_X (V), V) \in \partial  A_X$.  By (i)
above and (3.4) we have 
$$
u_X (V^\prime) - u_X (V) + \sum\limits_i P_i
(Y) (V^\prime_i - V_i) \geq 0   \eqno(3.18)
$$ 
for all $V^\prime \in B$.  Likewise, applying (i) above and (3.4) to
the point $Y^\prime := (u_X (V^\prime), V^\prime)$ we have
$$
u_X (V) - u_X (V^\prime) + \sum \limits_i P_i (Y^\prime) (V_i -
V^\prime_i) \geq 0 \ .   \eqno(3.19)
$$
As $V^\prime \rightarrow V, P(Y^\prime) \rightarrow P(Y)$, since $u_X$
is continuous and $P$ is continuous.  Thus, if $1 \leq j \leq n$ is
fixed and if $V^\prime_i := V_i$ for $i \not= j$, $V^\prime_j = V_j +
\varepsilon$ then, taking limits $\varepsilon \rightarrow 0$ in the two
inequalities above, we have that 
$$
{u_X (V^\prime) - u_X (V) \over
\varepsilon} \rightarrow - P_j (Y) \ , \eqno(3.20)
$$ 
which is precisely (3.17).

By assumption $P(Y)$ is continuous, so $u_X$ is continuously
differentiable, and hence locally Lipschitz continuous.  But then
$P(u_X (V), V)$ is locally Lipschitz continuous in $V$.

(v). The uniqueness is a standard application of Banach's contraction
mapping principle, given the important hypothesis that $P$ is locally
Lipschitz continuous and the connectedness of the open set $\uprho_{X}$.
$\rho_{X}$.\hfill\lanbox
\medskip
According to the last theorem the boundary of a forward sector 
is described by the unique solution of a system of differential 
equations. 
As a corollary it follows that all points on the boundary are 
adiabatically equivalent and thus have the same forward sectors:
\medskip

{\bf THEOREM 3.6 (Reversibility on the boundary).} {\it If $Y \in 
\partial 
A_X$, 
then $X \in \partial A_Y$ and hence $A_Y =  A_X$. }
\medskip
{\it Proof:} Assume $Y =(U^1,V^1)\in \partial  A_X$. The boundary 
$\partial 
A_{Y}$ is described by the function $u_{Y}$  which solves Eqs.\ (3.17) 
with the condition $u_{Y}(V^1)=U^1$. But $u_{X}$ , which describes the 
boundary $\partial A_{X}$, solves the same 
equation with the same initial condition. This solution is unique 
on $\rho_{Y}$ by Theorem 3.5(v), so we conclude that $\partial 
A_{Y}\subset \partial A_{X}$ and hence $\rho_{Y}\subset \rho_{X}$.
 The theorem will be proved if we show that $\uprho_X = \uprho_Y$.
Suppose, on the contrary, that $\uprho_Y$ is strictly smaller than
$\uprho_X$.  Then, since $\uprho_X$ is open, there is some point $V \in
\uprho_X$ that is in the boundary of $\uprho_Y$,  and hence $V \not\in
\uprho_Y$ since $  \uprho_Y$ is open.  We claim that $\partial  A_Y$ is
not relatively closed  in $\Gamma$, which is a contradiction since $
A_Y$ must be relatively closed. To see this, let $V^j$, for $j = 1,2,3,
\dots$ be in $\uprho_Y$ and $V^j \rightarrow V$ as $j \rightarrow
\infty$.  Then $u_X (V^j) \rightarrow u_X (V)$ since $u_X$ is
continuous.  But $u_Y (V^j) = u_X (V^j)$, so the sequence of points
$(u_Y (V^j), V)$ in $ A_X$ converges to $Z:= (u_X (V), V) \in \Gamma$.
Thus, $Z$ is in the relative closure of $\partial  A_Y$ but $Z \not\in
\partial  A_Y$ because $V \not\in \uprho_Y$, thereby establishing a
contradiction.  \hfill\lanbox
\medskip

We are now in a position to prove the main result in this section.  It 
shows that $\Gamma$ is foliated by the adiabatic surfaces $\partial 
A_X$, and that the points of $\Gamma$ are all comparable.  More 
precisely, $X\prec\prec Y$ if and only if $A_Y $ is contained in the 
interior of $A_X$, and $X \sima Y$ if and only if $Y \in \partial 
A_X$.  \medskip

{\bf THEOREM 3.7 (Forward sectors are nested).}  {\it With the above
assumptions, i.e., A1-A7 and S1-S3, we have the following.  If $A_X$
and $A_Y$ are two forward sectors in the state space, $\Gamma$, of a
simple system then exactly one of the following holds.

(a).  $A_X = A_Y$, i.e., $ X\sima Y$. 
\smallskip

(b).  $A_X \subset {\rm Interior}(A_Y)$, i.e., $Y \prec\prec X$.
\medskip

(c).  $A_Y \subset {\rm Interior}(A_X)$, i.e., $X \prec\prec Y$.
\medskip

\noindent
In particular, $\partial A_X$ and $\partial A_Y$ are either
identical or disjoint.}
\medskip

{\it Proof:\/} There are three (non-exclusive) cases:

Case 1. $Y\in A_X$

Case 2. $X\in A_Y$        

Case 3. $X\notin A_Y$ and $Y\notin A_X$ .

By transitivity, case 1 is equivalent to $A_Y \subset A_X$. Then, either
$Y \in \partial A_X$ (in which case $A_Y=A_X$ by Theorem 3.6) or $Y \in
{\rm Interior}(A_X)$. In the latter situation we conclude that $\partial
A_Y \subset {\rm Interior} (A_X)$, for otherwise $\partial A_Y \cap
\partial A_X$ contains a point $Z$ and Theorem 3.6 would tell us that
$\partial A_Y =\partial A_Z= \partial A_X$, which would mean that
$A_Y=A_X$.  Thus, case 1 agrees with the conclusion of our theorem.

Case 2 is identical to case 1, except for interchanging $X$ and $Y$.

Therefore, we are left with the case that $Y\notin A_X$ and $X\notin
A_Y$. This, we claim, is impossible for the following reason.

Let $Z$ be some point in the interior of $A_X$ and consider  the line
segment $L$ joining $Y$ to $Z$ (which lies in $\Gamma$ since $\Gamma$ is
convex). If we assume  $Y\notin A_X$ then part of $L$ lies outside
$A_X$, and therefore $L$ intersects $\partial A_X$ at some point $W\in
\partial A_X$. By Theorem 3.6, $A_X$ and $A_W$ are the same set, so
$W\prec Z$ (because $X\prec Z$).  By Lemma 3.1, $Y\prec Z$ also.  Since
$Z$ was arbitrary, we learn that ${\rm Interior} (A_X) \subset A_Y$. By
the same reasoning ${\rm Interior}(A_Y) \subset A_X$. Since $A_X$ and
$A_Y$ are both closed, the assumption that $Y\notin A_X$ and $X\notin
A_Y$ has led us to the conclusion that they are identical.  
\hfill\lanbox 
\bigskip

Figure 5 illustrates the content of Theorem 3.7. The end result is that
the forward sectors are nicely nested and thereby establishes the
comparison hypothesis for simple systems, among other things. 

\centerline{\sevenpoint ---- Insert Figure 5 here ----}

The adiabats $\partial A_{X}$ foliate $\Gamma$ and using Theorem 3.5 it
may be shown that there is always a continuous function $\sigma$ that
has exactly these adiabats as level sets.  (Such a function is usually
referred to as an `empirical entropy'.) But although the sets $A_X$ are
convex, the results established so far do not suffice to show that
there is a {\it concave} function with the adiabats as level sets. For
this and further properties of entropy we shall rely on the axioms
about {\it thermal equilibrium} discussed in the next section.

\bigskip

As a last topic in this section we would like to come back to the claim
made in Section II.A.2. that our operational definition of the
relation $\prec$ coincides with definitions in textbooks based on the
concept of `adiabatic process', i.e., a process taking place in an
'adiabatic enclosure'.  We already discussed  the connection from a
general point of view in Section II.C, and showed that both definitions
coincide. However, there is also another point of view that relates the
two, and which we now present. It is based on the idea that, quite
generally , if one relation is included in another then the two
relations must coincide for simple systems. This very general result is
Theorem 3.8 below.

Whatever `adiabatic process'  means, we consider it a minimal
requirement that the relation based on it is a subrelation of 
our $\prec$, according to the operational definition in Sect. 
II.A.
More precisely, denoting this hypothetical relation based on
`adiabatic process' by $\prec^*$, it should be true that $X\prec^* Y$
implies $X\prec Y$. Moreover, our motivations for the axioms A1-A6 and
S1-S3 for $\prec$ apply equally well to $\prec^*$, so we may assume that
$\prec^*$ also satisfies these axioms.  In particular, the forward
sector $A_X^*$ of $X$ with respect to $\prec^*$ is convex and closed
with a nonempty interior and with $X$ on its boundary. The following
simple result shows that $\prec$ and $\prec^*$ must then necessarily
coincide.
\medskip

{\bf THEOREM 3.8 (There are no proper inclusions).} {\it Suppose that
$\prec^{(1)}$ and  $\prec^{(2)} $ are two relations on multiple scaled
products of a simple system
$\Gamma$ satisfying axioms A1-A7 as well as S1-S3. If
$$
X\prec^{(1)} Y \quad\quad {\rm implies} \quad \quad X \prec^{(2)} Y
$$
for all $X, Y\in\Gamma$, then $\prec^{(1)} = \prec^{(2)} $ .  }

{\it Proof:} We use  superscripts $(1)$ and $(2)$ to denote the two
cases. Clearly, the hypothesis is equivalent to $A_X^{(1)} \subset
A_X^{(2)} $ for all $X\in \Gamma$. We have to prove $A_X^{(2)} \subset
A_X^{(1)}$. Suppose not. Then there is a $Y$ such that $X\prec^{(2)} Y$
but $X \not\prec^{(1)} Y$. By Theorem 3.7 for $\prec^{(1)}$ we have that
$Y\prec ^{(1)} X$. By our hypothesis, $Y\prec^{(2)} X$, and thus we have
$X \sima^{(2)} Y$.

Now we use what we know about the forward sectors of simple systems.
$A_X^{(2)}$ has a non-empty interior, so the complement of $A_X^{(1)}$
in $A_X^{(2)}$ contains a point $Y$ that is {\it not} on the
boundary of $A_X^{(2)}$. On the other hand, we just proved that
$X\sima^{(2)} Y$, which implies that $Y \in \partial A_X^{(2)}$.
This is a contradiction.     \hfill \lanbox

\vfill\eject
%%%%%%%%%%%%%%%%%%%%%%%%%%%%%%%%%%
%%%%%%%%%%%%%%%%%%%%
\noindent
{\tit IV.  THERMAL  EQUILIBRIUM}
\bigskip

In this section we introduce our axioms about thermal contact of 
simple systems.  We then use these assumptions to derive the 
comparison hypothesis for products of such systems.  This will be done 
in two steps.  First we consider scaled copies of a single simple 
system and then products of different systems.  The key idea is that 
two simple systems in thermal equilibrium can be regarded as a new 
simple system, to which Theorem 3.7 applies.  We emphasize that the 
word `thermal' has nothing to do with temperature---at this point in 
the discussion.  Temperature will be introduced in the next section, 
and its existence will rely on the properties of thermal contact, but 
thermal equilibrium, which is governed by the zeroth law, is only a 
statement about mutual equilibrium of systems and not a statement 
about temperature.

\bigskip

\noindent
{\subt A. Assumptions about thermal contact}
\bigskip

We assume that a relation $\prec$ satisfying axioms A1--A6 is given, 
but A7 and CH are {\it not } assumed here.  We shall make five 
assumptions about thermal equilibrium, T1-T5.  Our first axiom says 
that one can form new simple systems by bringing two simple systems 
into thermal equilibrium and that this operation is adiabatic (for the 
compound system, not for each system individually).  
\medskip

\item {\bf T1)} {\bf Thermal contact.} Given any two simple systems with 
state
spaces $\Gamma_1$ and $\Gamma_2$, there is another simple system, called 
the
{\bf the thermal join of} $\Gamma_1$ {\bf and} $\Gamma_2$, whose state 
space 
is
denoted by  $\Delta_{12}$. 
The  work coordinates in $\Delta_{12}$ are $(V_1,V_2)$ with $V_1$ the 
work 
coordinates of $\Gamma_1 $ and $V_2$ the work coordinates of  $\Gamma_2$.
The range of the (single) energy coordinate of $\Delta_{12}$ is the {\it 
sum} of
all possible energies in $\Gamma_1 $ and $\Gamma_2$ for the 
given values of the work coordinates.  In symbols: 
$$
\Delta_{12} = \{ (U,V_1,V_2) : U=U_1+U_2 \;{\rm with}\; (U_1,V_1)\in 
\Gamma_1,
(U_2,V_2)\in \Gamma_2\}.  \eqno (4.1)
$$ 
By assumption, there is always an adiabatic
process, called {\bf thermal equilibration} that takes a state in the 
compound
system, $\Gamma_1 \times \Gamma_2$, into a state in $\Delta_{12}$ which 
is
given by the following formula: 
$$ 
\Gamma_1 \times \Gamma_2 \ni ((U_1,V_1), (U_2,V_2))\prec
(U_1+U_2,V_1,V_2) \in \Delta_{12}. 
$$

\medskip

{}From the physical point of view, a state in $\Delta_{12}$ is a 
``black box" containing the two systems, with energies $U_1$ and 
$U_2$, respectively, such that $U_1+ U_2 = U$.  The values of $U_1$ 
and $U_2$ need not be unique, and we regard all such pairs (if there 
is more than one) as being equivalent since, by T2 below, they are 
adiabatically equivalent.  This state in $\Delta_{12}$ can be 
pictured, physically, as having the two systems side by side (each 
with its own pistons, etc.)  and linked by a copper thread that allows 
`heat' to flow from one to the other until thermal equilibrium is 
attained.  The total energy $U =U_1+ U_2$ can be selected at will 
(within the range permitted by $V_1$ and $V_2$), but the individual 
energies $U_1$ and $U_2$ will be determined by the properties of the 
two systems.  Note that $\Delta_{12}$ is convex---a fact that follows 
easily from the convexity of $\Gamma_1$ and $\Gamma_2$.

\medskip

The next axiom  simply declares the `obvious' fact that 
we can disconnect the copper thread, once equilibrium has been reached,
and restore the original two systems. \medskip

\item{\bf T2) } {\bf Thermal splitting.}
For any point $(U,V_1,V_2) \in \Delta_{12}$ there is at least
one pair of
states, $(U_1,V_1) \in \Gamma_1$, $(U_2,V_2))\in
\Gamma_2$, with $U=U_1+U_2$,  such that
$$
\Delta_{12} \ni (U,V_1,V_2)\sima ((U_1,V_1), (U_2,V_2)) \in 
\Gamma_1 \times \Gamma_2.
$$
In particular, the following is assumed  to hold:
If  $(U,V)$ is a state of a simple system $\Gamma$ and 
$\lambda\in[0,1]$ then
$$
(U,(1-\lambda)V,\lambda V) \sima
(((1-\lambda)U,(1-\lambda)V),(\lambda U,\lambda V)) \in 
\Gamma^{(1-\lambda)} \times \Gamma^{(\lambda)}.
$$
\medskip

We are now in a position to introduce another kind of  equivalence 
relation among states, in  addition to $\sima$. \medskip

{\bf Definition.} If $((U_1,V_1), (U_2,V_2))\sima
(U_1+U_2,V_1,V_2)$  we say that the states
$X=(U_1,V_1)$ and $Y=(U_2,V_2)$ are in {\bf thermal equilibrium} and
write 
$$
X\simt Y.
$$

It is clear that $X\simt Y$ implies $Y\simt X$. Moreover,  by axiom T2 
and
axioms A4 and A5 we always have $X\simt X$.

The next axiom implies that $\simt$ is, indeed, an equivalence relation. 
It
is difficult to overstate its importance since it is the key to 
eventually
establishing the fact that {\it entropy is additive not only with respect 
to
scaled copies of one system but also with respect to different kinds
of systems.} \medskip

\item{\bf T3)} {\bf Zeroth law of thermodynamics.} If $X\simt Y$ and if
$Y\simt Z$ then $X\simt Z$. \medskip

The equivalence classes w.r.t.\ the relation $\simt$ are called {\bf 
isotherms}.

The question whether the zeroth law is really needed as an independent
postulate or can be derived from other assumptions is the subject of some
controversy, see e.g., (Buchdahl, 1986), (Walter, 1989), (Buchdahl, 
1989).
Buchdahl (1986) derives it from his analysis of the second law for {\it 
three}
systems in thermal equilibrium. However, it is not clear whether the 
zeroth 
law
comes for free; if we really pursued this idea in our framework we should
probably find it necessary to invoke some sort of assumption about the
three-system equilibria.  

Before proceeding further let us point out a simple consequences
of T1-T3.  \medskip

{\bf THEOREM 4.1 (Scaling invariance of thermal equilibrium.)} {\it If 
$X$ 
and
$Y$ are two states of two simple systems (possibly the same or possibly
different systems) and if $\lambda, \mu >0$  then the relation $X\simt Y$
implies $\lambda X\simt \mu Y$.} \smallskip

{\it Proof:} $(X, \lambda X) = ((U_X,V_X), (\lambda U_X,\lambda V_X))
\sima ((1+\lambda)U_X,V_X,\lambda V_X)$ by axiom T2. But this means, by 
the above definition of thermal equilibrium,
that $X\simt \lambda X$. In the same way, $Y\simt \mu Y$. By the 
zeroth law, axiom T3, this implies $\lambda X\simt \mu Y$.
\hfill\lanbox
\medskip

Another simple consequence of the axioms for thermal contact concerns
the orientation of forward sectors with respect to the energy.  In
Theorem 3.3 in the previous section we had already showed that in a
simple system the forward sectors are either all on the positive energy
side or all on the negative energy side of the tangent planes to the
sectors, but the possibility that the direction is different for
different systems was still open.  The coexistence of systems belonging
to both cases, however, would violate our axioms T1 and T2.  The
different orientations of the sectors with respect to the energy
correspond to different signs for the temperature as defined in Section
V.  Our axioms are only compatible with systems of one sign. \medskip

{\bf THEOREM 4.2 (Direction of forward sectors).}{\it The forward 
sectors of all simple systems point the same way, i.e., they are 
either all on the positive energy side of their tangent planes or all 
on the negative energy side.}

{\it Proof:\/} This follows directly from T1 and T2, because a system
with sectors on the positive energy side of the tangent planes 
can never come to thermal equilibrium with a
system whose sectors are on the negative side of the tangent planes. To 
be precise, 
suppose that 
$\Gamma_{1}$ has positive
sectors, $\Gamma_{2}$ has negative sectors and that there are states
$X=(U_{1},V_{1})\in\Gamma_{1}$ and $Y=(U_{2},V_{2})\in\Gamma_{2}$ 
such that $X\simt Y$. (Such states exist by T2.) Then, for
any sufficiently small $\delta>0$,
$$
(U_{1},V_{1})\prec (U_{1}+\delta,V_{1})\qquad\hbox {\rm and}\qquad 
(U_{2},V_{2})\prec (U_{2}-\delta,V_{2})
$$
by Theorem 3.4 (Planck's principle). With $U:=U_1+U_2$ we then have the
two relations  
$$\eqalign{
(U,V_1,V_2) \sima ((U_1,V_1),\ (U_2,V_2))  &\prec 
((U_1+\delta,V_1),\ (U_2,V_2)) \prec (U+\delta,V_1,V_2) \cr
   \noalign{\smallskip}
(U,V_1,V_2) \sima ((U_1,V_1),\ (U_2,V_2))  &\prec 
((U_1,V_1),\ (U_2-\delta,V_2)) \prec (U-\delta,V_1,V_2). \cr}
$$ 
This means that starting from $(U,V_1,V_2)\in \Delta_{12}$ we can move
adiabatically both upwards and downwards in energy (at fixed work
coordinates), but this is impossible (by Theorem 3.3)
because $\Delta_{12}$ is a simple system, by Axiom T1.
\hfill\lanbox

%%%%%%%%%%%%%

For the next theorem we recall that an entropy function on $\Gamma$ is
a function that exactly characterizes the relation $\prec$ on multiple
scaled copies of $\Gamma$, in the sense of Theorem 2.2.  As defined in
Section II, entropy functions $S_1$ on $\Gamma_1$ and  $S_2$ on
$\Gamma_2$ are said to be {\it consistent} if together they
characterize the relation $\prec$ on multiple scaled products of
$\Gamma_1 $ and $\Gamma_2 $ in the sense of Theorem 2.5. The comparison
hypothesis guarantees the existence of such consistent entropy
functions, by Theorem 2.5, but our present goal is to derive the
comparison hypothesis for compound systems by using the notion of
thermal equilibrium.  In doing so, and also in Section V, 
we shall make use of the following 
consequence of consistent entropy functions.  \medskip

{\bf THEOREM 4.3 (Thermal equilibrium is characterized by
maximum entropy).} {\it If  $S$ is an entropy function on the state space 
of a simple system, then $S$ is a
 concave function of $U$ 
for fixed $V$. If $S_1$ and $S_2$ are consistent entropy functions on the 
state spaces $\Gamma_1$ and $\Gamma_2$ of two
simple systems and $(U_i,V_i)\in \Gamma_i$, $i=1,2$, then 
$(U_1,V_1)\simt (U_2,V_2)$
holds if and only if the sum of the entropies takes its maximum value at
$((U_1,V_1),(U_2,V_2))$ for fixed total energy and fixed work 
coordinates, 
i.e.,}
$$
\max_W\left[S_1(W,V_1)+S_2((U_1+U_2)-W),V_2)\right]=S_1(U_1,V_1)+S_2(U_2,
V
_2
).
\eqno(4.2)
$$ 
\smallskip

{\it Proof:} The concavity of $S$ is true for any simple system by 
Theorem 2.8, which 
uses the convex combination axiom A7. It is interesting to note, however, 
that
concavity in $U$ for fixed $V$ follows from axioms T1, T2  and A5 
alone, even 
if A7 is {\it not} assumed. In fact, by axiom T1 we have,
for states $(U,V)$ and $(U',V)$ of a simple system 
with the same work coordinates,
$$
(((1-\lambda)U,(1-\lambda)V),(\lambda U',\lambda V))\prec 
((1-\lambda)U+\lambda U',(1-\lambda)V,\lambda V).
$$
By T2, and with $U^{''} := (1-\lambda)U+\lambda U'$, this latter state 
is $\sima$ equivalent to
$$
((1-\lambda)U^{''},(1-\lambda) V), (\lambda U^{''},\lambda V)),
$$
which, by A5,  is $\sima$ equivalent to $(U^{''},V)$.
Since $S$ is additive and non decreasing under $\prec$ this implies
$$
(1-\lambda) S(U,V)+\lambda S(U',V)\leq S((1-\lambda)U+\lambda U',V).
$$
 
For  the second part of our  theorem, let $(U_1,V_1)$ and $(U_2,V_2)$ be
states of two simple systems. Then T1 says that for any $W$ such that
$(W,V_1)\in \Gamma_1$ and $((U_1+U_2-W),V_2)\in \Gamma_2$ one has
$$
((W,V_1),((U_1+U_2)-W),V_2))\prec (U_1+U_2,V_1,V_2) .
$$ 
The definition of
thermal equilibrium says that $(U_1+U_2,V_1,V_2)\sima 
((U_1,V_1)(U_2,V_2))$ 
if
and only if $(U_1,V_1)\simt (U_2,V_2)$. Since the sum of consistent 
entropies
characterizes the order relation on the product space the assertion of 
the
lemma follows.  \hfill\lanbox

\medskip

We come now to what we call the {\it transversality axiom}, which is
crucial for establishing the comparison hypothesis, CH, for products of
simple systems. \medskip

\item {\bf T4)} {\bf Transversality.} If $\Gamma$ is the state space of a
simple system and if $X \in \Gamma$, then there exist states $X_0\simt
X_1$ with
$X_0\prec\prec X\prec\prec X_1$.

To put this in words, the axiom requires that for every adiabat there 
exists at least one isotherm (i.e., an equivalence class w.r.t.  
$\simt$\ ), containing points on both sides of the adiabat.  Note that, 
for 
each given $X$, only two points in the entire state space $\Gamma$ are 
required to have the stated property.  See Figure 6.

\centerline{\sevenpoint ---- Insert Figure 6 here ----}

We remark that the condition $X \prec\prec X_1$ obviously implies 
axiom S1.  However, as far as the needs of this Section IV are 
concerned, the weaker condition $X_0\prec X\prec X_1$ together with 
$X_0\prec\prec X_1$ would suffice, and this would {\it not} imply S1.  
The strong version of transversality, stated above, will be needed in 
Section V, however.

At the end of this section we shall illustrate, by the example of 
`thermometers', the significance of axiom T4 for the existence of an 
entropy function.  There we shall also show how an entropy function 
can be defined for a system that violates T4, {\it provided} its 
thermal combination with some other system (that itself satisfies T4) 
does satisfy T4.

The final thermal axiom states, essentially, that the range of 
temperatures that a simple system can have is the same for all simple 
systems under consideration and is independent of the work 
coordinates.  In this section axiom T5 will be needed only for Theorem 
4.9.  It will also be used again in the next section when we establish 
the existence and properties of temperature.  (We repeat that the word 
`temperature' is used in this section solely as a mnemonic.)  \medskip

\item {\bf T5)} {\bf Universal temperature range.} If $\Gamma_1$ and 
$\Gamma_2$
are state spaces of simple systems then,  for every $X\in\Gamma_1$ and 
every
$V\in\uprho(\Gamma_2)$, where $\rho$ denotes the projection on the 
work coordinates, $\rho(U',V'):=V'$,  there is a $Y\in\Gamma_2$ with 
$\uprho (Y)=V$, such that 
$X\simt
Y$.
\smallskip

The physical motivation for T5 is the following.  A sufficiently large 
copy of the first system in the state $X \in \Gamma_1$ can act as a 
heat bath for the second, i.e., when the second system is brought into 
thermal contact with the first at fixed work coordinates, $V$, it is 
always possible to reach thermal equilibrium, but the change of $X$ 
will be very small since $X$ is so large.

This axiom is inserted mainly for convenience and one might weaken it 
and require it to hold only within a group of systems that can be 
placed in thermal contact with each other.  However, within such a 
group this axiom is really necessary if one wants to have a consistent 
theory.

\vfill\eject
\bigskip\noindent
{\subt B. The comparison principle in compound systems}

\bigskip\noindent
{\subsubt 1. Scaled copies of a single simple system}
\bigskip 

We shall now apply the thermal axioms, T4 in particular, to derive the
comparison hypothesis, CH, for multiple scaled copies of simple
systems. \medskip

{\bf THEOREM 4.4 (Comparison in multiple scaled copies of a simple
system).}
{\it Let $\Gamma$ be the state space of a simple system and
let $a_1, \dots , a_M, a^\prime_1, \dots , a^\prime_M$ be positive
real numbers with $a_1 + \cdots + a_N = a^\prime_1 + \cdots +
a^\prime_M$.  Then all points in $a_1 \Gamma \times \cdots \times a_N
\Gamma$ are comparable to all points in $a^\prime_1 \Gamma \times
\cdots \times a^\prime_M \Gamma$.}
\smallskip

{\it Proof:}  We may suppose that $a_1 + \cdots + a_N = a^\prime_1 +
\cdots + a^\prime_M = 1$.  We shall show that for any points $Y_1,
\dots, Y_N, Y^\prime_1, \dots ,Y^\prime_M \in \Gamma$ there exist points
$X_0 \prec\prec X_1$ in $\Gamma$ such that $(a_1 Y_1, \dots , a_N Y_N)
\sima ((1 - \alpha) X_0, \alpha X_1)$ and $(a^\prime_1 Y^\prime_1,
\dots , a^\prime_N Y^\prime_N) \sima ((1 - \alpha^\prime) X_0,
\alpha^\prime X_1)$ with $\alpha, \alpha^\prime \in \R$.  This will
prove the statement because of Lemma 2.2.  

By Theorem 3.7, the
points in $\Gamma$ are comparable, and hence there are points $X_0
\prec X_1$ such that all the points $Y_1, \dots , Y_N, Y^\prime_1,
\dots , Y^\prime_M$ are contained in the strip $\Sigma (X_0, X_1)
=\{X\in \Gamma:\ X_{0}\prec X\prec X_{1}\}$; in
particular, these $N+M$ points can be linearly ordered and $X_0$ and
$X_1$ can be chosen from this set.  If $X_0 \sima X_1$ then all the
points in the strip would be equivalent and the assertion would hold
trivially.  Hence we may assume that $X_0 \prec\prec X_1$.  Moreover,
it is clearly sufficient to prove that for each $Y \in \Sigma (X_0,
X_1)$ one has $Y \sima ((1 - \lambda) X_0, \lambda X_1)$ for some
$\lambda \in [0,1]$, because the general case then follows by the
splitting and recombination axiom A5 and Lemma 2.2.

If $X_0 \simt
X_1$ (or, if there exist $X^\prime_0 \sima X_0$ and $X^\prime_1 \sima
X_1$ with $X^\prime_0 \simt X^\prime_1$, which is just as good for the
present purpose) the existence of such a $\lambda$ for a given $Y$ can
be seen as follows.  For any $\lambda^\prime \in [0,1]$ the states $((1
- \lambda^\prime) X_0, \lambda^\prime X_1)$ and $((1 - \lambda^\prime)
Y, \lambda^\prime Y)$ are adiabatically equivalent to certain states in
the state space of a 
simple system, thanks to thermal
axiom T2.  Hence $((1 - \lambda^\prime) X_0, \lambda^\prime X_1)$
and $Y \sima ((1 - \lambda^\prime) Y, \lambda^\prime Y)$ are
comparable.  We define
$$
\lambda = \sup \{ \lambda^\prime \in [0,1]:
((1 - \lambda^\prime) X_0, \lambda^\prime X_1) \prec Y \}. \eqno(4.3)
$$
Since $X_0 \prec Y$ the set on the right of (4.3) is not empty (it
contains 0) and therefore $\lambda$ is well defined and $0 \leq \lambda
\leq 1$.  Next, one shows that $((1 - \lambda) X_0, \lambda X_1) \sima
Y$ by exactly the same argument as in Lemma 2.3.  (Note that this
argument only uses that $Y$ and $((1 - \lambda^\prime) X_0,
\lambda^\prime X^\prime)$ are comparable.)  Thus, our theorem is
established under the hypothesis that $X_0 \simt X_1$.

The following Lemma 4.1 will be needed to  show that we can, indeed,
always choose $X_0$ and $X_1$ so that $X_0 \simt X_1$.  %\hfill\lanbox

\bigskip
{\bf LEMMA 4.1 (Extension of strips).} 
{\it For any state space (of a simple or a compund system), if 
$X_0 \prec\prec
X_1, X'_0 \prec\prec X^\prime_1$ and if
$$
\eqalignno{X &\sima ((1 -
\lambda) X_0, \lambda X_1) \qquad&(4.4) \cr X_1 &\sima ((1 - \lambda_1)
X^\prime_0, \lambda_1 X^\prime_1) \qquad&(4.5)\cr X^\prime_0 &\sima ((1-
\lambda_0) X_0, \lambda_0 X_1) \qquad&(4.6) \cr}$$ then $$ X \sima ((1 -
\mu) X_0, \mu X^\prime_1) \eqno(4.7) 
$$ 
with 
$$ \mu = {\lambda \lambda_1
\over 1 - \lambda_0 + \lambda_0 \lambda_1}. 
$$}

{\it Proof:}  We first consider the special case $X = X_1$, i.e.,
$\lambda = 1$.  By simple arithmetic, using the cancellation law, one
obtains (4.7) from (4.5) and (4.6) with $\mu = \mu_1 = {\lambda_1 \over 1 
-
\lambda_0 + \lambda_0 \lambda_1}$.  The general case now follows by
inserting the splitting of $X_1$ into (4.4) and recombining.

\hfill\hfill\lanbox   

\medskip

{\it Proof of Theorem 4.4 continued:}  By the transversality
property, each point $X$ lies in some strip $\Sigma
(X_0, X_1)$ with $X_0
\prec\prec X_1$ and $X_0 \simt X_1$.  Hence the whole state space can be
covered by strips $\sum (X^{(i)}_0, X^{(i)}_1)$ with $X^{(i)}_0
\prec\prec X^{(i)}_0$ and $X^{(i)}_0 \simt X^{(i)}_1$.  Here $i$
belongs to some index set.  Since all adiabats $\partial A_X$ with $X
\in \Gamma$ are relatively closed in $\Gamma$ by axiom S3 we can
even cover each $X$ (and hence $\Gamma$) with {\it open} strips
$\mathop{{\sum}_i}\limits^o := \sum\limits^o (X^{(i)}_0, X^{(i)}_1) =
\{ X: X^{(i)}_0 \prec\prec X \prec\prec X^{(i)}_0 \}$ with $X^{(i)}_0
\simt X^{(i)}_1$.  Moreover, any compact subset, $C$, of $\Gamma$ is
covered by a finite number of such strips $\mathop{{\sum}_i}\limits^o,
i = 1, \dots , K$, and if $C$ is connected we may assume that
$\mathop{{\sum}_i}\limits^o \cap
\mathop{{\sum}_{i+1}}\limits^{o{\phantom{111}}} \not= \emptyset$.  If 
$\bar
X_0$ denotes the smallest of the elements $X^{(i)}_0$ (with respect to
the relation $\prec$) and $\bar X_1$ the largest, it follows from 
Lemma 2.3 that for any $X \in C$ we have $X \sima ((1 - \mu)
\bar X_0, \mu \bar X_1)$ for some $\mu$.  If a finite number of points,
$Y_1, \dots , Y_N, Y^\prime_1, \dots , Y^\prime_M$ is given, we take
$C$ to be a polygon connecting the points, which exists because $\Gamma$
is convex.  Hence each of the points $Y_1, \dots , Y_N, Y^\prime_1,
\dots , Y^\prime_M$ is equivalent to $((1 - \lambda) \bar X_0, \lambda
\bar X_1)$ for some $\lambda$, and the proof is complete. 
\hfill\lanbox
\medskip

The comparison hypothesis, CH, has thus been established for multiple 
scaled copies
of a single simple system. {}From Theorem 2.2 we then know 
that for such a system the relation $\prec$ is characterized by an
entropy function, which is unique up to an affine transformation
$S \rightarrow a S +B$. 

\bigskip\noindent
{\subsubt 2. Products of different simple systems}
\bigskip

Our next goal is to verify the comparison hypothesis for products of 
different
simple systems. For this task we shall appeal to the following:
\medskip

{\bf THEOREM 4.5 (Criterion for comparison in product spaces).}  {\it Let
$\Gamma_1$ and $\Gamma_2$ be two (possibly unrelated) state spaces.  
Assume
there is a relation $\prec$ satisfying axioms A1-A6 that holds for 
$\Gamma_1, \Gamma_2$ and their scaled products. 
Additionally, $\prec$ satisfies the comparison hypothesis CH on
$\Gamma_1$ and its multiple scaled copies and on 
$\Gamma_2$ and its multiple scaled
copies but, a-priori, not necessarily on $\Gamma_{1}\times\Gamma_{2}$ or 
any 
other
products involving both $\Gamma_1$ and $\Gamma_2$

If there are points $X_0, X_1 \in \Gamma_1$ and $Y_0, Y_1 \in \Gamma_2$
such that
$$
X_0 \prec\prec X_1, \quad Y_0 \prec\prec Y_1 \eqno(4.8)
$$
$$
(X_0, Y_1) \sima (X_1, Y_0), \eqno(4.9)
$$
then the comparison hypothesis CH holds on 
products of any number of scaled copies of $\Gamma_1$ and 
$\Gamma_2$.} \smallskip

{\it Proof:}  Since the comparison principle holds for $\Gamma_1$ and
$\Gamma_2$ these spaces have canonical entropy functions corresponding,
respectively, to the reference points $X_0, X_1$ and $Y_0, Y_1$. If $X 
\in
\Gamma_1$ and $\lambda_1 = S_1 (X \vert X_0, X_1)$ (in the notation of 
eq. (2.15)) then, by 
Lemma 2.3,
$$X \sima ((1 - \lambda_1) X_0, \lambda_1 X_1)$$
and similarly, for $Y \in \Gamma_2$ and $\lambda_2 = S_2 (Y \vert Y_0,
Y_1)$,
$$Y \sima ((1 - \lambda_2) Y_0, \lambda_2 Y_1).$$

Set $\lambda = \mfr1/2 (\lambda_1 + \lambda_2)$ and $\delta = \mfr1/2
(\lambda_1 - \lambda_2)$.  We then have
$$\eqalignii{(X, Y) &\sima ((1 - \lambda_1) X_0, \lambda_1 X_1, (1 -
\lambda_2) Y_0, \lambda_2 Y_1) \qquad&\hbox{by A3} \cr
&\sima ((1 - \lambda) X_0, - \delta X_0, \lambda X_1, \delta X_1, (1 -
\lambda) Y_0, \delta Y_0, \lambda Y_1, - \delta Y_1) \qquad&\hbox{by A5} 
\cr
&\sima ((1 - \lambda) X_0, - \delta X_0, \lambda X_1, \delta X_0, (1 -
\lambda) Y_0, \delta Y_1, \lambda Y_1, - \delta Y_1) \qquad&\hbox{by 
(4.9), A3, A4 }
\cr
&\sima ((1 - \lambda) (X_0, Y_0), \lambda (X_1, Y_1)) \qquad&\hbox{by 
A5}.
\cr}$$
Thus, every point in $\Gamma_{1}\times\Gamma_{2}=:\Gamma_{12}$ is 
equivalent 
to a point of the 
form
$((1 - \lambda) Z_0, \lambda Z_1)$ in $(1 - \lambda) \Gamma_{12} \times
\lambda \Gamma_{12}$ with $Z_0 = (X_0, Y_0)$ and $Z_1 = (X_1, Y_1)$ fixed
and $\lambda \in \R$.  But any two points of this form (with the same 
$Z_0,
Z_1$, but variable $\lambda$) are comparable by Lemma 2.2. 

A similar argument extends CH to multiple scaled copies of $\Gamma_{12}$.
Finally, by induction, CH extends to scaled products of
$\Gamma_{12}$ and $\Gamma_1$ and $\Gamma_2$, i.e., to scaled products 
of arbitrarily many
copies of $\Gamma_1$ and $\Gamma_2$. \hfill\lanbox

%%%%%%%%%%%%%%%%%%%%%
We shall refer to a quadruple of points satisfying (4.8) and (4.9) 
as an {\bf entropy calibrator}.  To establish the existence of such 
calibrators 
we need the following result.
\medskip

{\bf THEOREM 4.6 (Transversality and location of isotherms).}
{\it Let $\Gamma$ be the state space of a simple system that satisfies
the thermal axioms T1-T4.  Then either
\item{(i)}  All points
in $\Gamma$ are in thermal equilibrium, i.e., $X \simt Y$ for all $X,Y
\in \Gamma$.  \smallskip\noindent or
\item{(ii)}  There is at least one
adiabat in $\Gamma$ (i.e., at least one $\partial A_X$) that has at
least two points that are not in thermal equilibrium, i.e., $Z \simt Y$ 
is 
false for some pair of points $Z$ and $Y$ in $\partial A_X$.}

{\it Proof:}  Our proof will be somewhat indirect because it will use
the fact---which we already proved---that there is a concave entropy
function, $S$, on $\Gamma$ which satisfies the maximum principle,
Theorem 4.3 (for $\Gamma_1=\Gamma_2=\Gamma$).  This means that if $\sr
\subset \R$ denotes the range of $S$ on $\Gamma$ then the sets

$$
E_\sigma = \{ X \in \Gamma : S(X) =
\sigma \}, \qquad \sigma \in \sr
$$
are precisely the adiabats of
$\Gamma$ and, moreover, $X = (U_1, V_1), \  Y = (U_2, V_2)$ in $\Gamma$
satisfy $X \simt Y$ if and only if $W = U_2$, maximizes $S (U_1 + U_2
- W, V_1) + S(W, V_2)$ over all choices of $W$ such that $(U_1 + U_2 -
W, V_1) \in \Gamma$ and $(W, V_2) \in \Gamma$.  Furthermore, the
concavity of $S$ --- and hence its continuity on the connected open set
$\Gamma$ --- implies that $\sr$ is connected, i.e., $\sr$ is an
interval.

Let us assume now that (ii) is false. By the zeroth law, T3, $\simt$ is 
an
equivalence relation that divides $\Gamma$ into disjoint equivalence
classes.  Since (ii) is false, each such equivalence class must be a
union
of adiabats, which means that the equivalence classes are represented
by a
family of disjoint subsets of $\sr$.  Thus
$$
\sr = \bigcup \limits_{\alpha \in \I} \sr_\alpha
$$
where $\I$ is some index set, $\sr_\alpha$ is a subset of $\sr$,
$\sr_\alpha \cap \sr_\beta = 0$ for $\alpha \not= \beta$, and $E_\sigma
\simt E_\tau$ if and only if $\sigma$ and $\tau$ are in some common
$\sr_\alpha$.

We will now prove that each $\sr_\alpha$ is an open set.  It is then an
elementary topological fact (using the connectedness of $\Gamma$) 
that there can be
only one non-empty $\sr_\alpha$, i.e., (i) holds, and our proof is 
complete.

The concavity of $S(U,V)$ with respect to $U$ for each fixed $V$
implies
the existence of an upper and lower $U$-derivative at each point, which 
we 
denote by 
$1/T_+ $ and $1/T_- $, i.e., 
$$
(1/T_\pm) (U,V) = \pm \lim \limits_{\varepsilon \searrow 0}
\varepsilon^{-1} [S(U \pm \varepsilon, V) - S(U,V)].
$$
Theorem 4.3 implies that $X \simt Y$ if and only if the closed intervals
$[T_- (X), T_+ (X)]$ and $[T_- (Y), T_+ (Y)]$ are not disjoint.
Suppose
that some $\sr_\alpha$ is not open, i.e., there is $\sigma \in
\sr_\alpha$
and either a sequence $\sigma_1 > \sigma_2 > \sigma_3 \cdots$,
converging
to $\sigma$ or a sequence $\sigma_1 < \sigma_2 < \sigma_3 < \cdots$
converging to $\sigma$ with $\sigma_i \not\in \sr_\alpha$.  Suppose the
former (the other case is similar).  Then (since $T_\pm$ are
monotone
increasing in $U$ by the concavity of $S$) we can conclude that for
{\it
every} $Y \in E_{\sigma_i}$ and {\it every} $X \in E_\sigma$
$$
T_- (Y) > T_+ (X). \eqno(4.10)
$$
We also note, by the monotonicity of $T_\pm$ in $U$, that (4.10) 
necessarily holds if
$Y \in E_\mu$ and $\mu \geq \sigma_i$; hence (1) holds for all $Y \in
E_\mu$ for {\it any} $\mu > \sigma$ (because $\sigma_i \searrow
\sigma$).  On the other hand, if $\tau \leq \sigma$
$$
T_+ (Z) \leq T_-(X)
$$
for $Z \in E_\tau$ and $X \in E_\sigma$.  This contradicts
transversality, namely the hypothesis that there is $\tau < \sigma <
\mu$, $Z \in E_\tau, Y \in E_\mu$ such that $[T_- (Z), T_+ (Z)] \cap
[T_- (Y), T_+ (Y)]$ is not empty. \hfill\lanbox
\medskip

{\bf THEOREM 4.7 (Existence of calibrators).} 
{\it Let $\Gamma_1$
and
$\Gamma_2$ be state spaces of simple systems and assume the thermal
axioms, T1-T4, in particular the transversality property T4.  Then there 
exist
states $X_0, X_1 \in \Gamma_1$ and $Y_0, Y_1 \in \Gamma_2$ such that
$$
X_0 \prec\prec X_1 \qquad \hbox{and} \qquad Y_0 \prec\prec Y_1\ ,
\eqno(4.11)
$$
$$
(X_0, Y_1) \sima (X_1, Y_0). \eqno(4.12)
$$}

{\it Proof:}  Consider the simple system $\Delta_{12}$ obtained by 
thermally
coupling $\Gamma_1$ and $\Gamma_2$.  Fix some $\bar X = (U_{\bar X},
V_{\bar X})\in \Gamma_1$ and
$\bar Y = (U_{\bar Y},
V_{\bar Y})\in \Gamma_2$ with $\bar X \simt \bar Y$.  We form the 
combined
state $\phi (\bar X, \bar Y)= (U_{\bar X} +U_{\bar Y},
V_{\bar X}, V_{\bar Y}) \in \Delta_{12}$ and consider the adiabat 
$\partial
A_{\phi (\bar X, \bar Y)} \subset \Delta_{12}$.  By axiom T2
every point $Z \in \partial
A_{\phi (\bar X, \bar Y)}$ can be split in at least one way as 
$$
\psi (Z) = ((U_X, V_X), (U_Y, V_Y)) \in \Gamma_1 \times \Gamma_2, 
\eqno(4.13)
$$
where $(V_X, V_Y)$
are the work coordinates of $Z$ with $U_X + U_Y =
U_Z$ and where $X = (U_X, V_X), Y = (U_Y, V_Y)$ are in thermal 
equilibrium,
i.e., $X \simt Y$.  If the splitting in (4.13) is {\it not} unique, i.e.,
there exist $X^{(1)}, Y^{(1)}$ and $X^{(2)}, Y^{(2)}$ satisfying these
conditions, then we are done for the following reason:  First,
$(X^{(1)}, Y^{(1)}) \sima (X^{(2)}, Y^{(2)})$ (by axiom T2).  Second,
since $U_{X^{(1)}} + U_{Y^{(1)}} = U_{X^{(2)}} + U_{Y^{(2)}}$ we have
either $U_{X^{(1)}} < U_{X^{(2)}}, U_{Y^{(1)}} > U_{Y^{(2)}}$ or
$U_{X^{(1)}} > U_{X^{(1)}}, U_{Y^{(1)}} < U_{Y^{(2)}}$.  This implies,
by Theorem 3.4, that either $X^{(1)} \prec\prec X^{(2)}$ and $Y^{(2)}
\prec\prec Y^{(1)}$ or $X^{(2)} \prec\prec X^{(1)}$ and $Y^{(1)}
\prec\prec Y^{(2)}$.

Let us assume, therefore, that the thermal splitting (4.13) of each $Z 
\in
\partial A_{\phi (\bar X, \bar Y)}$ is unique so we can write $\psi (Z)
= (X,Y)$ with uniquely determined $X \simt Y$.  (This  means,
in particular, that alternative (i) in Theorem 4.6 is
excluded.) If some pair $(X,Y)$
obtained in this way does not satisfy $X \sima \bar X$ and $Y \sima
\bar Y$, e.g., $X \prec\prec \bar X$ holds, then it follows from axiom
A3 and the cancellation law that $\bar Y \prec\prec Y$, and thus we
have obtained points with the desired properties.

So let us suppose that $X \sima \bar X$ and $Y \sima \bar Y$ whenever
$(X,Y) = \psi (Z)$ and $Z \in \partial A_{\phi (\bar X, \bar Y)}$.  In
other words, $\psi (\partial A_{\phi (\bar X, \bar Y)}) \subset
\partial A_{\bar X} \times \partial A_{\bar Y}$.  We then claim that
all $Z \in \partial A_{\phi (\bar X, \bar Y)}$ are in thermal
equilibrium with each other.  By the zeroth law, T3, (and since $\uprho
(\partial A_{\phi (\bar X \bar Y)})$ is open and connected, by the
definition of a simple systems) it suffices to show that all points
$(U, V_1, V_2)$ in $\partial A_{\phi (\bar X, \bar Y)}$ with $V_1$
fixed are in thermal equilibrium with each other and, likewise, all
points $(U, V_1, V_2)$ in $\partial A_{\phi (\bar X, \bar Y)}$ with
$V_2$ fixed are in thermal equilibrium with each other.  Now each fixed
$V_1$ in $\uprho (A_{\bar X})$ determines a unique point $(U_1, V_1)
\in \partial A_{\bar X}$ (by Theorem 3.5 (iii)).  Since, by assumption, 
$\psi (U, V_1, V_2)
\subset \partial A_{\bar X} \times \partial A_{\bar Y}$ we must then
have
$$
\psi (U, V_1, V_2) = ((U_1, V_1)), (U_2, V_2)) \eqno(4.14)
$$
with $U_2 = U - U_1$.  But (4.14), together with the zeroth law, implies
that all points $(U, V_1, V_2) \in \partial A_{\phi (\bar X, \bar Y)}$
with $V_1$ fixed are in thermal equilibrium with $(U_1, V_1)$ (because
(4.14) shows that they all have the same $\Gamma_1$ component) and hence
they are in thermal equilibrium with each other.  The same argument
shows that all points with fixed $V_2$ are in thermal equilibrium.

We have demonstrated that the hypothesis $X \sima \bar X$ and $Y \sima
\bar Y$ for all $(X,Y) \in \psi (\partial A_{\phi (\bar X, \bar Y)})$
implies that all points in $\partial A_{\phi (\bar X, \bar Y}$ are in
thermal equilibrium.  Since, by Theorem 4.6, at least one adiabat in
$\Delta_{12}$ contains at least two points not in thermal equilibrium, 
the
existence of points satisfying (1) and (2) is established.
\hfill\lanbox
\smallskip

Having established the entropy calibrators we may now appeal to Theorem 
4.5
and summarize the discussion so far in the following theorem. \medskip

{\bf Theorem 4.8 (Entropy principle in products of simple
systems)} {\it Assume Axioms A1-A7, S1-S3 and T1-T4. Then the comparison
hypothesis CH is valid in arbitrary scaled products of simple systems. 
Hence, 
by
Theorem 2.5, the relation $\prec$ among states in such state spaces is
characterized by an entropy function $S$. The entropy function is unique, 
up 
to
an overall multiplicative constant and one additive constant for each 
simple
system under consideration.}
%\vfill \eject
%%%%%%%%%%%%%%%%%%%%

\bigskip\noindent
{\subt C. The role of transversality}
\bigskip

It is conceptually important to give an example of a state space $\Gamma$ 
of 
a
simple system and a relation $\prec$ on its multiple scaled copies, so 
that
all our axioms {\it except T4} are satisfied. In this example the 
comparison
hypothesis CH is violated for the spaces $\Gamma\times \Gamma$ and hence 
the
relation can {\it not} be characterized by an entropy function.  This 
shows
that the transversality axiom T4 is essential for the proof of Theorem 
4.8. 
The example we give is not entirely academic; it is based on the physics 
of
thermometers. See the discussion in the beginning of Section III.

For simplicity, we choose our system to be a degenerate simple system,
i.e., its state space is one-dimensional. (It can be interpreted as a
system with a work coordinate $V$ in a trivial way, by simply declaring
that everything is independent of $V$ and the pressure function is
identically zero). A hypothetical universe consisting only of scaled
copies of such a system (in addition to mechanical devices) might be
referred to as a `world of thermometers'.  The relation $\prec$ is
generated, physically speaking, by two operations:  ``rubbing'',
which increases the energy, and thermal equilibration of two scaled
copies of the system.

To describe this in a more formal way we take 
as our state space $\Gamma={\bf R}_+ =\{U: U>0\}$. 
Rubbing the system increases $U$ and we accordingly define $\prec$ on 
$\Gamma$ simply by the relation $\leq$ on the real numbers $U$.  On 
$\Gamma^{(\lambda_1)}\times \Gamma^{(\lambda_2)}$ we define the forward 
sector of $(\lambda_1U_1,\lambda_2 U_2)$ as the convex hull of the union 
$A\cup B$ of two sets of 
points, 
$$
A=\{(\lambda_1U_1',\lambda_2 U_2'):U_1\leq U_1',\,\,
U_2\leq U_2'\}
$$ 
and 
$$
B=\{(\lambda_1U_1^{\prime\prime},\lambda_2 
U_2^{\prime\prime}):\bar U\leq U_1^{\prime\prime},
\bar U\leq U_2^{\prime\prime}\}
$$ 
with 
$$\bar 
U=(\lambda_1+\lambda_2)^{-1}(\lambda_1U_1+\lambda_2 U_2).
$$ 
This choice of forward sector is minimally consistent with our axioms. 
The 
set 
$A$ 
corresponds to rubbing the individual thermometers while $B$ corresponds
to thermal equilibration followed by rubbing. 

The forward 
sector of a point $(\lambda_1U_1,\dots,\lambda_nU_n)$ 
in the product of more than two scaled copies of 
$\Gamma$ is then defined as the convex hull of all points of the form
$$(\lambda_1U_1,\dots,\lambda_iU_i',\dots \lambda_jU_j',\dots
\lambda_nU_n)\quad {\rm with}
\quad(\lambda_iU_i,\lambda_jU_j)\prec(\lambda_iU_i',\lambda_jU_j').$$
The thermal join of $\Gamma^{(\lambda_1)}$ and $\Gamma^{(\lambda_2)}$
is identified with $\Gamma^{(\lambda_1+\lambda_2)}$. Thermal 
equilibration is simply addition of the energies, and $\lambda_1U_1$
is in thermal equilibrium with $\lambda_2U_2$ if and only if $U_1=U_2$. 

Since the adiabats and isotherms in $\Gamma$ coincide (both consist
only of single points) axiom T4 is violated in this example.  The
forward sectors in $\Gamma\times\Gamma$ are shown in Figure 7. It is
evident that these sectors are not nested and hence {\it cannot be
characterized by an entropy function}.  This example thus illustrates
how violation of the  transversality axiom T4 can prevent the
existence of an entropy function for a relation $\prec$ that is well
behaved in other ways.

\centerline{\sevenpoint ---- Insert Figure 7 here ----}

On the other hand we may recall the usual entropy function for a body 
with
constant heat capacity, namely 
$$
S(U)=\ln U. \eqno (4.15)
$$ 
In the above example this function defines, by simple addition of 
entropies 
in
the obvious way, another relation, $\prec^*$, on the multiple scaled 
copies
of  $\Gamma$ which extends the relation $\prec$ previously defined.  On
$\Gamma$ the two relations coincide (since $S$ is a monotonous function 
of
$U$), but on $\Gamma\times \Gamma$ this is no longer the case:  The 
inequality
$S(U_1)+S(U_2)\leq S(U_1')+S(U_2')$, i.e., $U_1U_2\leq U_1'U_2'$, is only 
a
necessary but not a sufficient condition for $(U_1,U_2)\prec (U_1',U_2')$ 
to
hold.  The passage from $(U_1,U_2)$ to $(U_1',U_2')$ in the sense of the 
relation 
$\prec^*$ (but not  $\prec$) may,
however, be accomplished by coupling each copy of $\Gamma$ to another 
system,
e.g., to a Carnot machine that uses the two copies of $\Gamma$ as heat
reservoirs. {}From the relation $\prec^*$ one could then reconstruct $S$ 
in
(4.15) by the method of Section II. The lesson drawn is that even if T4 
fails
to hold for a system, it may be possible to construct an entropy function 
for
that system, provided its thermal join with some other system behaves
normally.

A precise version of this idea is given in the following theorem.
\medskip

{\bf THEOREM  4.9 (Entropy without transversality).} {\it Suppose $
\Gamma_1$ and $ \Gamma_2$ are normal or degenerate simple systems and
assume that axioms A1--A5, T1--T3 and T5 hold for the relation $ \prec$
on scaled products of $\Gamma_1$ and $\Gamma_2$. (They already hold
for $\Gamma_1$ and $ \Gamma_2$ separately---by definition.)
Let $\Delta_{12}$ be the thermal join of $ \Gamma_1$ and $
\Gamma_2$ and suppose that $\Delta_{12}$ and $ \Gamma_2$  have
consistent entropy functions $S_{12}$ and $S_2$, which holds, in
particular, if T4 is valid for $\Delta_{12}$ and $
\Gamma_2$. Then $ \Gamma_1$ has an entropy function $S_1$ that is
consistent with $S_2$ and satisfies
$$
S_{12}(\phi(X,Y)) = S_1(X) + S_2(Y)
$$
if $X\simt Y$,  where $\phi$ is the canonical map $ \Gamma_1 \times
\Gamma_2 \rightarrow \Delta_{12}$, given by $\phi(X,Y)= (U_X+U_Y,
V_X,V_Y)$ if $X= (U_X, V_X)$ and $ Y= (U_Y,V_Y)$.}

{\it Proof:} Given $X \in  \Gamma_1$ we can, by axiom T5, find a $Y \in 
 \Gamma_2$ with $X\simt Y$, and hence $Z:=\phi(X,Y) \sima (X,Y)$ by axiom
T2. If $Y^{\prime} \in  \Gamma_2$ is another point with $X\simt 
Y^{\prime}$
and $Z^{\prime}:=\phi(X,Y^{\prime})$ then, by axiom T2, 
$(Y^{\prime}, Z)\sima (Y^{\prime}, X,Y) \sima (Y,(X,Y^{\prime})) \sima 
(Y,Z^{\prime})$. Since $S_2$ and $S_{12}$ are consistent entropies, this
means that
$$
S_2(Y^{\prime}) + S_{12}(Z) = S_2(Y) + S_{12}(Z^{\prime}),
$$
or 
$$
S_{12}(Z) - S_2(Y) = S_{12}(Z^{\prime}) - S_2(Y^{\prime}).\eqno(4.16)
$$
We can thus {\it define} $S_1$ on $ \Gamma_1$ by 
$$
S_1(X) := S_{12}(\phi(X,Y)) - S_2(Y)\eqno(4.17)
$$
for each $X\in  \Gamma$ and for any $Y$ satisfying  $Y\simt X$,
because, according to (4.16),  the right side of (4.17) is independent of
$Y$, as long as $Y\simt X$.

To check that $S_1$ is an entropy on $ \Gamma_1$ we show first that 
the relation
$$
(X_1,X_2)\prec (X_1^{\prime},X_2^{\prime})
$$
with $X_1, X_2, X_1^{\prime}, X_2^{\prime} \in  \Gamma_1$ is 
equivalent to 
$$
S_1(X_1) + S_2(X_2) \leq S_1(X_1^{\prime}) + 
S_2(X_2^{\prime}).\eqno(4.18)
$$
We pick $Y_1, Y_2, Y_1^{\prime}, Y_2^{\prime} \in  \Gamma_2$ with
$Y_1\simt X_1, Y_2\simt X_2$, etc. and insert the definition (4.17) of 
$S_1$
into (4.18). We then see that (4.16) is equivalent to 
$$\eqalign{
&S_{12}(\phi(X_1, Y_1)) + S_2(Y_1^{\prime}) + S_{12}(\phi(X_2, Y_2))
+ S_2(Y_2^{\prime})\hfill\cr
&\leq 
S_{12}(\phi(X_1^{\prime}, Y_1^{\prime})) + S_2(Y_1) +
S_{12}(\phi(X_2^{\prime}, Y_2^{\prime})) + S_2(Y_2).\cr}
$$
Since  $S_{12}$ and $S_2$ are consistent entropies, this is equivalent to 
$$
(\phi(X_1, Y_1), Y_1^{\prime}, \phi(X_2, Y_2), Y_2^{\prime}) \prec
(\phi(X_1^{\prime}, Y_1^{\prime}), Y_1, \phi(X_2^{\prime}, Y_2^{\prime}),
Y_2).
$$
By the splitting axiom T2 this is equivalent to
$$
(X_1, Y_1, Y_1^{\prime}, X_2, Y_2, Y_2^{\prime}) \prec
(X_1^{\prime}, Y_1^{\prime}, Y_1, X_2^{\prime}, Y_2^{\prime}, Y_2).
$$
The cancellation law then tells us that this holds if and only if
$(X_1,X_2)\prec (X_1^{\prime},X_2^{\prime})$. 

To verify more generally that $S_1$ characterizes the relation on all
multiple scaled copies of $\Gamma_1$ one may proceed in exactly the same 
way,
using the scale invariance of thermal equilibrium (Theorem 4.1) and the
hypothesis that $S_{12}$ and $S_2$ are entropy functions, which means
that they characterize the relation on all products of scaled copies 
of $\Delta_{12}$ and $\Gamma_2$.  \hfill \lanbox

\medskip

%%%%%%%%%%%%%%%%%%%%%%%%%
\vfill\eject
%%%%%%%%%%%%%%%%%%%%

\bigskip\noindent
{\tit  V. TEMPERATURE AND ITS PROPERTIES}
\bigskip

Up to now we have succeeded in proving the existence of entropy functions
that do everything they should do, namely specify exactly the adiabatic
processes that can occur among systems, both simple and compound. The
thermal join was needed in order to relate different systems, or copies
of the same system  to each other, but temperature, as a numerical
quantifier of thermal equilibrium, was never used. Not even the concept
of `hot and cold' was used. In the present section we shall define
temperature and show that it has all the properties it is normally
expected to have. Temperature, then, is a corollary of entropy; it is
epilogue rather than prologue.

One of our main results here is equation (5.3): Thermal equilibrium and
equality of temperature are the same thing.  Another one is Theorem 5.3
which gives the differentiability of the entropy and which leads to
Maxwell's equations and other manipulations of  derivatives that are to 
be
found in the usual textbook treatment of thermodynamics.

Temperature will be defined {\it only}  for simple systems (because $1/
{\rm (temperature)}$ is the variable dual to energy and it is only the
simple systems that have only one  energy variable). 

\bigskip\noindent
{\subt A. Differentiability of entropy and the existence
of temperature}

\bigskip

The entropy function, $S$, defined on the (open, convex) state space,
$\Gamma$, of a simple system is concave (Theorem 2.8).  
Therefore (as already mentioned in the proof of 
Theorem 4.5) the upper and lower partial
derivatives of $S$ with respect to $U$ (and  also with respect to
$V$) exist at every point $X \in
\Gamma$, i.e., the limits
$$
\eqalignno{1/T_+ (X) &= \lim \limits_{\varepsilon \downarrow 0} {1 \over
\varepsilon} [S(U + \varepsilon, V) - S(U,V)] \cr
1/T_- (X) &= \lim \limits_{\varepsilon \downarrow 0} 
{1 \over \varepsilon}
[S (U,V) - S(U-\varepsilon, V)] \cr}
$$
exist for every $X = (U,V) \in \Gamma$.  The functions $T_+ (X)$
(resp.  $T_- (X))$ are finite {\it and positive} everywhere (since $S$
is strictly monotone increasing in $U$ for each fixed $V$
(by Planck's
principle, Theorem 3.4).  These functions are called, respectively,
the {\bf upper} and {\bf lower temperatures}.  Evidently, concavity
implies that if $U_1 < U_2$
$$
T_- (U_1,V) \leq T_+
(U_1, V) \leq T_- (U_2, V) \leq T_+ (U_2, V) \eqno(5.1)
$$
for all $V$.  The
concavity of $S$ {\it alone} does not imply continuity of these
functions.  Our goal here is to prove continuity by invoking some of our
earlier axioms.

First, we prove a limited kind of continuity.
\medskip

{\bf LEMMA 5.1 (Continuity of upper and lower temperatures on
adiabats).} {\it The temperatures $T_+$ and $T_-$ are locally Lipschitz
continuous along each adiabat $\partial A_X$.  I.e., for each $X \in
\Gamma$ and each closed ball $B_{X,r} \subset \Gamma$ of radius $r$ and
centered at $X$ there is a constant $c (X,r)$ such that
$$
\vert T_+ (X) - T_+ (Y) \vert \leq c(X,r) \vert X-Y
\vert$$
for all $Y \in \partial A_X$ with $\vert X-Y \vert < r$.  The
same inequality holds for $T_- (X)$. Furthermore, $c(X,r)$ is
a continuous function of $X$ in any domain $D\subset \Gamma$ such that
$B_{X,2r}\subset \Gamma$ for all $X\in D$.}
\medskip

{\it Proof:}  Recall that the pressure $P(X)$ is assumed to be locally
Lipschitz continuous and that $\partial U/\partial V_i = P_i$ on
adiabats.  Write $X = (U_0, V_0)$ and let the adiabatic surface through
$X$ be denoted by $(W_0 (V), V)$ where $W_0 (V)$ is the unique solution
to the system of equations
$${\partial W_0 (V) \over \partial V_i} = P_i (W_0 (V),
V)$$
with $W_0 (V_0) = U_0$.  (Thus $W_{0}$ is the function $u_{X}$ of 
Theorem 3.5.) Similarly, for $\varepsilon > 0$
we let $W_\varepsilon (V)$ be the solution to
$$
{\partial W_\varepsilon (V) \over \partial V_i} = P_i (W_\varepsilon (V),
V)$$
with $W_\varepsilon (V_0) = U_0 + \varepsilon$.  Of course all this makes
sense only if $\vert V - V_0 \vert$ and $\varepsilon$ are sufficiently
small so that the points $(W_\varepsilon (V), V)$ lie in $\Gamma$.  In 
this
region (which we can take to be bounded) we let $C$ denote the Lipschitz
constant for $P$, i.e. $\vert P(Z) - P(Z^\prime) \vert \leq C \vert Z -
Z^\prime \vert$ for all $Z,Z^\prime$ in the region.

Let $S_\varepsilon$ denote the entropy on $(W_\varepsilon (V),V)$; it is
constant on this surface by assumption.  By definition
$$
{1 \over T_+ (U_0, V_0)} = \lim \limits_{\varepsilon \downarrow 0}
{S_\varepsilon - S_0 \over \varepsilon},$$
and
$$
T_+ (W_0 (V), V) = \lim \limits_{\varepsilon \downarrow 0} {W_\varepsilon
(V) - W_0 (V) \over S_\varepsilon - S_0} = T_+ (U_0, V_0) \bigl[\lim
\limits_{\varepsilon \downarrow 0} G_\varepsilon (V) + 1\bigr],$$
where $G_\varepsilon (V) := {1 \over \varepsilon} [W_\varepsilon (V) - 
W_0
(V) - \varepsilon]$.  The lemma will be proved if we can show that there 
is
a number $D$ and a radius $R > 0$ such that $G_\varepsilon (V) \leq D 
\vert 
V
- V_0 \vert$ for all $\vert V - V_0 \vert < R$.

Let $v$ be a unit vector in the direction of $V - V_0$ and set $V(t) =
V_0 + t v$, so that $V(0) = V_0, V(t) = V$ for $t = \vert V - V_0
\vert$.  Set $W_\varepsilon (t) := W_\varepsilon (V(t))$ and $\Pi (U,t)
:= v \cdot P(U,V(t))$.  Fix $T > 0$ so that $CT \leq \mfr1/2$ and so
that the ball $B_{X,2T}$ with center $X$ and radius $2T$ satisfies 
$B_{X,2T} \subset \Gamma$.  Then,
for $0 \leq t \leq T$ and $\varepsilon$ small enough
$$
\eqalignno{W_0 (t) &= U_0 + \int^t_0 \Pi (W_0 (t^\prime), t^\prime)
\d t^\prime \cr W_\varepsilon (t) - \varepsilon &= U_0 + \int^t_0 \Pi
(W_\varepsilon (t^\prime) - \varepsilon + \varepsilon, t^\prime) \d
t^\prime. \cr}$$
Define
$$
g_\varepsilon = \sup \limits_{0 \leq t \leq T} {1 \over \varepsilon}
[W_\varepsilon (t) - \varepsilon - W_0 (t)] = \sup \limits_{0 \leq t
\leq T} G_\varepsilon (V(t)).$$ By subtracting the equation for $W_0$
from that of $W_\varepsilon$ we have that
$$
\vert G_\varepsilon (V(t)) \vert \leq \int \limits^t_0 C[1 +
g_\varepsilon] \d t^\prime \leq t C [1 + g_\varepsilon].$$ By taking
the supremum of the left side over $0 \leq t \leq T$ we obtain
$g_\varepsilon \leq TC [1 + g_\varepsilon]$, from which we see that
$g_\varepsilon \leq 1$ (because $TC\leq 1/2$).  But then $\vert G_\varepsilon 
(V(t) \vert \leq
2tC$ or, in other words, $\vert G_\varepsilon (V) \vert \leq 2 \vert V
- V_0 \vert C$ whenever $\vert V - V_0 \vert < T$, which was to be
proved.  \hfill\lanbox
\medskip

Before addressing our next goal---the equality of $T_+$ and $T_-$---let
us note the maximum entropy principle, Theorem 4.2,  and its
relation to $T_\pm$.  The principle states that if $X_1 = (U_1, V_1)$
and $X_2 = (U_2, V_2)$ are in $\Gamma$ then $X_1 \simt X_2$ if and only
if the following is true:
$$
S(X_1) + S(X_2) = \sup_W \{ S(U_1 + U_2 - W, V_1) + S (W, V_2) : 
(U_1 + U_2
- W, V_1) \in \Gamma \ \hbox{and} \ (W, V_2) \in \Gamma \}.
\eqno(5.2)
$$
Since $S$ is concave, at every point $X \in \Gamma$ there
is an upper temperature and lower temperature, as given in (5.1). This
gives us an  ``{\it interval-valued}'' function on $\Gamma$  which
assigns to each $X$ the interval
$$
T(X) = [T_- (X), T_+ (X)].
$$
If $S$ is differentiable at $X$ then $T_- (X) = T_+ (X)$
and the closed interval $T(X)$ is then merely the single
number $\left( {\partial S \over \partial U} \right) (X)$.  If $T_- (X)
= T_+ (X)$ we shall abuse the notation slightly by thinking of $T(X)$
as a number, i.e., $T(X) = T_- (X) = T_+ (X)$.

The significance of the interval $T(X)$ is that (5.2) is equivalent to:
$$
X_1 \simt X_2 \quad {\rm if \ and \ only \ if} \quad T (X_1) \cap T (X_2)
\not= \emptyset.
$$
In other words, if $\partial S/\partial U$ makes a jump at $X$ then one
should think of $X$ as having {\it all\/} the temperatures in the closed
interval $T(X)$.

In Theorem 5.1 we shall prove that the temperature is single-valued,
i.e., $T_- (X) = T_+ (X)$. Thus, we have the following
fact relating {\bf thermal equilibrium and temperature:}
$$
X_1 \simt X_2 \quad {\rm if \ and \ only \ if} \quad T (X_1)
= T (X_2). \eqno (5.3)
$$
\medskip

{\bf THEOREM 5.1 (Uniqueness of temperature).}  
{\it At every point $X$ in the state space of a simple system, $\Gamma$, 
we 
have
$$
T_+ (X) = T_- (X),
$$
i.e., $T(X)$ is the number $\left[\left( {\partial S \over \partial U} 
\right)
(X)\right]^{-1}$}.

\smallskip

{\it Proof:}  The proof will rely heavily on the zeroth law, on the
continuity of $T_\pm$ on adiabats, on transversality, on axiom T5 and
on the maximum entropy principle for thermal equilibrium,
Theorem 4.2.

Assume that $Z \in \Gamma$ is  a point for which $T_+ (Z) > T_- (Z)$.
We shall obtain a contradiction from this.

{\it Part 1:}  We claim that for every $Y \in \partial A_Z$, $T_+ (Y) =
T_+ (Z)$ and $T_- (Y) = T_- (Z)$.  To this end define the (conceivably
empty) set $K \subset \Gamma$ by $K = \{ X \in \Gamma : T_+ (X) = T_-
(X) \in T(Z)\}$.  If $X_1 \in K$ and $X_2 \in K$ then $T(X_1) = T(X_2)
\in T(Z)$ by the zeroth law (since $X_1 \simt Z$ and $X_2 \simt Z$, and
thus $X_1 \simt X_2$).  Therefore, there is a single number $T^* \in
T(Z)$ such that $T(X) = T^*$ for all $X \in K$.

Now suppose that $Y \in \partial A_Z$ and that $T_+ (Y) < T_+ (Z)$.  By
the continuity of $T_+$ on $\partial A_Z$ (Lemma 5.1) there is then 
another 
point $W
\in \partial A_Z$ such that $T_- (Z) \leq T_+ (W) < T_+ (Z)$, which 
implies that $W \simt Z$.  We write $W = (U_W, V_W)$ and consider $f_W
(U) = S(U, V_W)$, which is a concave function of one variable (namely
$U$) defined on some open interval containing $U_W$.  It is a general
fact about concave functions that the set of points at which $f_W$ is
differentiable (i.e., $T_+ = T_-$) is dense and that if $U_1 > U_2 > U_3
> \dots > U_W$ is a decreasing sequence of such points converging to
$U_W$ then $T (U_i)$ converges to $T_+ (U_W)$.  We denote the
corresponding points $(U_i, V_W)$ by $W_i$ and note that, for large $i$,
$T (W_i) \in T(Z)$.  Therefore $T(W_i) = T^*$ for all large $i$ and hence
$T_+ (W) = T^*$.

Now use continuity again to find a point $R \in \partial A_Z$ such that
$T^* = T_+ (W) < T_+ (R) < T_+ (Z)$.  Again there is a sequence $R_i =
(U^i, V_R)$ with $T_+ (R_i) = T_- (R_i) = T (R_i)$ converging downward to
$R$ and such that $T (R_i) \rightarrow T_+ (R) > T^*$.  But for large
$i$, $T(R_i) \in T(Z)$ so $T(R_i) = T^*$.  This is a contradiction, and
we thus conclude that
$$
T_+ (Y) = T_+ (Z)
$$
for all $Y \in \partial A_Z$ when $T_+ (Z) > T_- (Z)$.

Likewise $T_- (Y) = T_- (Z)$ under the same conditions.

{\it Part 2:}  Now we study $\uprho_Z \subset \R^n$, which is the
projection of $\partial A_Z$ on $\R^n$.  By Theorem 3.3, $\uprho_Z$ is
open and connected.  It is necessary to consider two cases.

{\it Case 1:}  $\uprho_Z$ is the projection of $\Gamma$, i.e.,
$\uprho_Z = \{ V \in \R^n :  (U,V) \in \Gamma$ for some $U \in \R \} =
\uprho(\Gamma)$.  In this case we use the {\it transversality axiom\/}
T4, according to which there are points $X$ and $Y$ in $\Gamma$ with $X
\prec\prec Z \prec\prec Y$, (and hence $S(X) < S(Z) < S(Y)$), but with
$X \simt Y$.  We claim that every $X$ with $S(X) < S(Z)$ has $T_+ (X)
\leq T_- (Z)$.  Likewise, we claim that $S(Y) > S(Z)$ implies that $T_-
(Y) \geq T_+ (Z)$.  These two facts will contradict the assumption that
$T(Y) \cap T(X)$ is not empty.  To prove that $T_+ (X) \leq T_- (Z)$ we
consider the line $(U, V_X) \cap \Gamma$.  As $U$ increases from the
value $U_X$, the temperature $T_+ (U, V_X)$ also cannot decrease (by
the concavity of $S$).  Furthermore, $(U_X, V_X) \prec (U, V_X)$ if and
only if $U \geq U_X$ by Theorem 3.4.  Since $\uprho_Z = \uprho
(\Gamma)$ there is (by Theorem 3.4) some $U_0 > U_X$ such that
$(U_0, V_X) \in \partial A_Z$.  But $T_- (U_0, V_X) = T_- (Z)$ as we
proved above.  However, $T_+ (X) \leq T_- (U_0, V_X)$ by (5.1).  A
similar proof shows that $T_- (Y) \geq T_+ (Z)$ when $S(Y) > S(Z)$.

{\it Case 2:}  $\uprho_Z \not= \uprho (\Gamma)$.  Here we use 
T5.  Both $\uprho_Z$ and $\uprho (\Gamma)$ are open sets and
$\uprho_Z \subset \uprho (\Gamma)$.  Hence, there is a point $V$ in
$\bar{\uprho}_Z$, the closure of $\uprho_Z$, such that $V \in \uprho
(\Gamma)$.  Let $l_V := L_V \cap \Gamma = \{ (U,V): U \in \R$ and
$(U,V) \in \Gamma \}$.  If $X \in l_V$ then either $Z\prec\prec X$ or
$X \prec\prec Z$.  (This is so because we are dealing with a simple
system, which implies that $X \succ Z$ or $X \prec Z$, but we cannot
have $X \sima Z$ because then $X \in \partial A_Z$, which is impossible
since $l_V \cap \partial A_Z$ is empty.) Suppose, for example, that $Z
\prec\prec X$ or, equivalently, $S(X) > S(Z)$.  Then $S(Y) > S(Z)$ for
all $Y \in l_V$ (by continuity of $S$, and by the fact that $S(Y) \not=
S(Z)$ on $l_V$).

Now $A_X$ has a tangent plane $\Pi_X$ at $X$, which implies that
$\uprho_X \cap \uprho_Z$ is {\it not\/} empty.  Thus there is a point
$$
W_1 = (U_1, V_1) \in \partial A_X \ {\rm with} \ V_1 \in \uprho_X
\cap \uprho_Z \ {\rm and} \ S(W_1) = S(X) > S(Z).
$$
By definition, there
is a point $(U_0, V_1) \in \partial A_Z$ with $U_0 < U_1$.  By
concavity of $U \mapsto S (U, V_1)$ we have that $T_- (W_1) \geq T_+
(U_0, V_1) = T_+ (Z)$.  By continuity of $T_-$ along the adiabat
$\partial A_X$ we conclude that $T_- (X) \geq T_+ (Z)$.  The same
conclusion holds for every $Y \in l_V$ and thus the range of
temperature on the line $l_V$ is an interval $(t_1, t_2)$ with $t_1 \geq
T_+ (Z)$.

By similar reasoning, if $R$ is in the set $\{ (U,V) : V \in \uprho_Z,
S(U,V) < S(Z) \}$ then $T_+ (R) \leq T_- (Z)$.  Hence the temperature
range on any line $l_{\widehat V}$ with $\widehat V \in \uprho_Z$
satisfies $t_1 \leq T_- (Z)$.  This contradicts T5 since $T_- (Z) < T_+
(Z)$.  A similar proof works if $X \prec\prec Z$.\hfill\lanbox

\bigskip

Having shown that the temperature is uniquely defined at each point of
$\Gamma$ we are now in a position to establish our  goal.

{\bf THEOREM 5.2  (Continuity of temperature).}  {\it The
temperature $T(X) = T_+ (X) = T_- (X)$ is a continuous function on the
state space, $\Gamma \subset \R^{n+1}$, of a simple system.}

{\it Proof:}  Let $X_\infty, X_1, X_2, \dots$ be points in $\Gamma$
such that $X_j \rightarrow X_\infty$ as $j \rightarrow \infty$.  We
write $X_j = (U_j, V_j)$, we let $A_j$ denote the adiabat $\partial
A_{X_j}$, we let $T_j = T(X_j)$ and we set $l_j = \{ (U, V_j): (U, V_j)
\in \Gamma \}$.  We know that $T$ is continuous and monotone along each
$l_j$ because $T_+ = T_-$ everywhere by Theorem 5.1. We also know that
$T$ is continuous on each $A_j$ by  Lemma 5.1.  In fact, if we assume
that all the $X_j$'s are in some sufficiently small ball, $B$ centered
at $X_\infty$, then by Lemma 5.1 we can also assume that for some $c <
\infty$
$$\vert T(X) -
T(Y) \vert \leq c \vert X-Y \vert$$
whenever $X$ and $Y$ are in $B$ and $X$ and $Y$ are on the same
adiabat, $A_j$. Lemma 5.1  also states that $c$ can be taken to be
independent of $X$ and $Y$ in the ball $B$.

By assumption, the slope of the tangent plane $\Pi_X$ is locally 
Lipschitz
continuous, i.e., the pressure $P(X)$ is locally Lipschitz continuous.
Therefore (again, assuming that $B$ is taken small enough) we can
assume that each adiabat $A_j$ intersects $l_\infty$ in some point,
which we denote by $Y_j$.  Since $\vert X_j - X_\infty \vert
\rightarrow 0$ as $j \rightarrow \infty$, we have that $Y_j
\rightarrow X_\infty$ as well.  Thus,
$$
\vert T(X_j) -
T(X_\infty) \vert \leq \vert T(X_j) - T(Y_j) \vert + \vert T(Y_j) -
T(X_\infty) \vert.$$
As $j \rightarrow \infty$, $T(Y_j) - T(X_\infty)
\rightarrow 0$ because $Y_j$ and $X_\infty$ are in $l_\infty$.  Also,
$T(X_j) - T(Y_j) \rightarrow 0$ because $\vert T(X_j) - T(Y_j) \vert <
c \vert X_j - Y_j \vert \leq c \vert X_j - X_\infty
\vert + c \vert Y_j - X_\infty \vert$. \hfill\lanbox
\medskip

{\bf THEOREM 5.3 (Differentiability of $S$).}  {\it The entropy,
$S$, is a continuously differentiable function on the state space
$\Gamma$ of a simple system.}
\smallskip

{\it Proof:}  The adiabat through a point $X \in \Gamma$ is characterized
by the once continuously differentiable function, $u_X (V)$, on $\R^n$. 
Thus, $S(u_X (V), V)$ is constant, so (in the sense of distributions)
$$
0 = \left( {\partial S \over \partial U} \right) \left( {\partial u_X
\over \partial V_j} \right) + {\partial S \over \partial V_j}.
$$
Since $1/T = \partial S /\partial U$ is continuous, and $\partial u_X
/\partial V_j = -P_j$ is Lipschitz continuous, we see that $\partial
S/\partial V_j$ is a continuous function
and we have the well known formula
$$
{\partial S \over {\partial V_j }}= {P_j \over T}
\eqno\lanbox
$$
\medskip

We are now in a position to give a simple proof of the most important
property of temperature, namely its role in determining the direction of
energy transfer, and hence, ultimately, the linear ordering of systems
with respect to heat  transfer (even though we have not defined `heat' 
and have no intention of doing so).  The fact that energy only flows
`downhill'  without the intervention of extra machinery was taken by
Clausius as the foundation of the second law of thermodynamics, as we
said in Section I.
\medskip

{\bf THEOREM 5.4 (Energy flows from hot to cold).} {\it Let $(U_1,
V_1)$ be a point in a state space $\Gamma_1$ of a simple system and let 
$(U_2, V_2) $ be a
point in a state space $\Gamma_2$ of another simple system.  Let $T_1$ 
and 
$T_2 $ be their
respective temperatures and assume that $T_1 > T_2$.  If $(U'_1, V_1) $
and $(U'_2, V_2) $ are two points 
with the same respective work coordinates as the original points, 
with the same total energy $U_1 + U_2 = U'_1 + U'_2$, 
and for which the temperatures are equal to a common value,
$T$ (the existence of such points is guaranteed by
axioms T1 and T2), then
$$
 U'_1 < U_1   \  {\sl and } \  U'_2 > U_2.
$$
}
\smallskip

{\it Proof: \/} By assumption $T_1 > T_2$ and  we claim that 
$$
T_1 \geq T \geq T_2 .  \eqno (5.4)
$$ 
(At least one of 
these inequalities is strict because of the uniqueness of temperature for
each state.)  Suppose that inequality (5.4) failed, e.g., 
$T > T_1 >T_2$. Then we would have that 
$U'_1 > U_1$ and $ U'_2 > U_2$ and at least one of these would be strict
(by the strict monotonicity of $U$ with respect to  $T$, 
which follows from the concavity and differentiability of $S$). 
This pair of inequalities is impossible in view of the condition
$U_1 + U_2 = U'_1 + U'_2$.

Since $T$ satisfies (5.4), the theorem now follows 
from the monotonicity of $U$ with respect to $T$. 
\hfill\hfill\lanbox
\medskip

%%%%%%%%%
{}From the entropy principle and the relation 
$$1/T=(\partial S/\partial U)^{-1}$$
between temperature and entropy 
we can now derive the usual formula for the {\bf Carnot efficiency} 
$$\eta_{\rm C}:=1-(T_0/T_1)\eqno(5.5)$$
as 
an upper bound for the efficiency of a `heat engine' that undergoes a 
cyclic process. Let us define a {\bf thermal
reservoir} to be a simple system whose work coordinates remains unchanged
during some process (or which has no work coordinates, i.e. is a 
degenerate simple system). Consider a combined system consisting of a 
thermal reservoir and
some machine, and an adiabatic process for this combined system.  
The entropy principle says that the total entropy change in this process 
is 
$$\Delta S_{\rm machine}+\Delta
S_{\rm reservoir}\geq 0.\eqno(5.6)$$  Let $-Q$ be the energy change of 
the 
reservoir,
i.e., if $Q\geq 0$, then the reservoir delivers energy, otherwise it
absorbs energy. If $T$ denotes the temperature of the reservoir {\it at 
the
end of the process}, then, by the convexity of $S_{\rm reservoir}$ in 
$U$, 
we have
$$\Delta S_{\rm reservoir}\leq -{Q\over T}.\eqno(5.7)$$ Hence 
$$\Delta S_{\rm machine}-{Q\over
T}\geq 0.\eqno(5.8)$$ 
Let us now couple the machine first to a `high temperature
reservoir' which delivers energy $Q_{1}$ and reaches a final temperature 
$T_1$,
and later to a "low 
temperature
reservoir" which absorbs energy $-Q_{0}$ and reaches a final temperature 
$T_0$. The whole process is 
assumed 
to be cyclic for the machine so the entropy changes for the machine in 
both steps cancel. (It
returns to its initial state.) Combining (5.6), (5.7) and (5.8) we 
obtain 
$$Q_1/T_1+Q_0/T_0\leq 0\eqno(5.9)
$$ 
which gives the usual inequality for the efficiency $\eta := 
(Q_{1}+Q_{0})/Q_{1}$:
$$\eta\leq 1-(T_0/T_1)=\eta_{\rm C}.\eqno(5.10)$$ 
In text book presentations it is usually assumed that the reservoirs are
infinitely large, so that their temperature remains unchanged, but 
formula
(5.10) remains valid for finite reservoirs, provided $T_1$ and $T_0$ are
properly interpreted, as above.  
%%%%%%%%
 
\bigskip\noindent
{\subt B. Geometry of isotherms and adiabats}
\bigskip

Each adiabat in a simple system is the boundary of a convex set and hence
has a simple geometric shape, like a `bowl'. It must be an object of
dimension $n$ when the state space in question is a subset of $\R^{n+1}$. 
In contrast, an isotherm, i.e., the set on which the temperature assumes 
a
given value $T$, can be more complicated. When $n=1$ ( with energy and
volume as coordinates) and when the system has a triple point, a portion 
of
an isotherm (namely the isotherm through the triple point) can be
two-dimensional. See Figure 8 where this isotherm is described 
graphically.

\centerline{\sevenpoint ---- Insert Figure 8 here ----}

One can ask whether isotherms can have other peculiar properties. Axiom
T4 and Theorem 4.5 already told us that an isotherm cannot 
coincide completely with an adiabat
(although they could coincide over some region).  If this were to happen
then, in effect, our state space would be cut into two non-communicating
pieces, and we have ruled out this pathology by fiat.  However, another
possible pathology would be that an isotherm consists of several
disconnected pieces, in which case we could not pass from one side of an
adiabat to another except by changing the temperature.  Were this to
happen then the pictures in the textbooks would really be suspect, but
fortunately, this perversity does not occur, as we prove next.  

There is one technical point that must first be noted. By concavity and
differentiability of the entropy, the range of the temperature function
over $\Gamma$ is always an interval. There are no gaps. But the range
need not go from $0$ to $\infty$ ---in principle. (Since we defined the
state spaces of simple systems to be open sets, the point $0$ can never
belong to the range.) Physical systems ideally always cover the entire
range $(0,\infty)$, but there is no harm, and perhaps even a whiff of
physical reality, in supposing that the temperature range of the world is
bounded. Recall that in axiom T5 we said  that the range must be the same
for all systems and, indeed, for each choice of work coordinate within a
simple system.  Thus, for an
arbitrary simple system, $\Gamma$, and $V\in\rho(\Gamma)$ 
$$
T_{\rm min}: = \inf\{T(X)  : X\in \Gamma\}
= \inf\{T(U,V)  : U\in \R\ \hbox{\rm such that } (U,V)\in\Gamma\}$$
and 
$$T_{\rm max} := \sup\{T(X)  : X\in \Gamma \}
= \sup\{T(U,V)  : U\in \R\ \hbox{\rm such that } (U,V)\in\Gamma\}.$$
\medskip

{\bf THEOREM 5.5 (Isotherms cut adiabats)} {\it Suppose $X_0  \prec
X \prec X_1$ and $X_0$ and $X_1$ have equal temperatures, $T(X_0) =
T(X_1)=T_0$. \smallskip

(1). If $T_{\rm min}<T_0<T_{\rm max}$ then there is a point $X' \sima X$
with $T(X')=T_0$. 
In other words: The isotherm through $X_0$ cuts every adiabat
between $X_0$ and $X_1$.  \smallskip

(2). If $T_0=T_{\rm max}$, then either there is an $X' \sima X$
with $T(X')=T_0$, or, for any $T_0'<T_0$ there exist points $X_0'$, $X'$ 
and $X_1'$ with $X_0'\prec X'\sima X\prec X_1'$ and 
$T(X_0')=T(X')=T(X_1')
=T_0'$.  \smallskip

(3). If $T_0=T_{\rm min}$, then either there is an $X' \sima X$
with $T(X')=T_0$, or, for any $T_0'>T_0$ there exist points $X_0'$, $X'$ 
and $X_1'$ with $X_0'\prec X'\sima X\prec X_1'$ 
and $T(X_0')=T(X')=T(X_1')
=T_0'$.}
\medskip

{\it Proof:} {\sl Step 1.}  First we show that for every $T_0$
with $T_{\rm min}<T_0<T_{\rm max}$
the  sets  $ \Omega_> := \{ Y : T(Y) > T_0 \}$
and $\Omega_< :=\{ Y : T(Y) < T_0 \}$ are open and connected. 
The openness follows from the continuity of $T$. Suppose that 
$\Omega_1$ and $\Omega_2$ are non empty, open  sets satisfying 
$ \Omega_> = \Omega_1 \cup \Omega_2$. We shall show that
$\Omega_1 \cap \Omega_2$ is not empty, thereby showing that $
\Omega_>$ is connected. By axiom T5, the range of $T$,
restricted to points $(U,V) \in \Gamma$, with $V$ fixed, is
independent of $V$, and hence $\uprho ( \Omega_>) = \uprho
(\Gamma)$, where $\uprho $ denotes the projection $(U,V)
\mapsto V$. It follows that $\uprho ( \Omega_1) \cup \uprho
( \Omega_2) = \uprho ( \Gamma)$ and, since $\uprho $ is an open
mapping and $\uprho ( \Gamma)$ is connected, we have that 
$\uprho ( \Omega_1) \cap \uprho ( \Omega_2)$ is not empty. Now
if $(U_1,V) \in \Omega_1 \subset \Omega_>$ and if $(U_2,V) \in
\Omega_2 \subset \Omega_>$, then, by the monotonicity of 
$T(U,V)$ in $U$ for fixed $V$, it follows that the line joining
$(U_1,V) \in \Omega_1 $ and  $(U_2,V) \in \Omega_2$ lies entirely
in $ \Omega_> = \Omega_1 \cup \Omega_2$. Since $\Omega_1$ and
$\Omega_2$ are open, $\Omega_1 \cap \Omega_2$ is not empty and 
$ \Omega_>$ is connected. Similarly, $ \Omega_<$ is connected. \smallskip

{\sl Step 2.} We show that if $T_{\rm min}<T_0<T_{\rm max}$, then there 
exist points $X_>$, $X_<$, with $X_>\sima X\sima X_<$ and $T(X_<)\leq 
T_0\leq T(X_>)$. We write the proof for $X_>$, the existence of $X_<$ 
is shown in the same way. In the case that
$V_{X_0}\in\rho(A_X)$ the existence of $X_>$ follows immediately from
the monotonicity of $T(U,V)$ in $U$ 
for fixed $V$. If $V_{X_0}\not\in\rho(A_X)$ we first remark that by axiom 
T5 and because $T_0<T_{\rm max}$ there exists $X_0'\prec X$ with 
$T_0<T(X_0')$. Also, by monotonicity of $T$ in $U$ there exists $X_1'$ 
with
$X\prec X_1\prec X_1'$ and $T(X_1')>T_0$. Hence $X_0'$ and $X_1'$ 
both belong to $\Omega_>$, and  $X_0'\prec X\prec X_1'$. Now $\Omega_>$ 
is nonempty, open and connected, and $\partial A_X$ splits 
$\Gamma\setminus \partial A_X$ into disjoint, open sets. Hence $\Omega_>$ 
must cut $\partial A_X$, i.e., there exists an $X_>\in \Omega_>\cap 
\partial A_X$.

Having established the existence of $X_>$ and $X_<$ we now appeal to 
continuity of $T$ and connectedness of $\partial A_X$ (axiom S4) to 
conclude that there is an $X'\in\partial A_X$ with $T(X')=T_0$. This 
completes the proof of assertion (1). \smallskip

{\sl Step 3.} If $T_0=T_{\rm max}$ and $V_{X_0}\in\rho(A_X)$, then the 
existence of $X'\in\partial A_X$ with $T(X')=T_0$ follows from 
monotonicity of $T$ in $U$. Let us now assume that all points on 
$\partial 
A_X$ have 
temperatures strictly less than $T_{\rm max}$. By axiom A5 and by 
continuity and monotonicity of $T$ in $U$, there is 
for every $T_0'<T_0$ an $X_0'\prec X_0$ with $T(X_0')=T_0'$. For 
the same reasons there is an $X_1'$ with $X\prec X_1'\prec X_1$ and 
$T(X_1')=T_0'$. By the argument of step 2 there is thus an 
$X'\in\partial A_X$ with $T(X')=T_0'$. Thus assertion (2) is established. 
The case $T_0=T_{\rm min}$ (assertion (3)) is treated 
analogously.\hfill\lanbox
\medskip
%\vfill\eject
\bigskip\noindent
{\subt C. Thermal equilibrium and uniqueness of entropy}
\bigskip

In Section II we have encountered two general uniqueness theorems for
entropy.  The first, Theorem 2.4, relies only on axioms A1-A6, and CH
for the double scaled copies of $\Gamma$, and states that an entropy
function on $\Gamma$ is uniquely determined, up to an affine
transformation of scale, by the relation $\prec$ on the double scaled
copies.  In the second, Theorem 2.10, it is further assumed that the
range of the entropy is connected which, in particular, is the case if 
the
convex combination axiom A7 holds. Under this condition the relation 
$\prec$ on
$\Gamma\times\Gamma$ determines the entropy. Both these uniqueness
results are of a very general nature and rely only on the structure
introduced in Section II. The properties of entropy and temperature that
we have now established on the basis of axioms A1--A7, S1--S3 and T1--T5,
allow us to supplement these results now with a uniqueness theorem of a
different kind.  
\medskip

{\bf THEOREM 5.6 (Adiabats and isotherms in $\Gamma$ determine the
entropy).}  {\it Let $\prec$ and $\prec^*$ be two relations on the 
multiple
scaled copies of a simple system $\Gamma$ satisfying axioms A1--A7, 
S1--S3
and T1--T5.  Let $\simt$ and $\simt^*$ denote the corresponding relations 
of
thermal equilibrium between states in $\Gamma$.  If $\prec$ and $\prec^*$
coincide on $\Gamma$ and the same holds for the relations $\simt$ and
$\simt^*$, then $\prec$ and $\prec^*$ coincide everywhere. In other 
words: 
The
adiabats in $\Gamma$ together with the isotherms determine the 
relation $\prec$ on all
multiple scaled copies of $\Gamma$ and hence the entropy is uniquely 
determined
up to an affine transformation of scale.} \smallskip

{\it Proof:}  Let $S$ and $S^*$ be (concave and continuously
differentiable) entropies characterizing respectively the relations
$\prec$ and $\prec^*$. (The existence follows from axioms A1-A7, S1-S3,
and T1-T4, as shown in the previous sections.)  For points $X,Y\in
\Gamma$ we have $S(X) = S(Y)$ if and only if $X \sima Y$, which holds if
and only if $S^* (X)= S^* (Y)$, because $\prec$ and $\prec^*$ coincide on
$\Gamma$ by assumption.  Hence $S$ and $S^*$ have the same level sets,
namely the adiabats of the simple system.  Thus, we can write
$$
S^* (X) = f(S(X))
$$
for some strictly monotone function, $f$, defined on the range of
$S$---which is some interval $I \subset \R$.  We claim that $f$ is
differentiable on $I$ and therefore
$$
{\partial S^* \over \partial U} (X) = f^\prime (S(X)) {\partial
S \over \partial U} (X).\eqno(5.11)
$$
To prove the differentiability note that $\partial S /\partial U$ is
never zero (since $S$ is strictly monotonic in $U$ by Planck's
principle, Theorem 3.4). This implies that for each fixed $V$ in
$\uprho (\Gamma)$ the function $U \mapsto S(U,V)$ has a continuous
inverse $K (S,V)$.  (This, in turn, implies that $I$ is open.)  Thus, if
$X = (U, V)$ and $S(U,V) = \sigma$ and if $\sigma_1, \sigma_2, \dots$ is
{\it any\/} sequence of numbers converging to $\sigma$, the sequence of
numbers $U_j := K (\sigma_j, V)$ converges to $U$.  Hence 
$$
{S^* (U_j, V)
- S^* (U,V) \over U_j - U} = \left[ {f (\sigma_j) - f(\sigma) \over
\sigma_j - \sigma} \right] \left[ {S(U_j, V) - S(U_j, V) \over U_j - U}
\right],
$$ 
from which we deduce the differentiability of $f$ and the
formula (5.11).

Now consider the function 
$$
G(X) = \left( {\partial S^* \over \partial U} \right) \biggl/ \left(
{\partial S \over \partial U} \right),
$$
which is continuous because $S$ and $S^*$ are 
continuously differentiable and $\left( {\partial S \over \partial U}
\right) \not= 0$.  By Eq.\ (5.11), with $g = f^\prime$, 
$$
G(X) =g(S(X)),
$$
and we now wish to prove that $g: I \rightarrow \R$ is a constant 
function
(call it $a$).  This will prove our theorem because it implies that
$$
S^* (U,V) = aS (U,V) +B(V),
$$
This, in turn, implies that $B(V)$ is constant on adiabats.  However, the
projection of an adiabat, $\partial A_X$, on $\R^n$ is an open set
(because the pressure, which defines the tangent planes, is finite
everywhere).  Thus, the projection $\uprho (\Gamma)$ is covered by open
sets on each of which $B(V)$ is constant.  But $\uprho (\Gamma)$ is
connected (indeed, it is convex) and therefore $B(V)$ is constant on all
of $\uprho (\Gamma)$.

To show that $g$ is constant, it suffices to show this locally.  We 
know that $X \mapsto G(X)= g(S(X))$ is constant on adiabats, and it is 
also constant on isotherms because the level sets of $\partial 
S/\partial U$ and $\partial S^*/\partial U$ both coincide with the 
isotherms.  We now invoke the transversality property and Theorem 5.5.
Let $\widehat\sigma$ be any fixed point in the range $I$ of 
$S$, i.e., $\widehat \sigma=S(\widehat X)$ for some $\widehat 
X\in\Gamma$.  By the transversality property there are points $X_0, 
X_1$ such that 
$$
\sigma_{0}=S(X_0) < \widehat \sigma < 
S(X_1)=\sigma_{1}
$$ 
and $X_0 \simt X_1$.  Now let $\sigma=S(X)$ be any 
other point in the open interval $(\sigma_{0},\sigma_{1})$. By 
Theorem 5.5 there are points $\widehat X'\sima \widehat X$ and 
$X'\sima X$ such that $\widehat X'$ and $X'$ both lie on the same 
isotherm (namely the isotherm through $X_{0}$ and $X_{1}$). But this 
means that $g(\sigma)=G(S(X'))=G(S(\widehat X')=g(\widehat \sigma)$, 
so $g$ is constant.
\hfill\lanbox
\medskip

{\it Remark:}  The transversality property is essential for this
uniqueness theorem.  As a counterexample, suppose that every isotherm is
an adiabat.  Then {\it any\/} concave $S$ that has the adiabats as its
level sets would be an acceptable entropy.  \bigskip\noindent

%%%%%%%%%%%%%%%%%%%%%%%%%%%%%%%%%%%%%%%%%%%

\vfill\eject
%%%%%%%%%%%%%%%%%%%%%%%%%%%%%%%%%%
%\eqno\eqlbl\fermat This is an example of usage
\newcount\chno    \chno=6
\newcount\equno   \equno=0
%%%%%%%%%%%%%%%%%%%%%%%%%%%%%%%%%%

\noindent
{\tit VI. MIXING AND CHEMICAL REACTIONS } 
\bigskip

\noindent
{\subt A. The difficulty of fixing  entropy constants}
\medskip

We have seen in Sections II and IV that the entropies of all simple
systems can be calibrated once and for all so that the entropy of any
compound system made up of any combination of the basic simple systems is 
exactly the sum of the individual entropies. This global entropy works
(i.e., it satisfies the entropy principle  of 
Sect.\ II B  and tells us {\it
exactly} which processes can occur) in those cases in which the 
`masses' of the individual systems are conserved. That is, splitting and
recombination of simple systems is allowed, but not mixing of different
systems or (chemical or nuclear) reactions.  

Nature does allow us to mix the contents of different simple systems,
however, (which is not to be confused with the formation of a compound
system).  Thus, we can mix one mole of water and one mole of alcohol to
form two moles of whiskey. The entropy of the mixture is certainly not 
the
sum of the individual entropies, as would be the case if we were forming 
a
compound system.  Nevertheless, our previous analysis, namely Theorem 
2.5, 
does tell us the entropy of the mixture---{\it up to an additive constant 
!
\/} {\it The multiplicative constant can be, and will be henceforth, 
fixed
by the entropy function of one standard system}, e.g., one mole of 
mercury.
The reason   that the multiplicative constant is fixed for the mixture 
is,
as we have stressed, the notion of thermal equilibrium.  Another way to 
say
this is that once the unit of energy (say Joules) and of temperature (say
Kelvin) have been fixed, then  the entropy of every system, simple and
compound, is fixed up to an additive constant.  Our assumptions A1-A7,
S1-S3 and T1-T5 guarantee this.

A similar discussion applies to chemical reaction products. After all, 
the
solution of alcohol in water can be considered a chemical reaction if one
wishes. It requires a certain amount of chemical sophistication, which 
was
not available before the enlightenment, to distinguish a mixture from a
chemical compound.  

The question addressed in this section is this:   to what extent can the
{\it additive constants}  (denoted by the letter $B$, in conformity with
Theorems 2.3 and 2.5) be determined so that whenever a mixture or 
reaction
occurs adiabatically we can say that the entropy has not decreased?  To
what extent is this  determination unique? 

One thing that conceivably might have to be discarded, partially at
least, is the idea that comparability is an equivalence relation.  As
stated in Section I, to have an equivalence relation would require that
whenever $X\prec Z$ and $Y\prec Z$ then $X\prec Y$  or $ Y\prec X$ (and
similarly for $Z\prec X$ and $Z\prec Y$).  If one were to resort to the
standard devices of semi-permeable membranes and van t'Hofft boxes, as in
the usual textbooks, then it would be possible to maintain this 
hypothesis,
even for mixing and chemical reactions. In that case, one  would be able 
to
prove that the additive entropy constants are uniquely determined for all
matter, once they have been chosen for the 92 chemical elements.  

Alas, van t'Hofft boxes do not exist in nature, except in imperfect form.
For example, Fermi (1956, p.\ 101), in a discussion of the van t'Hofft
box, writes that ``The equilibria of gaseous reactions can be treated
thermodynamically by assuming the existence of ideal 
semi-permeable
membranes'', but then goes on to state that ``We should notice, finally,
that in reality no ideal semi-permeable membranes exist. The best 
approximation of such a membrane is a hot palladium foil, which behaves 
like a semi-permeable membrane for hydrogen.'' Nevertheless, the rest of
Fermi's discussion is based on the existence of such membranes! 

We are the not saying that the comparison hypothesis {\it must} be 
discarded for chemical reactions and mixtures; we are only raising the 
logical possibility.  As a result, we shall try to organize our 
discussion without using this hypothesis.

Therefore, we shall have to allow the possibility that if a certain 
kind of process is theoretically possible then entropy increase {\it 
alone} does not determine whether it will actually occur; in 
particular cases it {\it might} conceivably be necessary to have a 
certain minimum amount of entropy increase before a reaction can take 
place.  Moreover, the entropy principle of Section II.\ B conceivably 
might not hold in full generality in the sense that there could be 
irreversible processes for which entropy does not strictly increase.  
What we do show in this section is that it is possible, nevertheless, 
to fix the entropy constants of all substances in such a way, that the 
entropy never decreases in an adiabatic process.  This weak form of 
the entropy principle is stated in Theorem 6.2.  However, it is only 
because of a technicality concerned with uncountably many dimensions 
that we cannot prove the entropy principle in the strong form and 
there is no doubt that the `good case' mentioned at the end of this 
section actually holds in the real world.  For all practical purposes 
we do have the strong form because the construction of the constants 
is done inductively in such a way that at each stage it is not 
necessary to revise the constants previously obtained; this means that 
in the finite world in which we live we are actually dealing, at any 
given moment, with the countable case.

A significant point to notice about the additive constants, $B$, is 
that they must scale correctly when the system scales; a somewhat 
subtler point is that they must also obey the additivity law under 
composition of two or more systems, $\Gamma_1\times \Gamma_2$, in 
order that (2.4) holds.  As we shall see in Sect.\ B, this latter 
requirement will not be met automatically and it will take a bit of 
effort to achieve it.

As a final introductory remark let us mention a computational device 
that is often used, and which seems to eliminate the need for any 
special discussion about mixing, reactions or other variations in the 
amount of matter.  This device is simply to regard the amount of a 
substance (often called the `particle number' because of our 
statistical mechanical heritage) as just one more work coordinate.  
The corresponding `pressure' is called the {\it chemical potential} in 
this case.  Why does this not solve our problems?  The answer, equally 
simply, is that the comparison hypothesis will not hold within a state 
space since the extended state space will `foliate' into sheets, in 
each of which the particle number is fixed.  Axiom S2 will fail to 
hold.  {\it If particle number is introduced as a work coordinate then 
the price we will have to pay is that there will be no simple 
systems.} Nothing will have been gained.  The question we address here 
is a true physical question and cannot be eliminated by introducing a 
mathematical definition.

\bigskip\noindent
{\subt B. Determination of additive entropy constants}
\bigskip

Let us consider a collection of systems (more precisely, state 
spaces), containing simple and/or compound systems.  Certain adiabatic 
state changes are possible, and we shall be mainly interested in those 
that take us from one specified system to another, e.g., $X \prec Y$ 
with $X \in \Gamma$ and $Y \in \Gamma'$.  Although there are 
uncountably many systems (since, in our convention, changing the 
amount of any component means changing the system), we shall always 
deal in the following with processes involving only finitely many 
systems at one time.  In our notation the process of making one mole 
of water from hydrogen and oxygen is carried out by letting $X$ be a 
state in the compound system $\Gamma$ consisting of one mole of ${\rm 
H}_2 $ and one half mole of ${\rm O}_2$ and by taking $Y$ to be a 
state in the simple system, $\Gamma'$, consisting of one mole of 
water.

Each system has a well defined entropy function, e.g., for $\Gamma$ 
there is $S_\Gamma$, and we know from Section IV that these  can be
determined
in such a way that the sum of the entropies increases in any  adiabatic
process in any compound space
$\Gamma_1 \times  \Gamma_2 \times ...$. Thus, if
$X_i \in  \Gamma_i$ and $Y_i \in  \Gamma_i$ then
$$
(X_1,X_2,...) \prec (Y_1,Y_2,...) \quad {\rm if \ and \ only \ if} 
\quad S_1(X_1)+S_{2}(X_2)
+\cdots \leq S_1(Y_1)+S_{2}(Y_2)+\cdots \ . \eqno\eqlbl\fd$$
where we have denoted $S_{\Gamma_i}$ by $S_i$ for short. 
The additive entropy constants do not matter here since each 
function $S_i$ appears on both sides of this inequality.

Now we consider relations of the type
$$
X\prec Y  \quad\quad {\rm with } \quad\quad X \in 
\Gamma, \ Y \in \Gamma'. \eqno\eqlbl\fe
$$
Our goal is to find constants $B(\Gamma)$, one for each 
state space $\Gamma$, in such a way that the 
entropy defined by
$$
S(X) := S_\Gamma(X) + B(\Gamma)  \quad\quad {\rm for} \quad\quad X \in 
\Gamma    \eqno\eqlbl\ff
$$
satisfies
$$
S(X) \leq S(Y)   \eqno\eqlbl\fg
$$
whenever (\fe) holds.

Additionally, we require that the  newly defined entropy satisfies
scaling and additivity under composition. Since the initial entropies
$S_\Gamma(X)$ already satisfy them, these requirements become conditions 
on 
the
additive constants $B(\Gamma)$:
$$
B(t_1\Gamma_1\times 
t_2\Gamma_2)=t_1B(\Gamma_1)+t_2B(\Gamma_2)\eqno\eqlbl\linear
$$
for all state spaces $\Gamma_1$, $\Gamma_2$ under consideration and 
$t_1,t_2>0$.

As we shall see, the additivity requirement is
not trivial to satisfy, the reason being that a given substance,
say hydrogen, can appear in many different compound systems with many
different ratios of the mole numbers of the constituents of the compound
system.

The condition (\fg) means that 
$$
B(\Gamma) -B(\Gamma')\leq S_{\Gamma'}(Y)-S_{\Gamma}(X)
$$
whenever $X\prec Y$. Let us denote by $D(\Gamma,\Gamma')$ the 
minimal entropy difference for all  adiabatic  processes that can 
take us from $\Gamma$ to $\Gamma'$, i.e., 
$$
D(\Gamma,\Gamma') := \inf \{S_{\Gamma'}(Y)-S_{\Gamma}(X) \ : \  X \prec Y 
\}.
\eqno\eqlbl\fh
$$
It is to be noted that $D(\Gamma,\Gamma')$ can be positive or negative 
and
$D(\Gamma,\Gamma') \neq D(\Gamma',\Gamma)$ in general. Clearly 
$D(\Gamma,\Gamma)=0$.  Definition (\fh) makes sense only if there
is at least one adiabatic process that goes from $\Gamma$ to 
$\Gamma'$,
and it is convenient  to define $D(\Gamma,\Gamma')=+\infty$ if there is  
no 
such 
process. In terms of the $D(\Gamma,\Gamma')$'s  condition (\fg) means 
precisely  
that
$$
-D(\Gamma',\Gamma) \leq  B(\Gamma)-B(\Gamma')\leq D(\Gamma,\Gamma')    
\eqno\eqlbl\be
$$

Although $D(\Gamma,\Gamma')$ has no particular sign, we can assert the 
crucial
fact that
$$  
-D(\Gamma',\Gamma)\leq D(\Gamma,\Gamma')  \eqno\eqlbl\fff 
$$
This is trivially true if $D(\Gamma,\Gamma')=+\infty$ or 
$D(\Gamma',\Gamma)=+\infty$. If both are $<\infty$ the reason 
(\fff) is true is simply (\fd): 
By the definition
(\fh), there is  a pair of states 
$X \in \Gamma$ and $Y\in \Gamma' $ such that $X
\prec Y$
and
$S_{\Gamma'}(Y)-S_{\Gamma}(X) =D(\Gamma,\Gamma')$ 
(or at least as closely as we please).
Likewise, we can find $W \in \Gamma $ and $Z\in \Gamma' $, such
that $Z\prec W$ and
$S_{\Gamma}(W)-S_{\Gamma'}(Z) =D(\Gamma',\Gamma)$. Then, in the
compound system $\Gamma\times \Gamma'$ we have that 
$(X, Z) \prec (W, Y)$, and this, by (\fd), implies (\fff). Thus
$D(\Gamma,\Gamma') > -\infty$ if there is at least one adiabatic process
from $\Gamma'$ to $\Gamma$.

Some reflection shows us that consistency in the definition of the 
entropy
constants $B(\Gamma)$ requires us to consider all possible chains of 
adiabatic
processes leading from one space to another via intermediate steps.  
Moreover,
the additivity requirement leads us to allow the use of a `catalyst' in 
these
processes, i.e., an auxiliary system, that is recovered at the end,
although a state change {\it within} this system might take place.

For this reason we 
now define new quantities, $E(\Gamma,\Gamma')$ and $F(\Gamma,\Gamma')$,
in the following way. First, for any
given $\Gamma$ and $\Gamma'$ we consider all 
finite chains of state spaces, 
$\Gamma=\Gamma_1,\Gamma_2,\dots,\Gamma_N=\Gamma'$
such that $D(\Gamma_i,\Gamma_{i+1})<\infty$ for all i, and we define
$$
E(\Gamma,\Gamma'):=\inf\{D(\Gamma_1,\Gamma_{2})+
\cdots +D(\Gamma_{N-1},\Gamma_{N})
\}, 
\eqno\eqlbl\fl
$$
where the infimum is taken over all such chains linking $\Gamma$ with 
$\Gamma'$.
Note that $E(\Gamma,\Gamma')\leq D(\Gamma,\Gamma')$ and 
$E(\Gamma,\Gamma')$ could be $<\infty$ even if there is no direct 
adiabatic 
process linking $\Gamma$ and $\Gamma'$, i.e.,  
$D(\Gamma,\Gamma')=\infty$.
We then define
$$
F(\Gamma,\Gamma'):=\inf\{E(\Gamma\times\Gamma_0, \Gamma'\times\Gamma_0)\}
\}, 
\eqno\eqlbl\flx
$$
where the infimum is taken over all state spaces $\Gamma_0$. (These 
are the `catalysts'.)

The following properties of $F(\Gamma,\Gamma')$ are easily verified:
$$F(\Gamma,\Gamma)=0 \eqno\eqlbl\Fa$$
$$F(t\Gamma,t\Gamma')=tF(\Gamma,\Gamma') \quad\quad {\rm for\ }t>0
 \eqno\eqlbl\Fb$$
$$F(\Gamma_1\times \Gamma_2,\Gamma_1'\times \Gamma_2')\leq
F(\Gamma_1,\Gamma_1')+F(\Gamma_2,\Gamma_2') \eqno\eqlbl\Fc$$
$$F(\Gamma\times \Gamma_0,\Gamma'\times \Gamma_0)=
F(\Gamma,\Gamma')\quad\quad \hbox{\rm for all\ \ }\Gamma_0.
\eqno\eqlbl\Fd$$
In fact, (\Fa) and (\Fb) are also shared by the $D$'s and the $E$'s. The 
`subadditivity' (\Fc) holds also for the $E$'s, but the `translational 
invariance' (\Fd) might only hold for the $F$'s. 
 
{}From (\Fc) and (\Fd) 
it follows that the $F$'s  satisfy the `triangle inequality'
$$ 
F(\Gamma,\Gamma^{\prime\prime})\leq F(\Gamma,\Gamma')+
F(\Gamma',\Gamma^{\prime\prime})\eqno\eqlbl\Fe
$$
(put $\Gamma=\Gamma_1$, $\Gamma^{\prime\prime}=\Gamma_1'$, 
$\Gamma'=
\Gamma_2=\Gamma_2'$.) This inequality also holds for the $E$'s as is 
obvious from the definition (\fl).
A special case (using (\Fa)) is the analogue of (\fff):
$$
-F(\Gamma',\Gamma)\leq F(\Gamma,\Gamma')\eqno\eqlbl\Ff
$$
(This is trivial if $F(\Gamma',\Gamma)$ or $F(\Gamma',\Gamma)$ is 
infinite, 
otherwise use (\Fe) with $\Gamma=\Gamma^{\prime\prime}$.)

Obviously, the following inequalities hold:
$$
-D(\Gamma',\Gamma) \leq -E(\Gamma',\Gamma) \leq -F(\Gamma',\Gamma) \leq
F(\Gamma,\Gamma') \leq E(\Gamma,\Gamma') \leq D(\Gamma,\Gamma').
$$

The importance of the $F$'s for the determination of the additive 
constants 
is
made clear in the following theorem:
\medskip

{\bf THEOREM 6.1 (Constant entropy differences).} {\it
If $\Gamma$ and  $\Gamma'$ are two state spaces 
then for any two points $X\in \Gamma$ and  $ Y\in \Gamma'$ 
$$
X\prec Y \quad \hbox{\rm if and only if} \quad S_\Gamma(X) 
+F(\Gamma,\Gamma')
\leq S_{\Gamma'}(Y) . \eqno\eqlbl\lemma
$$}

\medskip
{\it Remarks:} (1). Since $F(\Gamma,\Gamma')\leq D(\Gamma,\Gamma')$
the theorem is trivially true when  $F(\Gamma,\Gamma')=+ \infty$, 
in the sense that there is then no adiabatic 
process from $\Gamma$ to $\Gamma'$. The reason  for
the title `constant entropy differences' is that the minimum jump between 
the
entropies $S_\Gamma(X)$ and $S_{\Gamma'}(Y)$ for $X\prec Y$ to be 
possible 
is independent of $X$.

\noindent
$\phantom {Remarks}$ (2). There is an interesting corollary of
Theorem 6.1.  We know, from the definition (\fh),  that $X\prec Y$ only
if $S_\Gamma(X) +D(\Gamma,\Gamma') \leq S_{\Gamma'}(Y)$. Since
$D(\Gamma,\Gamma') \leq F(\Gamma,\Gamma')$, Theorem 6.1 tells us two
things:
$$
X\prec Y \quad \hbox{\rm if and only if} \quad S_\Gamma(X)
+F(\Gamma,\Gamma')
\leq S_{\Gamma'}(Y) . \eqno\eqlbl\cora
$$
and
$$
S_\Gamma(X) +D(\Gamma,\Gamma')
\leq S_{\Gamma'}(Y)\quad \hbox{\rm if and only if} \quad
S_\Gamma(X) +F(\Gamma,\Gamma') \leq S_{\Gamma'}(Y) . \eqno\eqlbl\corb
$$
We {\it cannot} conclude from this, however, that $D(\Gamma,\Gamma') =
F(\Gamma,\Gamma')$.

\medskip

{\it Proof:}  The `only if' part is obvious because 
$F(\Gamma,\Gamma')\leq
D(\Gamma,\Gamma')$, and thus our goal is to prove the `if' part.  For
clarity, we begin by assuming that the infima in (\fh), (\fl) and (\flx) 
are
minima, i.e., there are state spaces $\Gamma_0$, $\Gamma_1$, 
$\Gamma_2$,...,
$\Gamma_N$ and states $X_i \in \Gamma_i$ and $Y_i \in  \Gamma_i$, for
$i=0,...,N$ and states $\tilde X\in \Gamma $ and $\tilde Y\in \Gamma'$ 
such
that
$$\eqalignno{
(\tilde X, X_0) &\prec Y_1 \cr
X_i &\prec Y_{i+1} \quad\quad {\rm for}\ i=1,...,N-1 \cr
X_N &\prec (\tilde Y, Y_0) &\eqlbl\order \cr}
$$
and $F(\Gamma, \Gamma')$ is given by
$$\eqalignno{
F(\Gamma, \Gamma')&= D(\Gamma \times \Gamma_0, \Gamma_1)+ D(\Gamma_1, 
\Gamma_2)
+\cdots +D(\Gamma_N,\Gamma' \times \Gamma_0) \cr
&= S_{\Gamma'}(\tilde Y) +
\sum_{j=0}^N S_j(Y_j) - S_{\Gamma}(\tilde X) -\sum_{j=0}^N S_j(X_j).
&\eqlbl\Fdef  \cr}
$$
In (\Fdef) we used the abbreviated notation $S_j$ for $S_{\Gamma_{j}}$ 
and 
we
used the fact  that  $S_{\Gamma \times \Gamma_0} = S_{\Gamma} + S_0$.

{F}rom the assumed inequality $S_\Gamma(X) +F(\Gamma,\Gamma')\leq 
S_{\Gamma'}(Y)$ and (\Fdef) we conclude that
$$
S_{\Gamma}(X)+S_{\Gamma'}(\tilde Y) +\sum_{j=0}^N S_j(Y_j) \leq 
S_{\Gamma}(\tilde X)+S_{\Gamma'}(Y) +\sum_{j=0}^N S_j(X_j). 
\eqno\eqlbl\ineq
$$
However, both sides of this inequality can be thought of as the entropy 
of 
a state in the compound space $\hat \Gamma := \Gamma \times \Gamma' 
\times
\Gamma_0 \times \Gamma_1 \times \cdots \times \Gamma_N$. The entropy
principle (\fd) for $\hat \Gamma$ then tell us that
$$
(X, \tilde Y, Y_0,\dots ,Y_N) \prec (\tilde X, Y, X_0,\dots ,X_N)
\eqno\eqlbl\Fpreca
$$
On the other hand, using (\order) and the axiom of consistency, we have 
that
$$
(\tilde X, X_0, X_1, ..., X_N) \prec (\tilde Y, Y_0, Y_1, ..., Y_N).
\eqno\eqlbl\Fprecb
$$
By the consistency axiom again, we  have from (\Fprecb) that 
$(\tilde X, Y, X_0,\cdots ,X_N)\prec $ \hfill\break
$(Y,\tilde Y, Y_0, Y_1, ..., Y_N)$. {}From transitivity we then
have 
$$
(X, \tilde Y, Y_0, Y_1, ..., Y_N) \prec (Y,\tilde Y, Y_0, Y_1, ..., Y_N),
$$ 
and the desired conclusion, $X\prec Y$, follows from the cancellation 
law.

If $F(\Gamma,\Gamma')$ is not a minimum, then, for every $\varepsilon > 
0$,
there is a chain of spaces $\Gamma_0$, $\Gamma_1$, $\Gamma_2$,..., 
$\Gamma_N$
and corresponding states as in (\order) such that (\Fdef) holds to within
$\varepsilon$ and (\ineq) becomes (for simplicity of notation we omit the
explicit dependence of the states and $N$ on $\varepsilon$)
$$
S_{\Gamma}(X)+S_{\Gamma'}(\tilde Y) +\sum_{j=0}^N S_j(Y_j) \leq 
S_{\Gamma}(\tilde X)+S_{\Gamma'}(Y) +\sum_{j=0}^N S_j(X_j)
+ \varepsilon. \eqno\eqlbl\ineqb
$$
Now choose any auxiliary state space $\widetilde \Gamma$, with entropy
function $\widetilde S$, and two states 
$ Z_0, Z_1\in \widetilde \Gamma$ with $Z_0 \prec\prec
Z_1$. The space $\Gamma$ itself could be used for this purpose, but for
clarity we regard $\widetilde \Gamma$ as distinct. Define
$\delta (\varepsilon) := [\widetilde S(Z_1) - \widetilde S(Z_0)]^{-1}
\varepsilon$.
Recalling that $\delta \widetilde S(Z)= \widetilde S(\delta Z)$ by 
scaling,
we see that (\ineqb) implies the following analogue of (\Fpreca).
$$
(\delta Z_0, X, \tilde Y, Y_0,\dots ,Y_N) \prec (\delta Z_1, \tilde X, 
Y, X_0,\dots ,X_N). \eqno\eqlbl\Fprecc
$$
Proceeding as before, we conclude that
$$
(\delta Z_0, X, \tilde Y, Y_0, Y_1, ..., Y_N) \prec 
(\delta Z_1, Y,\tilde Y, Y_0, Y_1, ..., Y_N),
$$
and thus $(X,\delta Z_0)\prec (Y,\delta Z_1)$ by the cancellation law.
However, $\delta \rightarrow 0$ as $\varepsilon \rightarrow 0$ and hence
$X\prec  Y$ by the stability axiom.  \hfill \lanbox

According to Theorem 6.1 the  determination of the entropy 
constants $B(\Gamma)$ amounts to satisfying the estimates
$$
-F(\Gamma',\Gamma)\leq B(\Gamma)-B(\Gamma')\leq F(\Gamma,\Gamma')
\eqno\eqlbl\Bcondition
$$
together with the linearity condition (\linear). 
It is clear that (\Bcondition) can only be satisfied with finite
constants $B(\Gamma)$ and $B(\Gamma')$, if $F(\Gamma,\Gamma')>-\infty$. 
While the assumptions made so far do not exclude
$F(\Gamma,\Gamma')=-\infty$ as a possibility, it follows from (\Ff) that
this can only be the case if at the same time
$F(\Gamma',\Gamma)=+\infty$, i.e., there is no chain of intermediate
adiabatic processes in the sense described above that allows a passage
from $\Gamma'$ back to $\Gamma$. For all we know this is not the
situation encountered in nature and we exclude it by an additional
axiom. Let us write $\Gamma\prec \Gamma'$ and say that $\Gamma$ is {\it
connected to} $\Gamma'$ if $F(\Gamma,\Gamma')<\infty$, i.e. if there
is a finite chain of state spaces, $\Gamma_0, \Gamma_1
,\Gamma_2,\dots,\Gamma_N$ and states such that (\order) holds with
$\tilde X\in \Gamma$ and $\tilde Y\in \Gamma'$.  Our new axiom is the
following:

\bigskip

\item{\bf M)} {\bf Absence of sinks.} If $\Gamma$ is
connected to $\Gamma'$ then $\Gamma'$ is connected to $\Gamma$, i.e.,
$\Gamma \prec \Gamma' \Longrightarrow \Gamma' \prec \Gamma$. 
\bigskip 

The introduction of this axiom may seem a little special, even 
artificial,
but it is not.  For one thing, it is not used in Theorem 6.1 which, like
the entropy principle itself, states the condition under which adiabatic
process from $X$ to $Y$ is possible.  Axiom M is only needed for setting
the additive entropy constants so that (\lemma) can be converted into a
statement involving $S(X)$ and $S(Y)$ alone, as in Theorem
6.2. Second, axiom M should not be misread as saying that if we can make
water from hydrogen and oxygen then we can make hydrogen and oxygen
directly from water (which requires hydrolysis). What it does require is
that water can eventually be converted into its chemical elements, but 
not
necessarily in one step  and not necessarily reversibly.  The 
intervention
of irreversible processes involving other substances is allowed. Were 
axiom
M to fail in this case then all the oxygen in the universe would 
eventually
turn up in water and we should have to rely on supernovae to replenish 
the
supply from time to time.

By axiom M (and the obvious transitivity of the relation $\prec$ for 
state 
spaces), connectedness defines an equivalence relation between
state spaces, and instead of $\Gamma \prec \Gamma'$ we can write
$$
\Gamma \ \sim \ \Gamma' \ \eqno\eqlbl\fsim
$$
to indicate that the $\prec$ relation among state spaces goes both ways.
As already noted, $\Gamma\sim\Gamma'$ is equivalent to 
$-\infty<F(\Gamma,\Gamma')<\infty$ and 
$-\infty<F(\Gamma',\Gamma)<\infty$.

Without further assumptions (note, in particular, that no assumptions 
about 
'semi-permeable membranes' have been made) we can now derive the entropy 
principle
in the following weak version: 
\medskip

{\bf THEOREM 6.2 (Weak form of the entropy principle).} {\it
Assume axiom M in addition to A1-A7, S1-S3, T1-T5. Then the entropy 
constants 
$B(\Gamma)$ can be chosen in such a way that the entropy $S$,
defined on all states of all systems by (\ff), satisfies additivity and
extensivity (2.4), (2.5), and moreover 
$$
X\prec Y \quad \hbox{\rm implies} \quad S(X)\leq S(Y). \eqno\eqlbl\finala
$$   }  
%\medskip
{\it Proof:\/} The proof is a simple application of the Hahn-Banach 
theorem
(see, e.g., the appendix to (Giles, 1964) and (Reed and Simon, 1972)).
Consider the set ${\cal S}$ of all pairs of state spaces 
$(\Gamma,\Gamma')$.
On ${\cal S}$ we define an equivalence relation by declaring
$(\Gamma,\Gamma')$ to be equivalent to $(\Gamma\times 
\Gamma_0,\Gamma'\times
\Gamma_0)$ for all $\Gamma_0$. Denote by $[\Gamma,\Gamma']$ the 
equivalence
class of $(\Gamma,\Gamma')$ and let ${\cal L}$ be the set of all these
equivalence classes.

On ${\cal L}$ we define  multiplication by scalars and addition in the 
following way:
$$\eqalign{
t[\Gamma,\Gamma']&:= [t\Gamma,t\Gamma']\qquad\quad\hbox{\rm for }  t> 0 
\cr
t[\Gamma,\Gamma']&:= [-t\Gamma',-t\Gamma]\qquad\hbox{\rm for } t< 0 \cr
0[\Gamma,\Gamma']&:= [\Gamma,\Gamma]= [\Gamma',\Gamma'] \cr
[\Gamma_1,\Gamma_1']+[\Gamma_2,\Gamma_2'] &:= 
[\Gamma_1\times\Gamma_2,\Gamma_1'\times\Gamma_2']. \cr  }
$$
With these operations ${\cal L}$
becomes a  vector space, which is 
infinite dimensional in general. The zero element 
is the class $[\Gamma,\Gamma]$ for any $\Gamma$, because by our 
definition 
of 
the equivalence relation $(\Gamma,\Gamma)$ is equivalent to 
$(\Gamma\times 
\Gamma',\Gamma\times \Gamma')$, which in turn is equivalent to 
$(\Gamma',\Gamma')$. Note that for the same reason $[\Gamma',\Gamma]$ is 
the 
negative of $[\Gamma,\Gamma']$.

Next, we define a function $H$ on ${\cal L}$ by
$$H([\Gamma,\Gamma']):=F(\Gamma,\Gamma')$$
Because of (\Fd), this function is well defined and it takes values in 
$(-\infty,\infty]$. Moreover, it follows from (\Fb) and (\Fc) that $H$ is 
homogeneous, i.e., $H(t[\Gamma,\Gamma'])=tH([\Gamma,\Gamma'])$, 
and subadditive, i.e., $H([\Gamma_1,\Gamma_1']+[\Gamma_2,\Gamma_2'])
\leq H([\Gamma_1,\Gamma_1']) + H([\Gamma_2,\Gamma_2'])$. 
Likewise, 
$$G([\Gamma,\Gamma']):=-F(\Gamma',\Gamma)$$
is homogeneous and superadditive, i.e., $G([\Gamma_1,\Gamma_1']+
[\Gamma_2,\Gamma_2']) \geq G([\Gamma_1,\Gamma_1']) 
+G([\Gamma_2,\Gamma_2'])$.
By (\Ff)
we have $G\leq F$ so, by the Hahn-Banach theorem, there exists a 
real-valued 
{\it linear} function $L$ on ${\cal L}$ lying between $G$ and $H$; that 
is
$$
-F(\Gamma',\Gamma) \leq L([\Gamma,\Gamma']) \leq F(\Gamma,\Gamma').
\eqno\eqlbl\between
$$

Pick any 
fixed $\Gamma_0$ and define 
$$B(\Gamma):=L([\Gamma_0\times\Gamma,\Gamma_0]).$$
By linearity, $L$ satisfies $L([\Gamma,\Gamma']) = -L(-[\Gamma,\Gamma'])
=-L([\Gamma',\Gamma])$. 
We then have 
$$B(\Gamma)-B(\Gamma')=L([\Gamma_0\times\Gamma,\Gamma_0])+
L([\Gamma_0, \Gamma_{0}\times \Gamma'])=L([\Gamma,\Gamma'])$$
and hence (\Bcondition) is satisfied.
\hfill\lanbox

{}From the proof of Theorem 6.2 it is clear that the indeterminacy of the
additive constants $B(\Gamma)$ can be traced back to the non uniqueness 
of
the linear function $L([\Gamma,\Gamma'])$ lying between
$G([\Gamma,\Gamma'])=-F(\Gamma',\Gamma)$ and
$H([\Gamma,\Gamma'])=F(\Gamma,\Gamma')$. This non uniqueness has two
possible sources: One is that some pairs of state spaces $\Gamma$ and
$\Gamma'$ may not be connected, i.e., $F(\Gamma,\Gamma')$ may be infinite
(in which case $F(\Gamma',\Gamma)$ is also infinite by axiom M).  The
other possibility is that there is a finite, but positive `gap' between
$G$ and $H$, i.e.,
$$
-F(\Gamma',\Gamma)<F(\Gamma,\Gamma')\eqno\eqlbl\truegap
$$
might hold for some state spaces, even if both sides are finite.
 
In nature only states containing the same amount of the chemical
elements can be transformed into each other. Hence
$F(\Gamma,\Gamma')=+\infty$ for many pairs of state spaces, in
particular, for those that contain different amounts of some chemical
element.  The constants $B(\Gamma)$ are therefore never unique: For
each equivalence class of state spaces (with respect to $\sim$) one can
define a constant that is arbitrary except for the proviso that the
constants should be additive and extensive under composition and
scaling of systems. In our world, where there are  92 chemical elements
(or, strictly speaking, a somewhat larger number, $N$, since one should
count different isotopes as different elements), and this leaves us
with at least 92 free constants that specify the entropy of one mole of
each of the chemical elements in some specific state.

The other possible source of non uniqueness, a non zero gap (\truegap)
is, as far as we know, not realized in nature, although it is a logical
possibility.  The true situation seems rather to be the following: The
equivalence class $[\Gamma]$ (with respect to $\sim$) of every state
space $\Gamma$ contains a
distinguished  state space
$$
\Lambda([\Gamma])=\lambda_1\Gamma_1\times\dots\times\lambda_N\Gamma_N
$$
where the $\Gamma_i$ are the state spaces of one mole of each of the 
chemical
elements, and the numbers $(\lambda_1,\dots,\lambda_N)$ specify the 
amount 
of
each chemical element in $\Gamma$. We have 
$$\Lambda([t\Gamma])=t\Lambda([\Gamma])
\eqno\eqlbl\la$$
and
$$ \Lambda([\Gamma\times\Gamma'])=\Lambda([\Gamma]) 
\times\Lambda([\Gamma']).
\eqno\eqlbl\lb$$
Moreover (and this is the crucial `experimental fact'), 
$$-F(\Lambda([\Gamma]),\Gamma])=F(\Gamma, \Lambda([\Gamma]))
\eqno\eqlbl\lc$$
for all $\Gamma$. Note that (\lc) is subject to experimental 
verification by measuring on the one hand entropy differences for  
processes 
that
synthesize chemical compounds from the elements 
(possibly through many intermediate steps and with the aid of catalysts), 
and on the other hand for processes
where chemical compounds are decomposed into the elements. 

It follows from (\Fe) (\Ff) and (\lc)
that 
$$F(\Gamma,\Gamma')=F(\Gamma,\Lambda([\Gamma]))+F(\Lambda([\Gamma]),
\Gamma')
\eqno\eqlbl\lddd$$
and
$$-F(\Gamma',\Gamma)=F(\Gamma,\Gamma')\eqno\eqlbl\ld
$$
for all $\Gamma'\sim\Gamma$.
Moreover, an explicit formula for $B(\Gamma)$ can be given in this 
good case:
$$B(\Gamma)=F(\Gamma, \Lambda([\Gamma]).
\eqno\eqlbl\le$$
If $F(\Gamma,\Gamma')=\infty$, then (\Bcondition) holds trivially, while 
for
$\Gamma\sim\Gamma'$ we have by (\lddd) and(\ld)
$$B(\Gamma)-B(\Gamma')=F(\Gamma,\Gamma')=-F(\Gamma',\Gamma),\eqno\eqlbl
\lee
$$
i.e., the inequality (\Bcondition) is saturated. It is also clear that 
in this case 
$B(\Gamma)$ is unique up to the choice of arbitrary constants for the 
fixed 
systems $\Gamma_1,\dots,\Gamma_N$. The particular choice (\le) 
corresponds 
to 
putting $B(\Gamma_i)=0$ for the chemical elements $i=1,\dots,N$.

{}From Theorem 6.1 it follows that in the good case just described the
comparison principle holds in the sense that {\it all states belonging to
systems in the same equivalence class are comparable}, and the relation
$\prec$ is {\it exactly} characterized by the entropy function, i.e., the
{\it full} entropy principle holds.

If there is a genuine gap, (\truegap), then for some pair of state spaces 
we
might have only the weak version of the entropy principle, 
Theorem 6.2. 
Moreover, it follows from Theorem 6.1 that 
in this case
there are no states 
$X\in\Gamma$
and $Y\in \Gamma'$ such that $X\sima Y$. Hence, in order for the full
entropy principle to hold as far as $\Gamma$ and $\Gamma'$ are concerned,
it is only necessary to ensure that $X\prec\prec Y$ implies $ S(X)<S(Y)$,
and this will be the case (again by Theorem 6.1) if and only if
$$
-F(\Gamma',\Gamma)<B(\Gamma)-B(\Gamma')<F(\Gamma,\Gamma').\eqno\eqlbl\ldf
$$

{\it In other words, we would have the full entropy principle, gaps
notwithstanding, if we could be sure that whenever (\truegap) holds then
the inequalities in (\between) are both strict inequalities.}

We are not aware of a proof of the Hahn-Banach theorem that will allow
us to conclude that (\between) is strict in all cases where (\truegap)
holds.  If, however, the dimension of the linear space ${\cal L}$
considered in the proof of Theorem 6.2 were finite then the Hahn-Banach
theorem would allow us to choose the $B$'s in this way. This is a
consequence of the following lemma.  \medskip

%%%%%%%%%%%%%%%%%%%%%%%%%%%%%%%%%%
{\bf LEMMA 6.1 (Strict Hahn-Banach).} {\it Let $V$ be a finite 
dimensional, real 
vector space and $p:
V\to \R$ subadditive, i.e., $p(x+y)\leq p(x)+p(y)$ for all $x,y\in V$, 
and 
homogenous, i.e., $p(\lambda x)=\lambda p(x)$ for all $\lambda\geq 0$, 
$x\in V$.
Then there is a linear functional $L$ on $V$, such that 
$-p(-x)\leq L(x)\leq p(x)$
for all $x\in V$. Moreover,  for those $x$ for which
$-p(-x)<p(x)$
holds we have the strict inequalities
$-p(-x)<L(x)<p(x).$
}

\medskip
{\it Proof:\/} Note first that subadditivity implies that
$p(x)-p(-y)\leq p(x+y)\leq p(x)+p(y)$ for all $x,\,y\in V$. 
Define $V_0=\{x: -p(-x)=p(x)\}$. If $x\in V$ and $y\in V_{0}$, then
$p(x)+p(y)=p(x)-p(-y)\leq p(x+y)\leq p(x)+p(y)$
and hence $p(x)+p(y)=p(x+y)$. (Note that $x$ need not belong to 
$V_{0}$.)
If $x\in V_0$ and $\lambda \geq 0$, then 
$p(\lambda x)=\lambda p(x)=\lambda (-p(-x))=-p(-\lambda x)$, and
if $\lambda<0$ we have, in the same way,
$p(\lambda x)=p((-\lambda)(-x))=(-\lambda)p(-x)=\lambda (-p(-x))=\lambda 
p(x)$. Thus $V_{0}$ is a linear space, and $p$ is a linear functional 
on it. We define $L(x)=p(x)$ for $x\in V_{0}$.

Let $V_0^\prime$ be an algebraic complement of $V_0$, i.e., all $x\in V$ 
can be 
written as $x=y+z$ with $y\in V_0$, $z\in V_0^\prime$ and the 
decomposition is 
unique if $x\neq 0$.
On $V_0^\prime$ the strict inequality
$-p(-x)<p(x)$
holds for all $x\neq 0$. If $L$ can be defined on $V_0^\prime$
such that $-p(-x)<L(x)<p(x)$ for all $V_0^\prime\ni x\neq 0$ we  
reach our goal by defining $L(x+y)=L(x)+L(y)$ for $x\in V_0^\prime$, 
$y\in
V_0$. Hence it suffices to consider  the case that $V_0=\{0\}$.

Now suppose $V_{1}\subset V$ is a linear space and and $L$ has 
been extended from $\{0\}$ to $V_{1}$ such that our requirements
are fulfilled on $V_1$, i.e., $-p(-x)<L(x)<p(x)$ for $x\in V_1$, $x\neq 
0$.  
Define, for $x\in V$
$$\bar p(x)=\inf_{y\in V_1}\{p(x+y)-L(y)\}.$$
By subadditivity it is clear that for all $x$
$$-p(-x)\leq -\bar p(-x)\leq \bar p(x)\leq p(x).$$ 
Since $V$ is finite 
dimensional (by assumption) and $p$ continuous (by
convexity) the infimum is, in fact, a minimum  for 
each 
$x$, i.e.,
$\bar p(x)=p(x+y)-L(y)$ with some $y\in V_1$, depending on $x$.

Suppose $V_{1}$ is not the whole of $V$. 
Pick $x_{2}$ linearly independent of
$V_{1}$. On the
space spanned by $V_{1}$ and $x_{2}$ we define
$$L(\lambda x_{2}+x_{1})=(\lambda/2) (\bar p(x_{2})-\bar p(-x_{2}))+
L(x_{1}).$$
if $x_{1}\in V_{1}$, $\lambda\in \R$.

Then
$$p(\lambda x_{2}+x_{1})-L(\lambda x_{2}+x_{1})=
p(\lambda x_{2}+x_{1})-L(x_{1})-L(\lambda x_{2})\geq 
\bar p(\lambda x_{2})-L(\lambda x_{2})\geq 0$$
and equality holds in the last inequality if and only if 
$\bar p(\lambda x_{2})=-\bar p(-\lambda x_{2})$, i.e., 
$$p(\lambda x_{2}+y)+p(-\lambda x_{2}+y')=L(y+y')\leq 
p(y+y').\eqno\eqlbl\nr$$
for some $y,\ y' \in V_1 $ (depending on $\lambda x_2$). 
On the other hand,
$$p(\lambda x_{2}+y)+p(-\lambda x_{2}+y')\geq p(y+y')$$
by subadditivity, so (\nr) implies
$$L(y+y')=p(y+y')\eqno\eqlbl\nrr$$
By our assumption about $V_{1}$ this hold only if $y+y'=0$.
But then 
$$p(-\lambda x_{2}+y')=p(-\lambda x_{2}-y)$$
and from (\nr) and (\nrr) we get $-p(-\lambda x_{2}-y)=p(\lambda 
x_{2}+y)$
and hence $\lambda x_{2}=-y\in V_1$.
Since $x_{2}\notin V_1$  this is only possible for $\lambda=0$, in which 
case 
$p(x_{1})=L(x_{1})$ and hence (by our assumption about
$V_{1}$), $x_{1}=0$. Thus the statement $L(x)=p(x)$  for
some $x$ lying in the span of $V_{1}$ and $x_{2}$ implies
that $x=0$. 
In the same way one shows
that $L(x)=-p(-x)$ implies $x=0$.
Thus, we have succeeded in extending $L$ from
$V_1$ to the larger space  ${\rm span}\{ V_1, x_2\}$. 
Proceeding by induction we 
obtain $L$ satisfying our requirements on all $V$.
\hfill\lanbox
 
Since the proof of the above version of the Hahn-Banach theorem proceeds
inductively over subspaces of increasing dimension it generalizes in a 
straightforward way to spaces of countable
algebraic dimension. Morover, in such spaces the condition (\ldf) could 
be 
fulfilled at any induction step
{\it without modifying the constants previously defined.} Hence, even in
cases where (\ld) is violated, this hypothetical weakening of the full
entropy principle could never be detected in real experiments involving
only finitely many systems.

\vfill\eject
%%%%%%

%%%%%%%%%%
\newcount\chno    \chno=7
\newcount\equno   \equno=0

%%%%%%%%%%%%%%%%%%%%%%
\noindent
{\tit VII. SUMMARY AND CONCLUSIONS } 
\bigskip

In this final section we recall our notation for the convenience of the
reader and collect all the axioms introduced in Sects. 2, 3, 4 and 6. 
We  then review the logical structure of the paper and the main
conclusions.

Our axioms concern equilibrium states, denoted by $X, Y$ etc., and the 
relation
$\prec$ of adiabatic accessibility between them. If $X\prec Y$ and 
$Y\prec X$ we write  $X\sima Y$, while $X\prec\prec Y$ means that $X\prec 
Y$, 
but not $Y\prec X$. States belong to state spaces 
$\Gamma,\Gamma',\dots$ of
systems, that may be simple or compound. The composition of two state 
spaces 
$
\Gamma,\Gamma'$ is the Cartesian product $\Gamma\times \Gamma'$ 
(the order of the factors is unimportant); the composition of 
$X\in\Gamma$ 
and 
$Y\in\Gamma'$ is denoted $(X,Y)\in\Gamma\times \Gamma'$.
A state $X\in\Gamma$ may be scaled by a real parameter $t>0$, 
leading to  a state $t X$ in a scaled state space $\Gamma^{(t)}$, 
sometimes 
written 
$t\Gamma$. 
For simple systems the states are
parametrized by the energy coordinate $U\in{\bf R}$ and the work 
coordinates 
$V\in{\bf R}^n$.

The axioms
are grouped as follows:
\bigskip

\noindent
{\bf A. GENERAL AXIOMS}
\medskip

\item{{\bf A1)}}  {\bf Reflexivity.} $X \sima X$.

\item{{\bf A2)}}  {\bf Transitivity.}  $X \prec Y$ and $Y \prec Z$
implies $X \prec Z$.

\item{{\bf A3)}} {\bf Consistency.}  $X \prec X^\prime$ and $Y 
\prec Y^\prime$ implies $(X,Y) \prec
(X^\prime, Y^\prime)$.

\item{{\bf A4)}} {\bf Scaling invariance.}  If $X\prec Y$, then 
$tX \prec tY$ for all $t>0$.

\item{{\bf A5)}}  {\bf Splitting and recombination.}   For $0 < 
t < 1$,  $X \sima (t X, (1-t) X)$.

\item{{\bf A6)}}  {\bf Stability.}   If
$(X, \varepsilon Z_0) \prec (Y, \varepsilon Z_1)$
holds for a sequence of $\varepsilon$'s tending to zero and some states 
$Z_0$, $Z_1$, then $X \prec Y$.

\item{\bf A7)}  {\bf Convex combination.}
Assume $X$ and $Y$ are states in the same state space, $\Gamma$,
that has a convex structure. If $t \in [0,1]$ then 
$ (t X, (1-t) Y) \prec t X + (1-t)Y\ $.

%%%%%%%%%%%%%%%%%%%
\medskip
\noindent
{\bf B. AXIOMS FOR SIMPLE SYSTEMS}
\medskip

Let $\Gamma$, a convex subset of ${\bf R}^{n+1}$ for some $n>0$, 
be the state space of a simple system.

\item{{\bf S1)}} {\bf Irreversibility.} For each $X \in \Gamma$ there
is a point $Y \in \Gamma$ such that $X \prec\prec Y$. (Note: This
axiom is implied by T4, and hence it is not really independent.)

\item{{\bf S2)}} {\bf Lipschitz tangent planes.} For each $X\in \Gamma$
the forward sector $A_X=\{Y\in\Gamma:X\prec Y\}$ has a {\it unique} 
support 
plane at $X$ (i.e.,
$ A_X$ has a {\it tangent plane} at $X$).
The slope of the tangent plane is
assumed to be a {\it locally Lipschitz continuous} function of $X$.
\item{{\bf S3)}} {\bf Connectedness of the boundary.} 
The boundary $\partial A_X$  of a forward sector is connected. 

%%%%%%%%%%%%%%%%%%%%%
\medskip
\noindent
{\bf C. AXIOMS FOR THERMAL EQUILIBRIUM}
\medskip

\item {\bf T1)} {\bf Thermal contact.} For any two simple systems with 
state
spaces $\Gamma_1$ and $\Gamma_2$, there is another simple system, the {
\it thermal join}
of $\Gamma_1$ and $\Gamma_2$,  with
state space
$$
\Delta_{12} = \{ (U,V_1,V_2) : U=U_1+U_2 \;{\rm with}\; (U_1,V_1)\in 
\Gamma_1,
(U_2,V_2)\in \Gamma_2\}. 
$$
Moreover, 
$$
\Gamma_1\times \Gamma_2 \ni ((U,V_1), \ (U_2,V_2)) \prec
(U_1+U_2, V_1,V_2) \in \Delta_{12}.
$$

\item{\bf T2) } {\bf Thermal splitting.}
For any point $(U,V_1,V_2) \in \Delta_{12}$ there is at least
one pair of
states, $(U_1,V_1) \in \Gamma_1$, $(U_2,V_2))\in
\Gamma_2$, with $U=U_1+U_2$,  such that
$$
(U,V_1,V_2)\sima ((U_1,V_1), (U_2,V_2)).
$$
In particular, if
$(U,V)$ is a state of a simple system $\Gamma$ and 
$\lambda\in[0,1]$ then
$$
(U,(1-\lambda)V,\lambda V) \sima
(((1-\lambda)U,(1-\lambda)V),(\lambda U,\lambda V)) \in 
\Gamma^{(1-\lambda)} \times \Gamma^{(\lambda)}.
$$

If $(U,V_1,V_2)\sima ((U_1,V_1), (U_2,V_2))$ we write
$(U_1,V_1)\simt (U_2,V_2)$.

\item{\bf T3)} {\bf Zeroth law.} If $X\simt Y$ and if 
$Y\simt Z$
then $X\simt Z$.

\item {\bf T4)} {\bf Transversality.} If $\Gamma$ is the state space of a
simple system and if $X \in \Gamma$, then there exist states $X_0\simt
X_1$ with
$X_0\prec\prec X\prec\prec X_1$.

\item {\bf T5)} {\bf Universal temperature range.} If $\Gamma_1$ and 
$\Gamma_2$
are state spaces of simple systems then,  for every $X\in\Gamma_1$ and 
every
$V$ in the projection of $\Gamma_2$ onto the space of its 
work
coordinates,  there is a $Y\in\Gamma_2$ with work coordinates $V$ such 
that 
$X\simt Y$.

\medskip
\noindent
{\bf D. AXIOM FOR MIXTURES AND REACTIONS}
\medskip

Two state spaces, $\Gamma$ and $\Gamma'$ are said to be connected, 
written
$\Gamma \prec \Gamma'$, 
if there are state spaces $\Gamma_0$, $\Gamma_1$, $\Gamma_2$,...,
$\Gamma_N$ and states $X_i \in \Gamma_i$ and $Y_i \in  \Gamma_i$, for
$i=1,...,N$ and states $\tilde X\in \Gamma $ and $\tilde Y\in \Gamma'$ 
such
that
$(\tilde X, X_0) \prec Y_1 $, 
$X_i \prec Y_{i+1} $ for $i=1,...,N-1$, and 
$X_N \prec (\tilde Y, Y_0) $.
\medskip

\item{\bf M)} {\bf Absence of sinks.} If $\Gamma$ is
connected to $\Gamma'$ then $\Gamma'$ is connected to $\Gamma$, i.e.,
$\Gamma \prec \Gamma' \Longrightarrow \Gamma' \prec \Gamma$. 
\bigskip 

The main goal of the paper is to derive the {\bf entropy principle}
(EP) from these properties of $\prec \ $:
\medskip
 
{\it There is a function, called {\bf entropy} and denoted by $S$, 
on all states of all simple and compound systems,  such that

\item{a)} \underbar{{\tt Monotonicity:}} 
If $X\prec\prec Y$, then $S(X)<S(Y)$, and if $X\sima Y$, then 
$S(X)=S(Y)$.
\item{b)} \underbar{\tt Additivity and extensivity:} 
$S((X,X')) = S(X) + S(X')$ and $S(t X)=t S(X)$.}
\medskip

Differentiability of $S$ as function of the energy and work 
coordinates of simple systems is also proved and  {\it 
temperature} is derived from entropy.
\smallskip

A central result on our road to the EP is a proof, from our axioms, of 
the {\it comparison hypothesis} (CH) for simple and compound systems, 
which says that for any two states $X, Y$ in the {\it same} state 
space either $X\prec Y$ or $Y\prec X$ holds.  This is stated in 
Theorem 4.8.  The existence of an entropy function is discussed 
already in Section II on the basis of Axioms A1-A6 alone {\it 
assuming} in addition CH.  In the subsequent sections CH is {\it derived} 
from 
the other axioms.  The main steps involved in this derivation of CH 
are as follows.

The comparison hypothesis (which, once proved,  is more 
appropriately called the {\it comparison principle}) is first derived for
simple systems in Theorem 3.7 in Sect.~III.  This proof uses both the 
special
axioms S1-S3 of Sect.\ III  and the general axioms A1-A7 introduced in 
Sect.\ 
II.
On the other hand, it should be stressed that Theorem 3.7 is
independent of the discussion in Sect.\ II D-E, where an entropy function 
is
constructed, assuming the validity of CH.

The extension of CH to compound systems relies heavily on the axioms
for thermal equilibrium that are discussed in Sect.\ IV.  The key point
is that by forming the thermal join of two simple systems we obtain a
new simple system to which Theorem 3.7 can be applied.  The extension of
CH from simple to compound systems is first carried out for products
of scaled copies of the {\it same} simple system (Theorem 4.4). Here the
transversality axiom T4 plays an essential role by reducing the
consideration of states of the compound system that are not in thermal
equilibrium to states in the thermal join.

The proof of CH for products of {\it
different} simple systems requires more effort. The main step here is to
prove the existence of \  `entropy calibrators'\   (Theorem 4.7). This 
says 
that for
each pair of simple systems $\Gamma_1,\Gamma_2$ there are exist four 
states,
$X_0, X_1\in\Gamma_1$, $Y_0, Y_1\in\Gamma_2$ such that $X_0\prec\prec 
X_1$,
$Y_0\prec\prec Y_1$, but $(X_0, Y_1)\sima (X_1, X_0)$. In establishing 
this
property,  we find it convenient to make use of the existence of an 
entropy
function for each of the spaces $\Gamma_1$ and $\Gamma_2$ separately, 
which, 
as
shown in Sects. II D-E, follows from axioms A1-A6  and the already 
established
property CH for products of scaled copies of the {\it
same} simple system. 

Once CH has been established for arbitrary products of
simple systems the entropy principle for all 
adiabatic state changes, except
for mixing of different substances and chemical reactions, follows from
the considerations of Sects.\ II D-E. An explicit formula for $S$
is given in Eq. (2.20): We pick a reference system with two states
$Z_0\prec\prec Z_1$, and for each system $\Gamma$ a reference point
$X_\Gamma\in\Gamma$ is chosen in such a way that $X_{t\Gamma}=tX_\Gamma$
and $X_{\Gamma_1\times \Gamma_2}=(X_{\Gamma_1}, X_{\Gamma_2})$. Then, for
$X\in \Gamma$, $$ S(X) = \sup \{ \lambda \, \, : \, \, (X_{\Gamma},
\lambda Z_1) \prec (X , \lambda Z_0) \}.  $$
(For $\lambda <0$, $(X_{\Gamma}, \lambda Z_1)\prec (X , \lambda Z_0)$ 
means,
per definition, that $(X_{\Gamma}, -\lambda Z_0)\prec (X , -\lambda 
Z_1)$,
and for $\lambda=0$ that $X_{\Gamma}\prec X$.)

In Section V we prove that for a simple system the entropy function is a 
once continuously differentiable function of the energy and the work 
coordinates.  The convexity axiom A7, which leads to concavity of the 
entropy, and the axiom S2 (Lipschitz tangent planes) are essential here.  
We prove that the usual thermodynamic relations hold, in particular 
$T=(\partial S/ \partial U)^{-1}$ {\it defines} the absolute temperature.  
Up to this point neither temperature nor hotness and coldness have 
actually 
been used.  In this section we also prove (in Theorem 
5.6) that the entropy for every simple 
system is uniquely determined, up to an affine change of scale, by the 
level sets of $S$ and $T$, i.e., by the adiabats and isotherms regarded
only as sets, and without numerical values.

In the final Section VI we discuss the problem of fixing the additive 
entropy constants when processes that change the system by mixing and 
chemical reactions are taken into account. We show that, even without 
making any assumptions about the existence of unrealistic semi-permeable 
membranes, it is always possible to fix the constants in such a way that 
the entropy remains additive, and never decreases under adiabatic 
processes. This is not quite the full entropy principle, since there 
could 
still be states with $X\prec\prec Y$, but $S(X)=S(Y)$. This abnormal 
possibility, however, is irrelevant in practice, and we give a necessary 
and
sufficient condition for the situation to occur that 
seems to be realized in nature: The entropy of every substance is 
uniquely determined once an arbitrary entropy constant has been fixed for 
each of the chemical elements, and $X\prec\prec Y$ implies that 
$S(X)<S(Y)$. 

After this summary of the logical structure of the paper we 
add some remarks on the relation of our treatment of the second law 
and more conventional formulations, e.g., the classical statements of 
Kelvin, Clausius and Carath\'eodory paraphrased in Sect.\ I.A. What 
immediately strikes the eye is that these classical formulations are 
{\it negative} statements: They claim that certain processes are {\it 
not} possible.  Thus, the Clausius formulation essentially says that 
thermal contact leads to an irreversible process.  On the other hand, 
what the founding fathers seem to have taken for granted, is that 
there also exist {\it reversible} processes.  Thus the Clausius 
inequality, $\int\delta Q/T\leq 0$, which ostensibly follows from his 
version of the second law and is the starting point for most textbook 
discussions of entropy, does not by itself lead to an entropy function.  
What is needed in this formulation is the existence of reversible 
processes, where {\it equality} holds (or at least processes that 
approximate equality arbitrarily closely).  One might even question 
the possibility of attaching a precise meaning to `$\delta Q$' and 
`$T$' for irreversible processes. (See, however, Eq.\ (5.8) and the 
discussion preceding it, where the symbols are given a precise 
meaning in a concrete situation.)

The basic question we set out to examine is this: Why can adiabatic
processes within a system be exactly characterized by the increase (more
precisely, non-decrease) of an additive entropy function?  In Section
II, where the comparison principle CH is {\it assumed}, an answer is 
already given:
It is because all reasonable notions of adiabatic
accessibility should satisfy axioms A1-A6, and these axioms, {\it
together with CH}, are {\it equivalent} to the existence of an additive
entropy function that characterizes the relation.  This is expressed in
Theorem 2.2.  If we now look at axioms A1-A6 and the comparison
principle we see that these are all {\it positive} statements about the
relation $\prec$: They all say that certain elementary processes are
{\it possible} (provided some other processes are possible), and none of
them says that some processes are {\it impossible}.  In particular, the
trivial case, when everything is accessible form everything else, is not
in conflict with A1-A6 and the comparison principle: It corresponds to a
constant entropy.

{}From this point of view the existence of an entropy function is an
issue that can, to a large extent, be discussed independently of the
second law, as originally formulated by the founders (as given in
Section I.A).  {\it The existence of entropy has more to do with
comparability of states and reversibility than with irreversibility.\/}
In fact, one can conceive of mathematical examples of a relation 
$\prec$ that is characterized by a function $S$ and satisfies A1-A6 
and CH, but $S$ is a constant in a whole neighbourhood of some 
points---and the Clausius inequality fails. 
Conversely, the example of the `world of
thermometers', discussed in Sect.\ IV. D and Fig.\ 7.\  is relevant in
this context.  Here the second law in the sense of Clausius holds, but
the Clausius {\it equality} $\int\delta Q/T=0$ cannot be achieved and
there is no entropy that characterizes the relation for compound
systems!

In our formulation the reversibility required for the definition of
entropy is a consequence of the comparison principle and the stability
axiom A3.  (The latter allows us to treat  reversible processes as
limiting cases of irreversible processes, which are, strictly speaking
the only processes realized in nature.)  This is seen most directly in
Lemma 2.3, which characterizes the entropy of a state in terms of
adiabatic ${\it equivalence}$ of this state with another state in a
compound system.  This lemma depends crucially on CH (for the compound
system) and A3.

So one may ask what, in our formulation, corresponds to the negative 
statements in the classical versions of the second law.  The answer 
is: It is axiom S1, which says that from every state of a simple 
system one can start an irreversible adiabatic process.  In 
combination with A1-A6 and the convexity axiom A7, this is {\it 
equivalent} to Carath\'eodory's principle.  Moreover, together with 
the other simple system axioms, in particular the assumption about the 
pressure, S2, it leads to Planck's principle, Theorem 3.4, which 
states the impossibility of extracting energy adiabatically from a 
simple system at fixed work coordinates.  Hence, the entropy not only 
exists, but also {\it it is nowhere locally constant}.  This 
additional property of entropy is a precise version of the classical 
statements of the second law.  By contrast, an entropy having level 
sets like the temperature in Fig.\ 8 would allow the construction of a 
perpetual motion machine of the second kind.

It would be mistake, however, to underestimate the role played by the 
axioms other than S1.  They are all part of the structure of 
thermodynamics as presented here, and conspire to produce an entropy 
function that separates precisely the possible from the impossible and 
has the convexity and regularity properties required in the
practical application of thermodynamics .

\vfill\eject
%%%%%%%%%%%%
\noindent
{\tit LIST OF SYMBOLS  }
\bigskip
\bigskip
{\parindent = 0pt  
{\subt A. Some Standard Mathematical Symbols}
\bigskip

\settabs \+ sBor aBasupset Aa & Functions that vanish
at infinity aaaaaaaaaaaaaaiaaaaaaaaaaaaaaaaa &\qquad (Sect. II.A.1) \cr

\+ $a\in A$ \ or $A\ni a$ &means `the point $a$ is an element of
          the set $A$'. &\cr
\smallskip
\+ $a \not \in A$ \  &means `the point $a$ is not an element of
          the set $A$'. &\cr
\smallskip
\+ $A\subset B$ or $B\supset A$ &means `the set $A$ is in the set $B$'. 
&\cr
\smallskip
\+ $A \cap B$ & is the set of objects that are in the set 
$A$ and in the set $B$.  \cr
\smallskip
\+ $A \cup B$ & is the set of objects that are either in the set 
$A$ or in the set $B$ or in both sets.  \cr
\smallskip
\+ $A \times B$  &is the set consisting of pairs $(a,b)$ with
                  $a\in A$ and $b \in B$. & \cr
\smallskip
\+ $\{ a:P\}$ &means the set of objects $a$ having property $P$. &\cr
\smallskip
\+ $a:=b$ or $b=:a$ &means `the quantity $a$ is defined by $b$'. &\cr 
\smallskip
\+ $P\Rightarrow Q$ &means `$P$ implies $Q$'. &\cr
\smallskip
\+ $\R^n$ &is $n$-dimensional Euclidean space whose points are &\cr
\+ &$n$-tuples $(x_1,...,x_n)$ of real numbers.  &\cr
\smallskip
\+ [s,  t] & means the closed interval $s\leq x \leq t$. \cr
\smallskip
\+ $\partial A$ & means the boundary of a set $A$. \cr
\smallskip

\bigskip
\bigskip
{\subt B. Special Symbols}
\bigskip

\+ $X\prec Y$ &(`$X$ precedes $Y$') means that the state $Y$ is &\cr
\+ &adiabatically accessible from the state $X$. &(Sect. II.A.2)\cr
\smallskip
\+ $X \not\prec Y$ &(`$X$ does not precede $Y$') means that $Y$ is 
    { \it not} adiabatically &\cr
\+ & accessible from $X$. &(Sect. II.A.2) \cr
\smallskip
\+ $X\prec\prec Y$ &(`$X$ strictly precedes $Y$') means that
         $Y$ is adiabatically &\cr
\+ & accessible from $X$, but $X$ is not accessible from $Y $.
&(Sect. II.A.2)\cr
\smallskip
\+ $X\sima Y$ &(`$X$ is adiabatically equivalent to $Y$') means that &\cr
\+ & $X\prec Y$ {\it and } $Y\prec X$. &(Sect. II.A.2) \cr
\smallskip
\+ $X\simt Y$ &means that the states $X$ and $Y$ are in thermal 
equilibrium.
&(Sect. IV.A)   \cr
\smallskip
\+ $A_X$ &the `forward sector' of a state $X\in \Gamma$, i.e., $\{Y\in
\Gamma : X\prec Y\} $. & (Sect. II.F)  \cr
\smallskip
\+ $tX$ & a copy of the state $X$, but scaled by a factor $t$.  &
    (Sect. II.A.1)  \cr
\smallskip
\+ $\Gamma^{(t)}$    & the state space consisting of scaled states $tX$,
          with $X\in \Gamma$.    &(Sect. II.A.1) \cr
\smallskip
\+ $tX \ + \ (1-t)Y$ & a convex combination of states $X$ and $Y$ in a 
\cr
\+ &   state space with a convex structure. &(Sect. II.F) &\cr
\smallskip
\+ $\Sigma(X_{0},X_{1})$ & the \lq strip\rq\ $\{X\in\Gamma:\ X_{0}\prec 
X\prec X_{1}\}$ between the adiabats 
\cr
\+ & through $X_{0}$ and 
$X_{1}\in\Gamma$, $X_{0}\prec X_{1}$.&(Sect. II.D) \cr
\smallskip
\+ $\uprho_{X}$ & the projection of $\partial A_{X}$ onto 
the space of work coordinates,\cr
\+ &  for $X$ in the state space of a simple 
system $\Gamma\subset \R^{n+1}$,\cr
\+ &   i.e., $\uprho_{X}=\{V\in \R^{n}: 
(U,V)\in \partial A_{X}\ \hbox{\rm for some }U\in\R\}.$
  &(Sect. III.C) \cr
\smallskip
\+ $\rho$ &  the projection onto the space of work coordinates 
 of a simple system $\Gamma$,  \cr
\+ &  i.e., if $X=(U,V)\in\Gamma$, then 
$\rho(X)=V$.
  &(Sect. IV.A) \cr
}

\vfill\eject
%%%%%%%%%%%%%%%%%%%%%%%%%%
\settabs \+ Composition of systems 
aaaaaaaaaaaaaaiaaaaaaaaaaaaaaaaaaaaaaaaaaaaaa
aaa &\qquad (Sect. II.A.1) \cr
\noindent
{\tit INDEX OF TECHNICAL TERMS  }
\bigskip
\bigskip
{\parindent = 0pt  

\+ Additivity of entropy \dotfill\ \
&(Sect. II.B)   \cr
\+ Adiabat\dotfill\ \
&(Sect. III.B)   \cr
\+ Adiabatic accessibility \dotfill\ \
&(Sect. II.A.2)   \cr
\+ Adiabatic equivalence\dotfill\ \
&(Sect. II.A.2)   \cr
\+ Adiabatic process \dotfill\ \
&(Sect. II.A.1)   \cr
\+ Boundary of a forward sector\dotfill\ \
&(Sect. III.B)   \cr
\+ Canonical entropy\dotfill\ \
&(Sect. II.D)   \cr
\+ Cancellation law\dotfill\ \
&(Sect. II.C)   \cr
\+ Carath\'eodory's principle\dotfill\ \ 
&(Sect. II.G)   \cr
\+ Carnot efficiency\dotfill\ \
&(Sect. V.A)   \cr
\+ Comparable states \dotfill\ \
&(Sect. II.A.2)   \cr
\+ Comparison hypothesis (CH)\dotfill\ \
&(Sect. II.C)   \cr
\+ Composition of systems\dotfill\ \
&(Sect. II.A.1)   \cr
\+ Consistent entropies\dotfill\ \
&(Sect. II.E)   \cr
\+ Convex state space\dotfill\ \
&(Sect. II.F)   \cr
\+ Degenerate simple system (=thermometer)\dotfill\ \
&(Sect. III.A)   \cr
\+ Entropy \dotfill\ \
&(Sect. II.B)   \cr
\+ Entropy calibrator\dotfill\ \
&(Sect. IV.A)   \cr
\+ Entropy constants\dotfill\ \
&(Sect. II.E)   \cr
\+ Entropy function on a state space\dotfill\ \
&(Sect. II.D)   \cr
\+ Entropy principle (EP)\dotfill\ \
&(Sect. II.B)   \cr
\+ Extensivity of entropy \dotfill\ \
&(Sect. II.B)   \cr
\+ First law of thermodynamics\dotfill\ \
&(Sect. III.A)   \cr
\+ Forward sector\dotfill\ \
&(Sect. II.F)   \cr
\+ Generalized ordering\dotfill\ \
&(Sect. II.D)   \cr
\+ Internal energy\dotfill\ \
&(Sect. III.A)   \cr
\+ Irreversible process\dotfill\ \
&(Sect. II.G)   \cr
\+ Isotherm\dotfill\ \
&(Sect. IV.A)   \cr
\+ Lipschitz continuity\dotfill\ \
&(Sect. III.B)   \cr
\+ Lower temperature\dotfill\ \
&(Sect. V.A)   \cr
\+ Multiple scaled copy\dotfill\ \
&(Sect. II.A.1)   \cr
\+ Planck's principle\dotfill\ \
&(Sect. III.C)   \cr
\+ Pressure\dotfill\ \
&(Sect. III.B)   \cr
\+ Reference points for entropy\dotfill\ \
&(Sect. II.D)   \cr
\+ Second law of thermodynamics\dotfill\ \
&(Sect. II.B)   \cr
\+ Scaled copy\dotfill\ \
&(Sect. II.A.1)   \cr
\+ Scaled product\dotfill\ \
&(Sect. II.A.1)   \cr
\+ Simple system \dotfill\ \
&(Sect. III)   \cr
\+ Stability\dotfill\ \
&(Sect. II.C)   \cr
\+ State\dotfill\ \
&(Sect. II.A.1)   \cr
\+ State space\dotfill\ \
&(Sect. II.A.1)   \cr
\+ Subsystem\dotfill\ \
&(Sect. II.A.1)   \cr
\+ System\dotfill\ \
&(Sect. II.A.1)   \cr
\+ Temperature\dotfill\ \
&(Sect. V.A)   \cr
\+ Thermal contact\dotfill\ \
&(Sect. IV.A)   \cr
\+ Thermal equilibration\dotfill\ \
&(Sect. IV.A)   \cr
\+ Thermal equilibrium\dotfill\ \
&(Sect. IV.A)   \cr
\+ Thermal join\dotfill\ \
&(Sect. IV.A)   \cr
\+ Thermal reservoir\dotfill\ \
&(Sect. V.A)   \cr
\+ Thermal splitting\dotfill\ \
&(Sect. III.C)   \cr
\+ Thermometer (=degenerate simple system)\dotfill\ \
&(Sect. III.A)   \cr
\+ Transversality\dotfill\ \
&(Sect. IV.A)   \cr
\+ Upper temperature\dotfill\ \
&(Sect. V.A)   \cr
\+ Work coordinate\dotfill\ \
&(Sect. III.A)   \cr
\+ Zeroth law of thermodynamics\dotfill\ \
&(Sect. IV.A)   \cr

\vfill\eject

%%%%%%%%%%%%%%%
\leftline{{\tit REFERENCES} }
\parindent=0pt
\bigskip

Arens, R.,  1963, {\it An axiomatic basis for classical thermodynamics,}
J.~Math.~Anal.~and Appl.~{\bf 6}, 207-229.

Bazarow, I.P., 1964, {\it Thermodynamics,} (Pergamon Press, Oxford).

Bernstein, B., 1960, {\it Proof of Carath\'eodory's local theorem
and its global application to thermodynamics,} Jour.~Math.~Phys.
{\bf 1}, 222-264. 

%Bridgman, P.W., 1941, {\it The Nature of Thermodynamics} (Harvard
%University Press, Cambridge)

%Bryan, G.H., 1907, {\it Thermodynamics, an introductory treatise dealing
%mainly with first principles and their applications}, Teubner,
%Leipzig. 

Borchers, H.J., 1981, {\it Some remarks on  the second law of
themodynamics,} Rep.~Math.~Phys.~{\bf 22}, 29-48.

Born, M., 1964, {\it Natural Philosophy of Cause and Chance,} (Dover,
New York).

Born, M., 1921, {\it Kritische Bemerkungen zur traditionellen
Darstellung der Thermodynamik,} Phys. Zeits. {\bf 22},
218-224, 249-254 and 282-286.

Boyling, J.B., 1968, {\it Carath\'eodory's principle and
the existence of global integrating factors,} \hfill\break 
Commun.~Math.~Phys. {\bf 10}, 52-68.

Boyling, J.B., 1972, {\it An axiomatic approach to classical
thermodynamics,} Proc.~Roy.~Soc.~London {\bf A329}, 35-70.

%Bridgman, P.W., 1961, {\it The Nature of Thermodynamics,}
%(Harper Torchbooks).

Buchdahl, H.~A., 1958, {\it A formal treatment of the consequences of
the second law of thermodynamics in Carath\'eodory's formulation,}
Zeits.~f.~Phys. {\bf 152},  425-439.

Buchdahl, H.~A., 1960, {\it The concepts of classical thermodynamics}
Am.~J.~Phys. {\bf 28},  196-201.

Buchdahl, H.~A., 1962, {\it Entropy concept and ordering of states. I,}
Zeits.~f.~Phys. {\bf 168}, 316-321.

Buchdahl, H.~A.~and Greve, W., 1962, {\it Entropy concept and 
ordering of states. II,} Zeits.~f.~Phys. {\bf 168}, 386-391.

Buchdahl, H.~A., 1966, {\it The Concepts of Classical Thermodynamics,}
(Cambridge University Press, Cambridge). 

Buchdahl, H.~A., 1986, {\it On the redundancy of the zeroth law of
thermodynamics,} J.~Phys.~A: Math.~Gen.~ {\bf 19}, L561-L564.

Buchdahl, H.~A., 1989, {\it Reply to commment by J.~Walters
on `On the redundancy of the zeroth law of
thermodynamics',} J.~Phys.~A: Math.~Gen.~ {\bf 22}, 343.

Callen, H.B., 1985, {\it Thermodynamics and an Introduction to 
Thermostatistics,}  (Wiley, New York).

Carnot, S., 1824, {\it Reflexions sur la puissance motrice du feu,}
(Bachelier, Paris). Engl.~transl.,~Fox, R., 1986 {\it 
Reflexions on the motive power of fire ,} (Manchester Univ.~Press).

Carath\'eodory, C., 1909, {\it  Untersuchung \"uber die Grundlagen der
Thermodynamik,} Math.~Annalen {\bf 67}, 355-386.

Carath\'eodory, C., 1925, {\it \"Uber die  Bestimmung der Energie und
der absoluten Temperatur mit Hilfe von reversiblen Prozessen,}
Sitzungsber.~Preuss.~Akad. Wiss., Phys.~Math.~Kl., 39-47.

Clausius, R., 1850, {\it \"Uber die bewegende Kraft der W\"arme und die
Gesetze, welche sich daraus f\"ur die W\"armelehre selbst ableiten
lassen} Annalen der Physik Und Chemie {\bf 79}, 368-397. English
tr. in Kestin, 1976. 
 
Coleman, B.D.~and Owen, D.R., 1974, {\it A mathematical foundation for
thermodynamics,} Arch. Rat. Mech. Anal. {\bf 54}, 1-104.

Coleman, B.D.~and Owen, D.R., 1977, {\it On the thermodynamics of
semi-systems with restrictions on the accessibility of states,} 
Arch. Rat. Mech. Anal. {\bf 66}, 173-181.

Coleman, B.D., Owen, D.R.~and Serrin, J., 1981, {\it The second law
of thermodynamics for systems with approximate cycles,}
Arch. Rat. Mech. Anal. {\bf 77}, 103-142.

Cooper, J.L.B., 1967, {\it The foundations of thermodynamics,} 
Jour.~Math.~Anal.~and Appl. {\bf 17}, 172-193. 

Dafermos, C., 1979 {\it The second law of thermodynamics and stability,}
Arch. Rat. Mech. Anal. {\bf 70}, 167-179.

Day, W.~A., 1987, {\it A comment on a formulation of the second law of
thermodynamics,}  Arch. Rat. Mech. Anal. {\bf 98}, 211-227.

Day, W.~A., 1988, {\it A Commentary on Thermodynamics,} (Springer, New
York).

Dobrushin, R.L.,  and Minlos, R.A., 1967, {\it Existence and continuity
of pressure in classical statistical mechanics,}
English trans.: Theory of Probability and its Applications, {\bf 12},
535-559.

Duistermaat, J.~J., 1968, {\it Energy and entropy as  real
morphisms for addition and order,} Synthese {\bf 18}, 327-393.

Falk G.~and Jung H., 1959, {\it Axiomatik der Thermodynamik} in
{\it Handbuch der Physik,} {\bf III/2}, S.~Fl\"ugge ed.,  pp.~199-175.

Feinberg, M.~and Lavine, R., 1983, {\it Thermodynamics based on the
Hahn-Banach theorem: the Clausius inequality,} 
Arch.~Rat.~Mech.~Anal. {\bf 82}, 203-293. 

Fermi, E., 1956, {\it Thermodynamics,} (Dover, New York).

Fisher, M.E. and Milton, G.W., 1983, {\it Continuous fluids with a
discontinuity in the pressure,} J. Stat. Phys. {\bf 32}, 413-438.
See also {\it Classifying first order phase transitions,} Physica
{\bf 138A}, 22-54 (1986).  

%Fong, P., 1963, {\it Foundations of Thermodynamics,} (Oxford
%Univ.~Press, New York).

%Fuchs, L., 1963, {\it Partially Ordered Algebraic Systems,}
%(Addison-Wesley, Reading).

Gibbs, J.W., 1928, {\it On the equilibrium of heterogeneous
substances} in {\it Collected Works of  J. Willard Gibbs, Vol. 1}
(Longmans, New York), pp.~55-349.

Giles, R., 1964, {\it Mathematical Foundations of Thermodynamics,}
(Pergamon, Oxford).

%Ginibre, J., 1970, {\it Some applications of functional integration in
%statistical mechanics,} in {\it Statistical Mechanics and Quantum Field
%Theory}, Les Houches lectures,  C. De Witt and R. Stora eds. pp.
%385-390 (Gordon and Breach). 

Green, A.H.~and Naghdi, P.M., 1978, {\it The second law of thermodynamics 
and cyclic processes,} J. Appl. Mech. {\bf 45}, 487-492.

R.B. Grifiths, 1972, {\it Rigorous results and theorems,} in
{\it Phase Transitions and Critical Phenomena,} vol. 1, Sec. IX.A,
C. Domb and J. Lebowitz eds.,  (Academic Press, NY). 

Guggenheim, E.A., 1933, {\it Modern Thermodynamics by the Methods of
Willard Gibbs} (Methuen, London). 

Gurtin, M.E., 1975, {\it Thermodynamics and stability,} 
Arch. Rat. Mech. Anal. {\bf 59}, 63-96.

Hardy, G.H.,  Littlewood, J.E.,  and Polya, G., 1934,
{\it Inequalities}, Cambridge University Press.

Hornix, W. J. 1970, {\it An axiomatization of classical
phenomenological thermodynamics,} in {\it A Critical Review 
of Thermodynamics}, A.J.~Brainard, E.B.~Stuart, B.~Gal-Or eds. 
(Mono Book Corp., Baltimore) pp.~235-253.

Kestin, J., 1976, {\it The Second Law of Thermodynamics,} 
Benchmark Papers on Energy/5,
(Dowden, Hutchinson and Ross, Stroudsburg, PA).

Landau, L.D. and Lifschitz, E.M., 1969, {\it Course of 
Theoretical Physics, vol. 5, Statistical Physics}, trans. by
E.M. Sykes and M.J. Kearsley,  (Addison Wesley, Reading). 

Landsberg, P.T., 1956, {\it Foundations of thermodynamics,} 
Rev.~Mod.~Phys.~{\bf 28}, 363-392.

%Landsberg, P.T., 1961, {\it On suggested simplifications of
%Carath\'eodory's thermodynamics,} Phys.~Status Solidi   {\bf 1},
%120-126.

%Landsberg, P.T., 1983, {\it Born's centenary: remarks about classical 
%thermodynamics,} Am.~J.~Phys. {\bf 51} 842-845.

Lewis, G.~N. and Randall, M., 1923, {\it Thermodynamics and
the Free Energy of Chemical Substances,} (McGraw-Hill, New York).

Lieb, E.H. and Yngvason, J., 1998, {\it A guide to entropy and the second
law of thermodynamics,} Notices of the Amer. Math. Soc. {\bf 45}
571-581.

Macdonald, A., 1995, {\it A new statement of the second law of 
thermodynamics}
Am.\ J.\ Phys.\ {\bf 63}, 1122-1127 (1995)

Man, C.-S., 1989, {\it Thermodynamics based on a work inequality,}
Arch. Rat. Mech. Anal. {\bf 106}, 1-62.

Owen, D.R., 1984, {\it A First Course in the Mathematical Foundations of 
Thermodynamics,} (Springer, Heidelberg).

Pitteri, M., 1982, {\it Classical thermodynamics of homogeneous
systems based upon Carnot's general axioms,} 
Arch.~Rat.~Mech.~Anal.~{\bf 80}, 333-385.

Planck, M., 1926, {\it \"Uber die Begr\"undung des zweiten Hauptsatzes 
der
Thermodynamik,}
Sitzungsber.~Preuss.~Akad. Wiss., Phys.~Math.~Kl.,
453-463.

%Planck, M., 1954, {\it Thermodynamik} (de Gruyter, Berlin).

Rastall, P., 1970, {\it Classical thermodynamics simplified}
J.~Math.~Phys. {\bf 11}, 2955-2965.

%Redlich, O. 1968, {\it  Fundamental thermodynamics since Carath\'eodory}
%Rev.~Mod.~Phys. {\bf 40} 556-563. 

Reed, M. and Simon, B., 1972, {\it Methods of Modern Mathematical
Physics}, volume 1, (Academic Press, New York). 

Roberts, F.~S. and Luce, R.~D., 1968, {\it Axiomatic thermodynamics
and extensive measurement}, Synthese {\bf 18}, 311-326.

Serrin, J., 1983, {\it The structure and laws of thermodynamics,}
Proceedings of Inter. Congress of Math., Warsaw, Aug. 16-24, 1717-
1728.

Serrin, J., ed., 1986, {\it New Perspectives in Thermodynamics,}
(Springer, Berlin). 

Serrin, J., 1979, {\it Conceptual analysis of the classical second
laws of thermodynamics,} Arch. Rat. Mech. Anal. {\bf 70},
355-371 Arch. Rat. Mech. Anal. {\bf 80}, 333-385.

\v Silhav\'y, M., 1997, {\it The Mechanics and Thermodynamics of
Continuous Media,} (Springer, Heidelberg).

%ter Horst, H.~J., 1987, {\it Fundamental functions in equilibrium 
%thermodynamics,} Ann. of Phys., NY {\bf 176} 183-217.

Thirring, W., 1983, {\it A course in mathematical physics,}   vol. 4,
Remark 2.3.30, (Springer, New York). 

Thomson, W., (Lord Kelvin), 1849, {\it An account of Carnot's theory of
the motive power of heat; with numerical results deduced from
Regnault's experiments on steam,} Trans. Roy. Soc. Edinburgh, {\bf 
16},
541-574.

Tisza, L., 1966, {\it Generalized Thermodynamics,} (M.I.T.~Press,
Cambridge).

Truesdell, C.A.~and Bharata, S., 1977, {\it The Concepts and Logic  of
Classical Thermodynamics as a Theory of Heat Engines,}
(Springer, Heidelberg).

Truesdell, C.A., 1980, {\it The Tragicomical History of Thermodynamics,
 1822-1854,} (Springer, New York).

Truesdell, C.A., 1984, {\it Rational Thermodynamics,} (Springer, New
York).

%Turner, L.A., 1960, {\it Simplification of Carath\'eodory's treatment of
%thermodynamics,} Amer. J. Phys. {\bf 28} 781-786.

Walter, J., 1989, {\it On H.~Buchdahl's project of a thermodynamics
without empirical temperature as a primitive concept,} 
J. Phys. A: Math. Gen. {\bf 22}, 341-342.

Wightman, A.S., 1979, {\it Convexity and the notion of equilibrium
states in thermodynamics and statistical mechanics} in R.H. Israel,
{\it Convexity in the Theory of Lattice Gases}, (Princeton University
Press, Princeton).

Zeleznik, F.J., 1976, {\it Thermodynamics,} J. Math. Phys. {\bf 17},
1579-1610.

\vfill\eject

%%%%%%%%%%%%%
%%%%%%%%%%%%%

\vbox to \vsize{
\noindent
%{\tit FIGURE CAPTIONS}
\bigskip

\vfill
\hbox to \hsize{\hfill
\epsfxsize=\hsize
\epsffile{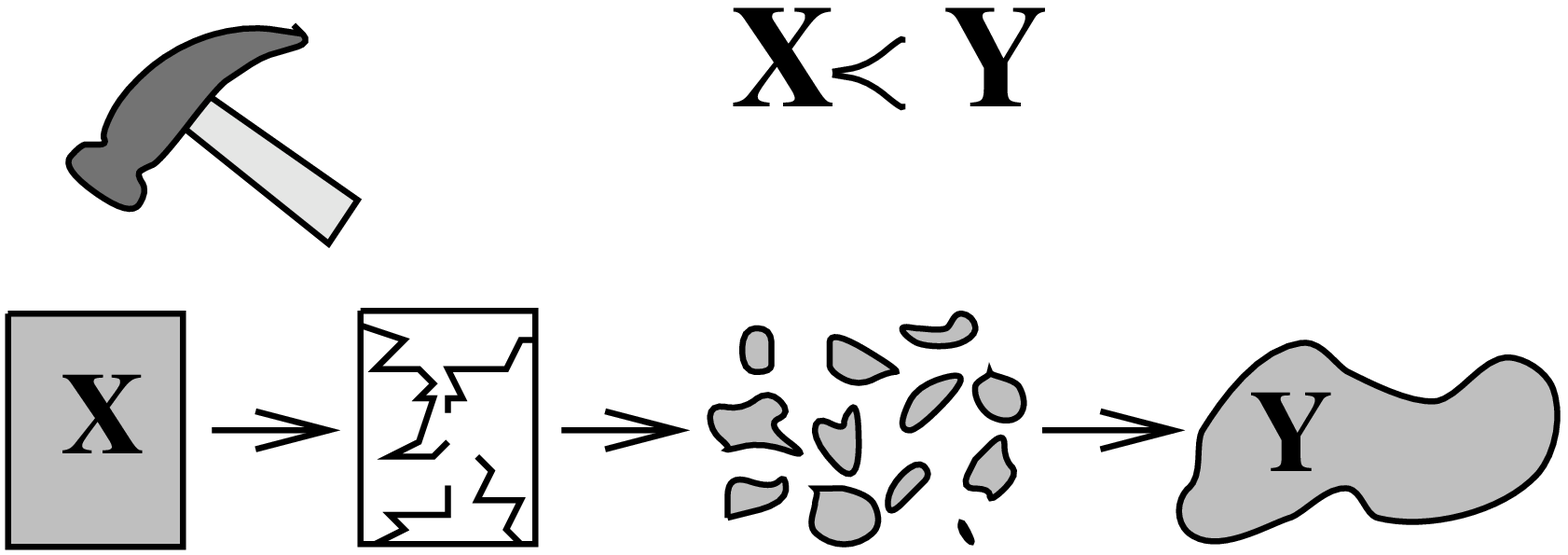}\hfill}
\vfill
\noindent
{\bf Figure 1.} An example of a violent adiabatic process. The system in an
equilibrium state $X$ is transformed by mechanical means to another
equilibrium state $Y$.}
\eject

\vbox to \vsize{
\vfill
\epsfxsize=\hsize
\epsffile{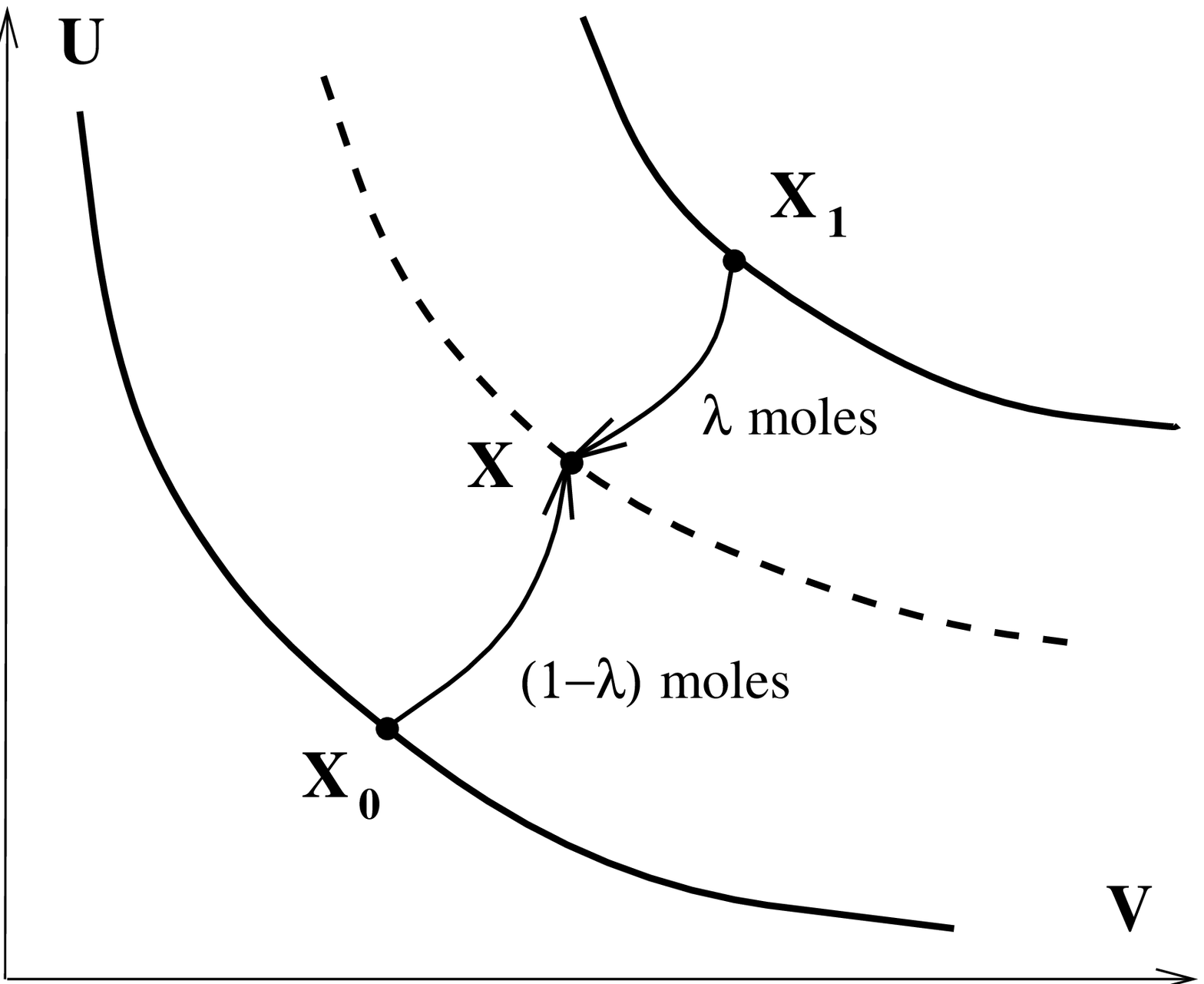}
\vfill
\noindent
{\bf Figure 2.} The entropy of a state $X$ is determined, according to 
formula 2.14, by the amount of substance in state $X_1$ that can be
transformed to $X$ with the aid of a complementary amount of substance
in the state $X_0$.}
 
\vbox to \vsize{
\vfill
\epsfxsize=\hsize
\epsffile{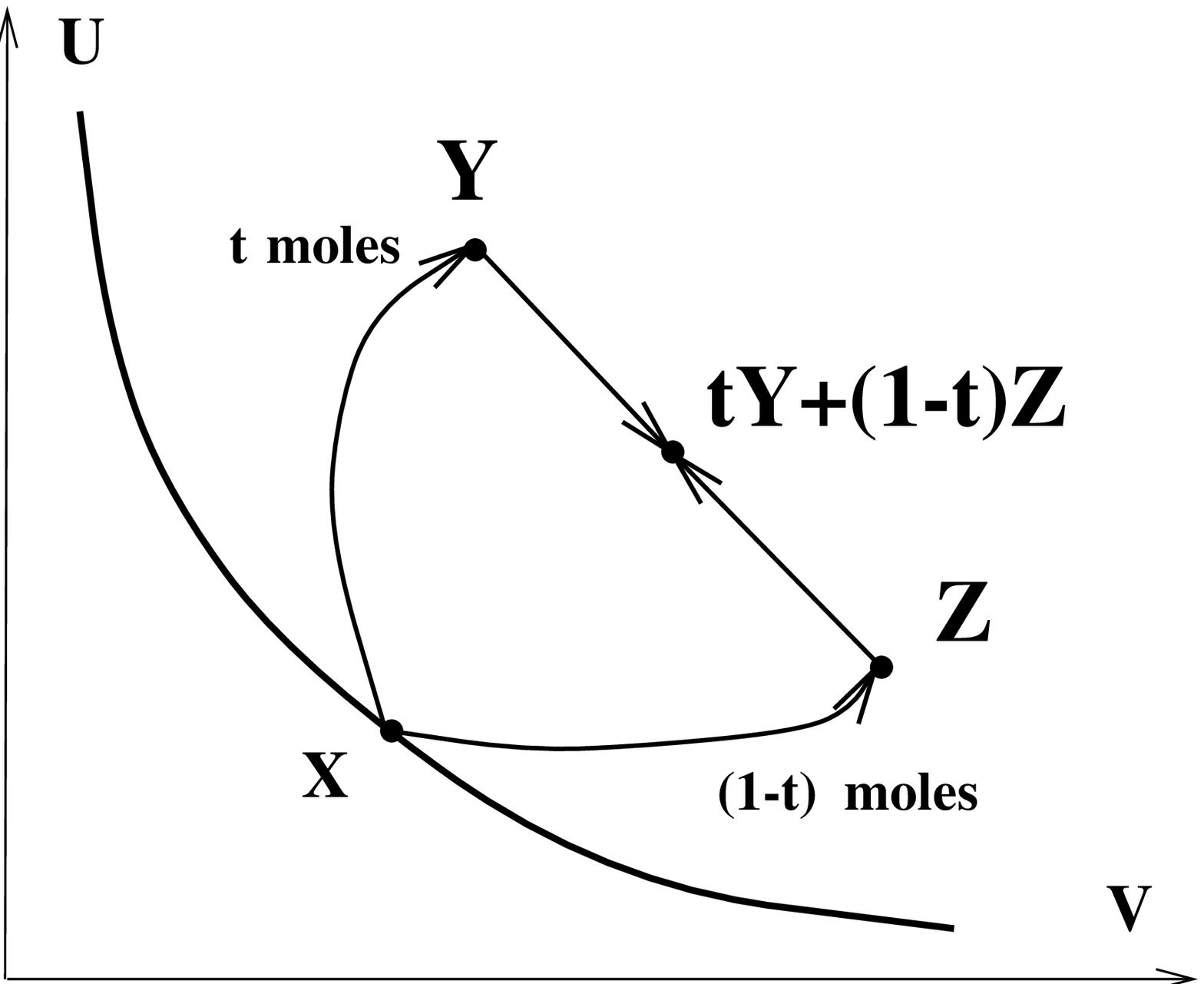}
\vfill
\noindent {\bf Figure 3.} This illustrates axiom A7 and Theorem 2.6
which says that if states $Y$ and $Z$ can be reached adiabatically from
a state $X$ and if the state space has a convex structure then convex
combinations of $Y$ and $Z$ are also in the forward sector of $X$.}  

\vbox to \vsize{
\vfill
\epsfxsize=\hsize
\epsffile{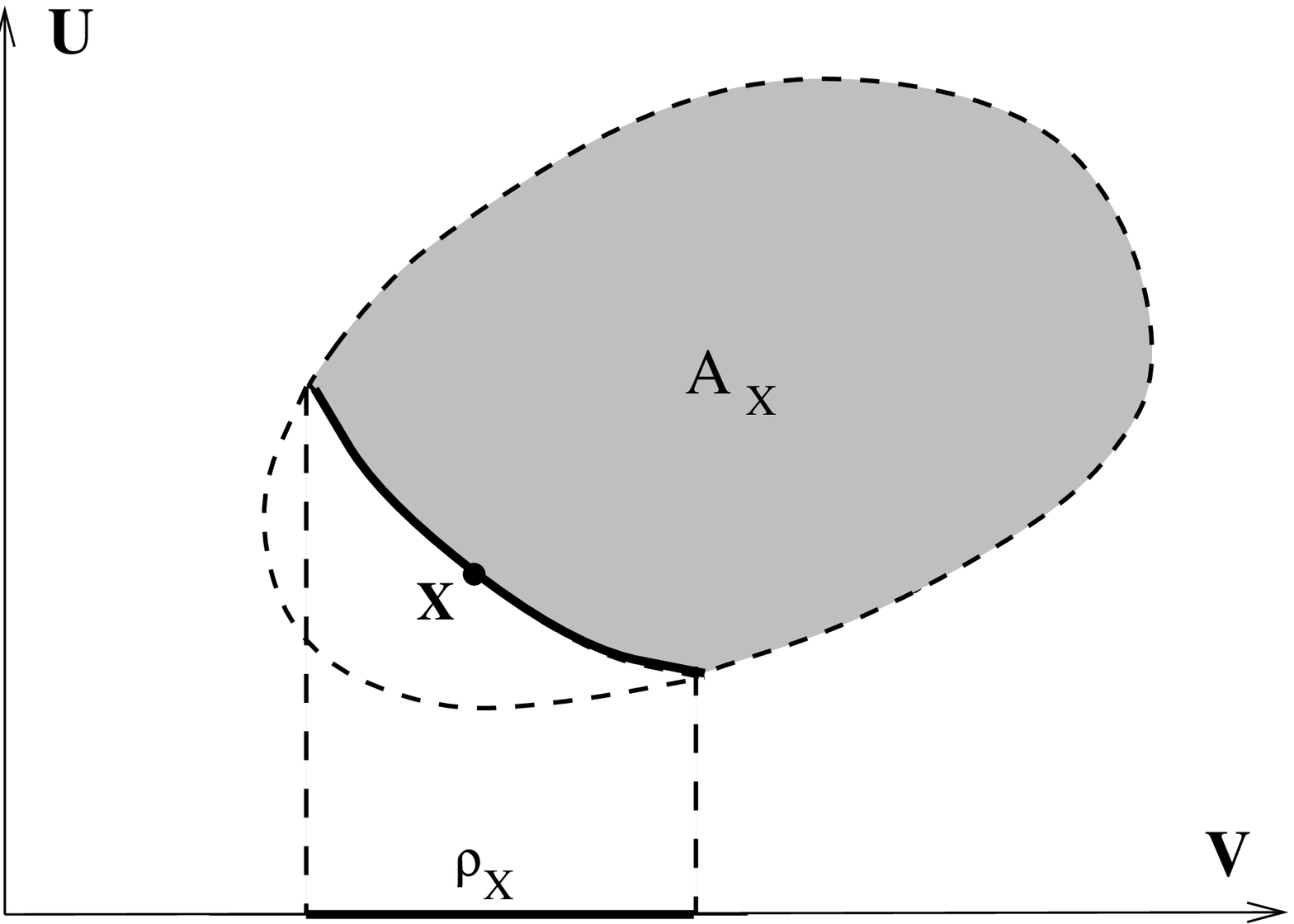}
\vfill
\noindent
{\bf Figure 4.} This illustrates the energy $U$ and work coordinates
$V$ of a simple system. The state space (dashed line) is always a convex
set and the forward sector $A_X$ of any point $X$ is always a convex
subset of the state space. The heavy dark curve denotes the 
boundary $\partial A_X$ of $A_X$ and consists of points that are
adiabatically equivalent to $X$ (as Theorem 3.7 states). The projection
of this boundary on the work coordinates is $\uprho_X$ which can be
strictly smaller than the projection of $A_X$.} 

\vbox to \vsize{
\vfill
\hbox to\hsize{\epsfysize=0.8\vsize
\hfil\epsffile{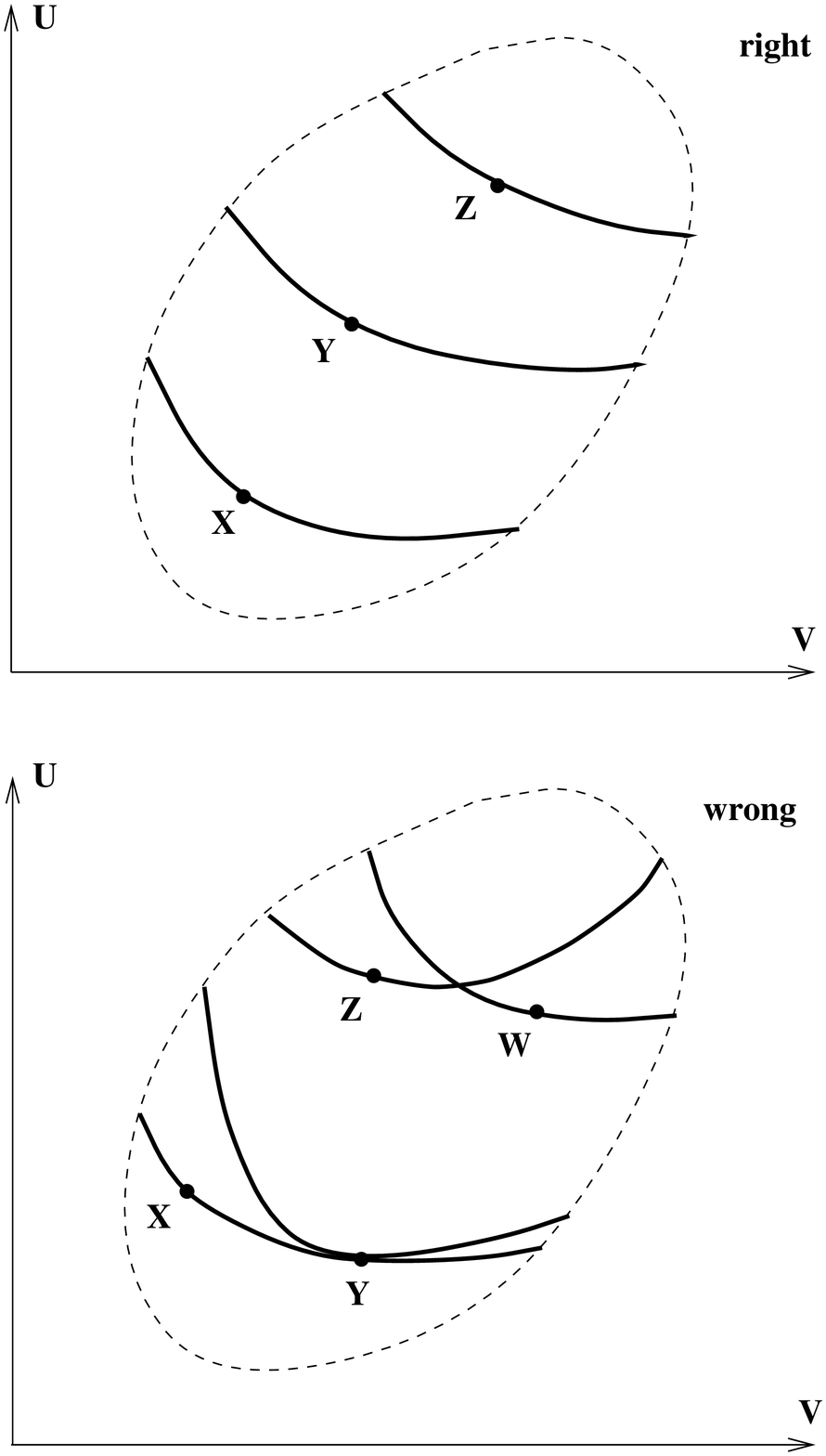}\hfil}
\vfill
\noindent {\bf Figure 5.} The top figure illustrates how the forward
sectors of a simple system are nested. The adiabats (i.e., the
boundaries of the forward sectors) do not overlap. The 3 points are
related by $X\prec\prec Y\prec\prec Z$. The lower figure shows what, in
principle, could go wrong---but doesn't, according to Theorem 3.7.  The
top pair of adiabats have a point in common but neither $W\prec Z $ nor
$Z\prec W$ holds.  The bottom pair is a bit more subtle; $X\prec Y$ and
$Y$ is on the boundary of the forward sector of $X$, but $X$ is not in
the forward sector of $Y$.}  

\vbox to \vsize{
\vfill
\epsfxsize=\hsize
\epsffile{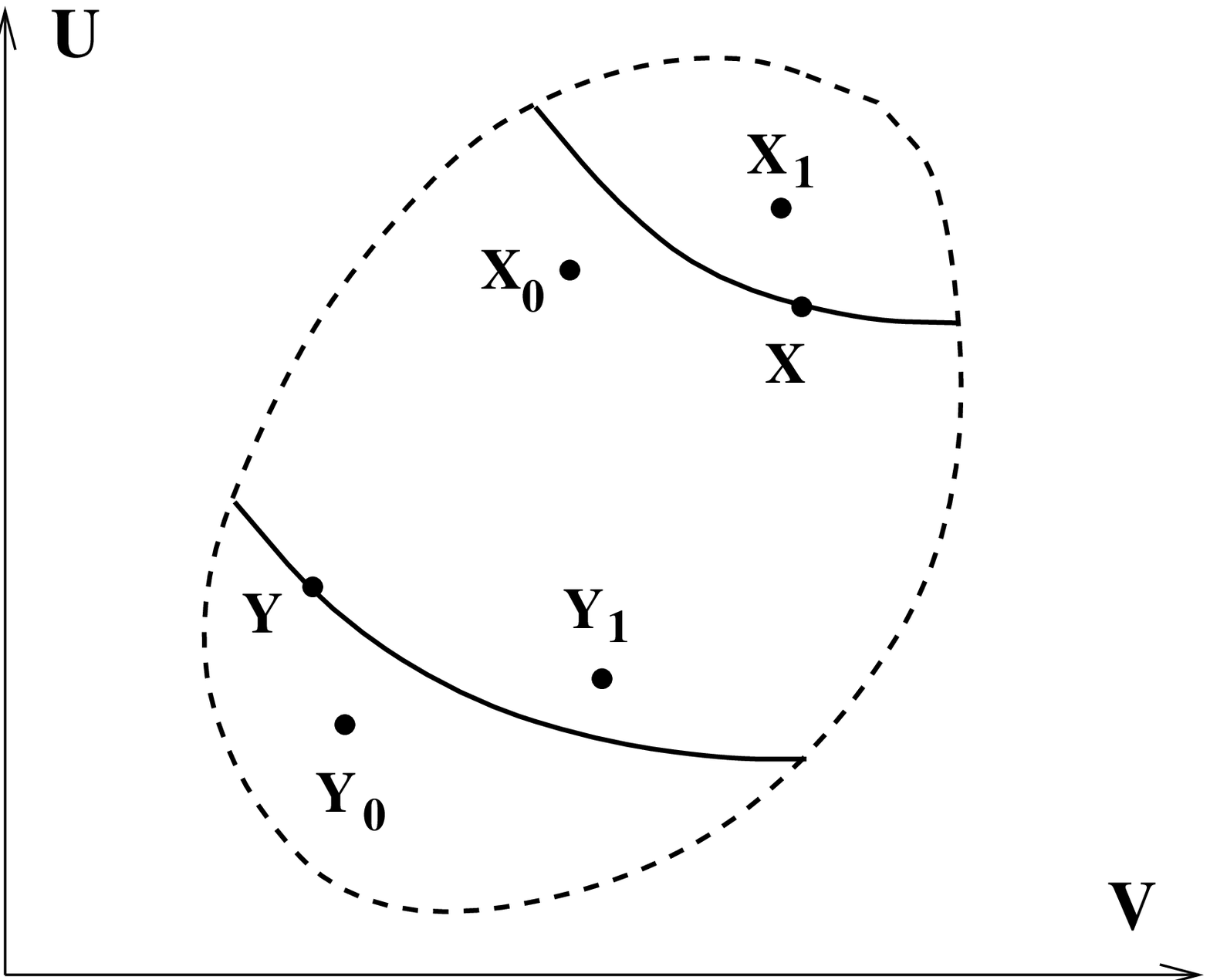}
\vfill
\noindent
{\bf Figure 6.}  This illustrates the transversality
axiom T4. For every state $X$ there are points $X_0 $ and 
$X_1$ on both sides of the adiabat through $X$ that are in 
thermal equilibrium with each other. The points $Y_0$ and 
$Y_1$ (corresponding to some other point $Y$) need not be in 
thermal equilibrium with $X_0$ and $X_1$.}

\vbox to \vsize{
\vfill
\epsfxsize=\hsize
\epsffile{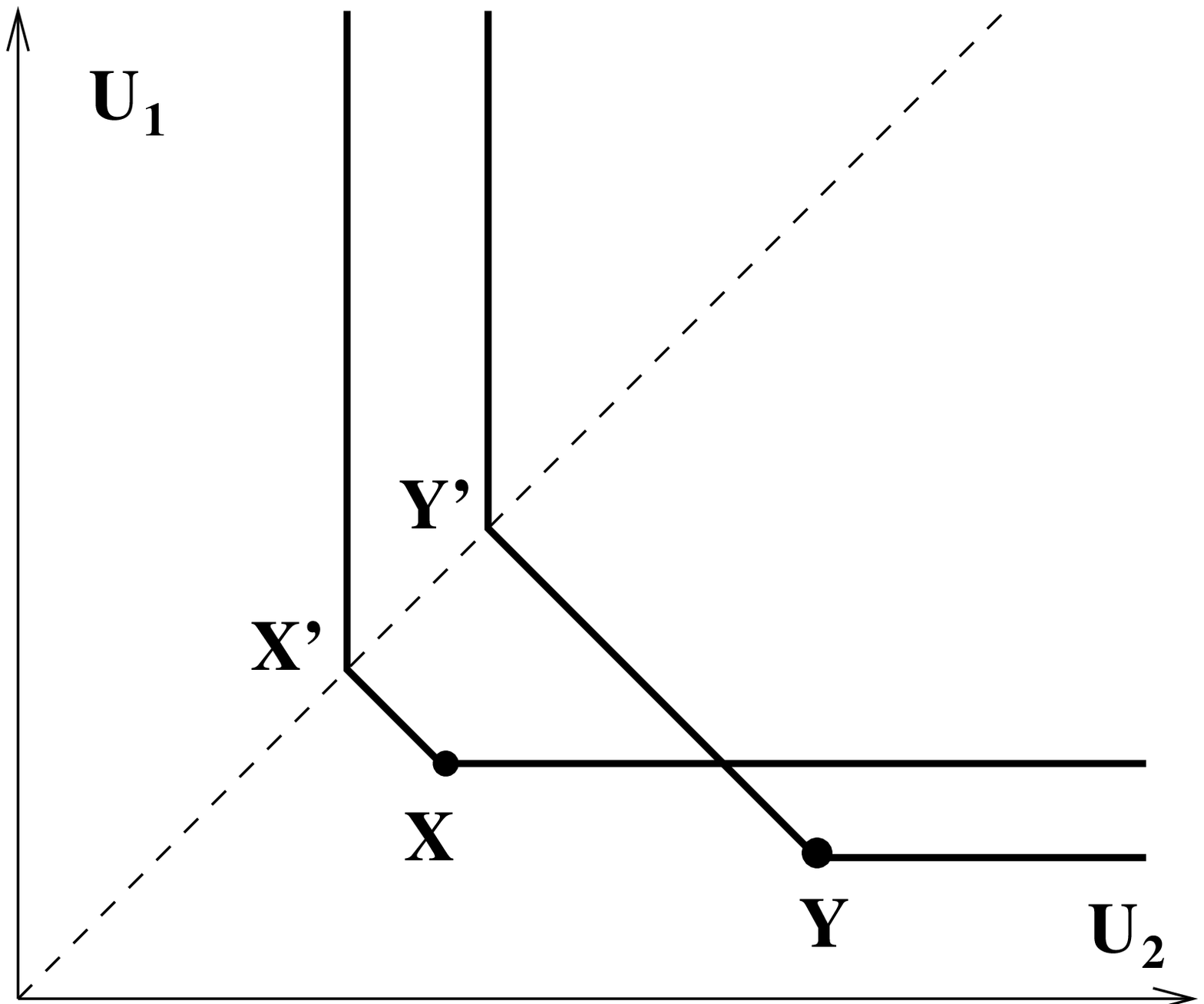}
\vfill
\noindent 
{\bf Figure 7.} This shows the state space of two
`thermometers', which means that there are only energy coordinates. The
forward sectors of $X$ and $Y$ are shown under the assumption that the
only allowed adiabatic operations are thermal equilibration (which moves
$X$ to $X'$ and $Y$ to $Y'$) and rubbing (which increases, but never
decreases the energy).  We see clearly that these sectors are not nested
(i.e., one does not lie inside the other), as they are for compounds of
simple systems, satisfying the transversality axiom T4.}

\vbox to \vsize{
\vfill
\epsfxsize=\hsize
\epsffile{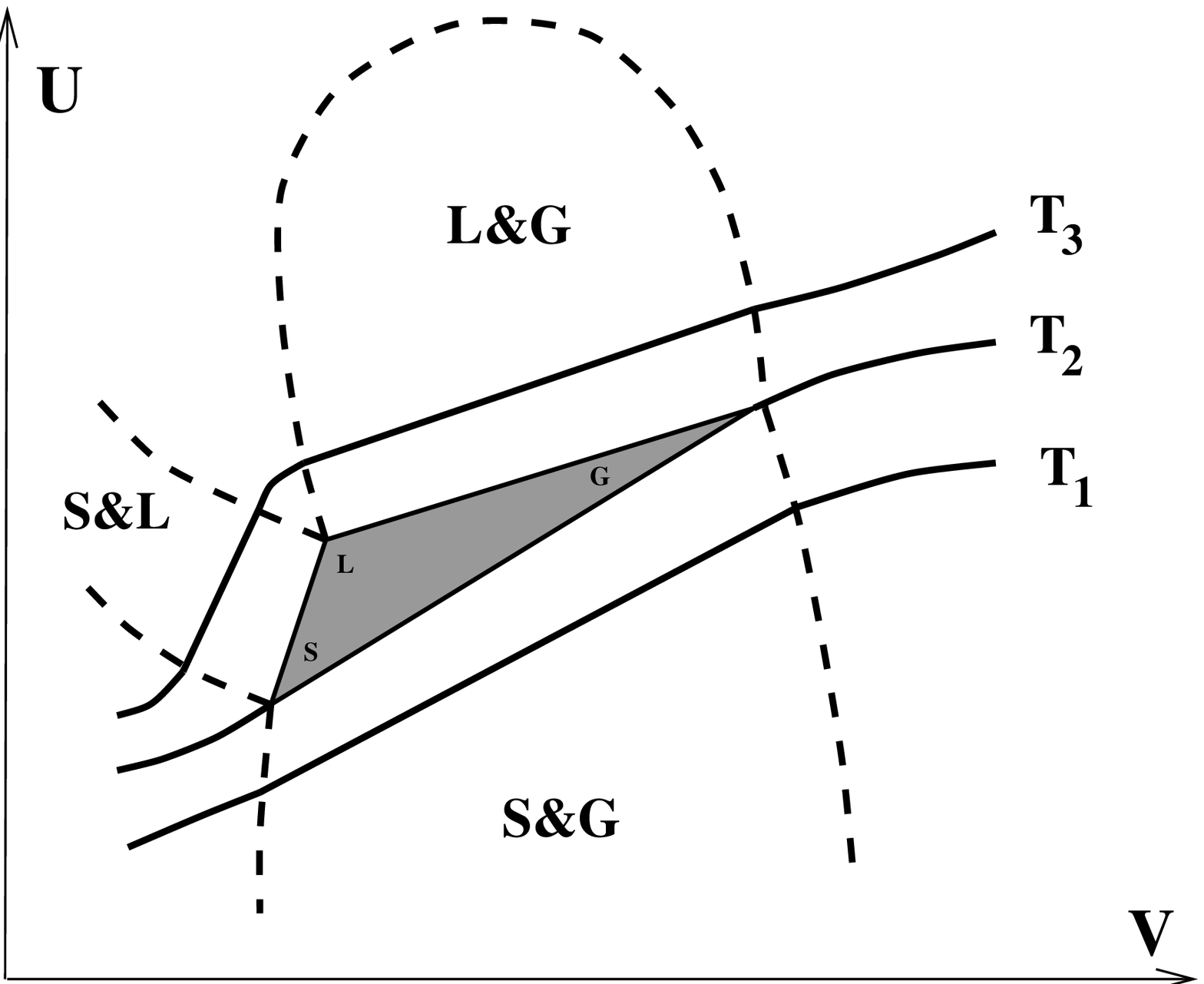}
\vfill
\noindent 
{\bf Figure 8.} This shows isotherms in the $(U,V)$ plane near
the triple point of a simple system. If one substituted pressure or
temperature for $U$ or $V$ as a coordinate then the full two-dimensional
region would be compressed into a one-dimensional region.  In the triple
point region the temperature is constant, which shows that isotherms
need not be one-dimensional curves.} 
%%%%%%%%%

\end